\documentclass[10pt,twocolumn,aps,superscriptaddress,showpacs,floatfix,prd,longbibliography]{revtex4-2}
\usepackage[body={19.cm,24.cm}]{geometry}     
\usepackage{mathrsfs}
\usepackage{amsmath}
\usepackage{array}
\makeatletter
\@addtoreset{equation}{section}
\renewcommand{\theequation}{\arabic{section}.\arabic{equation}}
\makeatother
\usepackage{graphicx}
\usepackage[normalem]{ulem}
\usepackage{xcolor}
\usepackage{bm}
\usepackage{longtable}
\usepackage{slashed}
\usepackage{enumitem}
\usepackage{empheq}
\usepackage{booktabs}
\usepackage{multirow}

\usepackage{titlesec}
\renewcommand{\thesection}{\Roman{section}}
\titleformat{\section}{\small\bfseries\centering}{\thesection.}{0.5em}{}

\usepackage[T1]{fontenc}
\usepackage[utopia]{mathdesign}

\usepackage[breaklinks=true]{hyperref}
\hypersetup{
  colorlinks=true,
  citecolor=brown,
  linkcolor=black,
  urlcolor=magenta,
}

\renewcommand\sout{\bgroup \color{red} \ULdepth=-.5ex \ULset}

\renewcommand{\v}[1]{\textbf{#1}}
\renewcommand{\rm}[1]{\textrm{#1}}
\renewcommand{\d}{\mathrm{d}}

\usepackage{tikz}

\definecolor{lime}{HTML}{A6CE39}
\DeclareRobustCommand{\orcidicon}{
	\begin{tikzpicture}
	\draw[lime, fill=lime] (0,0) 
	circle [radius=0.16] 
	node[white] {{\fontfamily{qag}\selectfont \tiny ID}};
	\draw[white, fill=white] (-0.0625,0.095) 
	circle [radius=0.007];
	\end{tikzpicture}
	\hspace{-2mm}
}
\foreach \x in {A, ..., Z}{%
	\expandafter\xdef\csname orcid\x\endcsname{\noexpand\href{https://orcid.org/\csname orcidauthor\x\endcsname}{\noexpand\orcidicon}}
}

\DeclareSymbolFont{cmsymbols}{OMS}{cmsy}{m}{n}
\DeclareSymbolFontAlphabet{\mathcal}{cmsymbols}

\begin{document}

\title{Nucleon Short-Range Correlations and Supersoft Nuclear Symmetry Energy:\\Compatibility with Massive Neutron Stars and GW170817}

\author{Hai-Xiang Zhu\orcidD{}}
\affiliation{Shanghai Research Center for Theoretical Nuclear Physics, NSFC and Fudan University, Shanghai 200438, China}
\author{Bao-Jun Cai\orcidA{}}\email{bjcai@fudan.edu.cn}
\affiliation{Shanghai Research Center for Theoretical Nuclear Physics, NSFC and Fudan University, Shanghai 200438, China}
\affiliation{Key Laboratory of Nuclear Physics and Ion-beam Application (MOE), Institute of Modern Physics, Fudan University, Shanghai 200433, China} 
\author{Yu-Gang Ma\orcidC{}}\email{mayugang$@$fudan.edu.cn}
\affiliation{College of Physics, East China Normal University, Shanghai 200241, China}
\affiliation{Shanghai Research Center for Theoretical Nuclear Physics, NSFC and Fudan University, Shanghai 200438, China}
\affiliation{Key Laboratory of Nuclear Physics and Ion-beam Application (MOE), Institute of Modern Physics, Fudan University, Shanghai 200433, China}

\date{\today}

\newcommand{\x}{\mathrm{X}}
\newcommand{\y}{\mathrm{Y}}
\newcommand{\hr}{\widehat{r}}
\newcommand{\hP}{\widehat{P}}
\newcommand{\heps}{\widehat{\varepsilon}}
\newcommand{\hrho}{\widehat{\rho}}

\fontdimen2\font=1.5pt

\begin{abstract}

Massive neutron stars (NSs) require a sufficiently stiff core equation of state (EOS), while the relatively small tidal deformability inferred from GW170817 favors softer nuclear matter at intermediate densities. Given the importance of the nuclear symmetry energy in determining the EOS of neutron-rich matter, an intriguing question arises: {Can the symmetry energy exhibit supersoft behavior at high densities while remaining compatible with these seemingly competing constraints?} Within the nonlinear Walecka framework, we demonstrate that nucleon short-range correlations (SRCs), manifested as a high-momentum tail (HMT) in the nucleon momentum distribution, can accommodate a supersoft symmetry energy compatible with both constraints without introducing additional interaction parameters. In particular, we derive for the first time analytical expressions for the symmetry energy curvature $K_{\rm{sym}}$ for both free Fermi gas (FFG) and HMT nucleon momentum distributions. Using model parameter sets that reproduce empirical nuclear matter properties around saturation density, we find that SRC-HMT yields substantially more negative values of $K_{\rm{sym}}$ than the FFG momentum distribution, allowing the symmetry energy to reach a maximum at low supra-saturation densities and subsequently decrease, while the total pressure remains sufficient to support massive NSs. The resulting EOSs also predict smaller NS radii and tidal deformabilities consistent with GW170817. Our findings establish a direct connection between the microscopic momentum distribution of nucleons and the macroscopic properties of NSs, demonstrating how SRCs can shape the density dependence of the symmetry energy while satisfying astrophysical constraints.
\end{abstract}

\maketitle

\section{Introduction}
\label{sec:intro}

What can the momentum distribution of nucleons tell us about the most massive neutron stars (NSs) in our Universe and the tidal deformability inferred from gravitational wave (GW) observations? At first sight, these phenomena appear to probe very different aspects of nuclear physics. Nucleon short-range correlations (SRCs) arise from strong interactions between nucleons at short distances, whereas NS masses, radii, and tidal deformabilities reflect the collective behavior of matter under extreme gravity. Yet they are connected through the equation of state (EOS) of dense nuclear matter\,\cite{Ma2026NST,Danielewicz2002,Drischler2021,Baym2018,Oertel2017,LiA25,Chat25,LattimerPrakash2016,Lattimer2021,Soren23,LiCaiXieZhang2021,Burgio2021,BaldoBurgio2016,LiRamosVerdeVidana2014,Yang2025CPL}. In particular, the nuclear symmetry energy is crucial for understanding the properties of neutron-rich matter in NSs\,\cite{LiA25,Chat25,LiCaiXieZhang2021,BaldoBurgio2016,LiRamosVerdeVidana2014,Burgio2021,Xu2026SCPMA,Xie2026SCPMA,Xie2026}. 

The challenge is to understand how the EOS of dense nuclear matter can satisfy two key astrophysical constraints probing different density regimes. The existence of NSs with masses around two solar masses, established through various observations including those from NICER, requires sufficient pressure at high densities to resist gravitational collapse\,\cite{Demorest2010,Antoniadis2013,Cromartie2020,Fonseca2021,Romani2022,Miller21,Riley21,Salmi22,Salmi24,Ditt24,Mauviard26}. Meanwhile, the relatively small tidal deformability inferred from GW170817 favors more compact NS configurations and constrains the EOS at intermediate densities\,\cite{Abbott2017,Abbott2018}. Accommodating both observations requires an appropriate density dependence of the pressure, rather than an EOS that is uniformly stiff or soft. This naturally leads to a crucial question:
{\it Can the nuclear symmetry energy reach a maximum at relatively low supra-saturation densities and then decrease as the density increases further, a behavior referred to here as supersoft, while the full EOS remains sufficiently stiff to support massive NSs constrained by radio pulsar timing and NICER mass-radius observations and simultaneously remains consistent with the tidal deformability inferred from GW170817?}

The nuclear symmetry energy plays a central role in this question, as it governs the energy cost of changing the neutron proton asymmetry\,\cite{Baran2005,Steiner2005,Li2008,Li2026Sym,Stone2024,Miyatsu2025}. Its density dependence near saturation density is commonly characterized by the slope $L$ and curvature  $K_{\rm{sym}}$, which describe its first- and second-order density variations, respectively. While considerable progress has been made in constraining the slope $L$\,\cite{Li2013,ZhangChen2013,ZhangChen2014,Reed2021,Essick2021,LattimerLim2013}, the magnitude and even the sign of $K_{\rm{sym}}$ remain uncertain\,\cite{LiMagno2020,NewtonCrocombe2021,RichterLi2023,LiGrundler2024,WangJM2025}. This uncertainty is particularly important for NSs, where strongly neutron-rich matter extends over a broad range of densities. A sufficiently negative $K_{\rm{sym}}$ can lead to a supersoft symmetry energy that decreases rapidly with increasing density. Such behavior can reduce the pressure of neutron-rich matter and favor smaller NS radii and tidal deformabilities, but it also raises the question of whether sufficient pressure can be maintained to support massive NSs\,\cite{Li2022}. The answer depends not only on the symmetry energy itself but also on the underlying microscopic description of nuclear matter.

One important ingredient is the nucleon momentum distribution. In the conventional free Fermi gas (FFG) momentum distribution, nucleons occupy momentum states below their respective Fermi surfaces\,\cite{AGD1975}. In this work, FFG refers only to the momentum distribution within the quasiparticle picture and does not imply the absence of nucleon interactions. In realistic nuclear matter, however, SRCs deplete the occupation below the Fermi surface and generate a high-momentum tail (HMT) above it\,\cite{Bethe1971,Subedi2008,Hen2014,Ciofi2015,Hen2017,CaiLiMa2026SRC,Arrington2022SRC,Fomin2017SRC,Dalal2024SRC,Duer2018,Schmidt2020,Nguyen2026,Schmidt2024,Korover2023,Cao2023,Shen2022,Dick04,Arr12,Aum21,Fra08,Fomin2026,Hen2026,Sargsian2026}. This redistribution of nucleons in momentum space changes their kinetic and interaction contributions to the EOS, particularly in asymmetric nuclear matter (ANM)\,\cite{Vidana2011,Carbone2012,Hen2015,Cai2015,Cai2016}. Consequently, the symmetry energy and its density derivatives may differ substantially from those obtained using the FFG distribution, even within the same interaction framework\,\cite{Cai2016,Cai2025Review}. Previous studies have shown that SRCs can influence NS masses, radii, and tidal deformabilities\,\cite{Cai2016,Cai2025Review,Souza2020,Lourenco2022a,Lourenco2022b,Lourenco2026,Hong2025,Lu2021SRC,Lu2022SRC,LiRen2019,HongRen2022,ShangDong2020,WangDong2024,LiuDong2024}. A key issue, therefore, is whether SRC-induced modifications can provide a mechanism for reconciling a supersoft symmetry energy with the constraints from massive NSs and GW170817.

The nonlinear Walecka framework provides a useful setting for investigating this possibility\,\cite{Walecka1974,Chin1977,Serot1986,Mueller1996,Long2004,Meng2006,MengZhou2015}. By incorporating scalar and vector meson interactions and their nonlinear couplings, it offers a relativistic description of nuclear matter that reproduces empirical properties around saturation density while allowing considerable flexibility in the high density EOS\,\cite{Serot1986,Mueller1996}. In particular, the isovector interactions govern the density dependence of the symmetry energy\,\cite{Horowitz2001,ToddRutel2005,Chen2007}, making this framework well suited for exploring the interplay between SRC-HMT and $K_{\rm{sym}}$. An explicit analytical expression for $K_{\rm{sym}}$ with both FFG and HMT momentum distributions is particularly valuable, as it allows us to identify how SRCs modify the curvature within a given interaction framework and determine whether the resulting high density symmetry energy can become supersoft without violating astrophysical constraints.

In this work, we derive for the first time within this framework analytical expressions for $K_{\rm{sym}}$ incorporating both FFG and HMT nucleon momentum distributions. We construct EOSs consistent with empirical nuclear matter properties around saturation density and examine their compatibility with the tidal deformability constraint from GW170817 and the observed masses of massive NSs. We find that SRC-HMT permits substantially more negative values of $K_{\rm{sym}}$ than the FFG momentum distribution, allowing the symmetry energy to decrease strongly and become supersoft at high densities. Importantly, such behavior does not necessarily prevent the EOS from supporting massive NSs. The resulting EOSs can simultaneously accommodate large maximum NS masses and the relatively small radii and tidal deformabilities favored by GW170817, without introducing additional interaction parameters. Our results thus extend the known astrophysical implications of SRCs by demonstrating that even an extreme softening of the nuclear symmetry energy can remain compatible with these observational constraints. We further investigate the consequences for the proton fraction, core-crust transition, EOS-parameter, and sound speed, providing a broader picture of how SRC-HMT influences dense NS matter.

The remainder of this paper is organized as follows. In Section \ref{sec:framework}, we introduce the theoretical framework, including the SRC-modified nucleon momentum distributions and their implementation in the nonlinear Walecka model. In Section \ref{sec:Ksym_analytical}, we derive analytical expressions for the symmetry energy curvature $K_{\rm{sym}}$ for both FFG and HMT distributions. We then examine the resulting constraints on $K_{\rm{sym}}$ and their implications for the EOS of ANM in the subsequent section. In Section \ref{sec:NS_properties}, we investigate the effects of SRC-HMT on NS properties and confront the predictions with astrophysical observations. Finally, Section \ref{sec:summary} summarizes our main findings. The appendices provide technical details and derivations of several quantities and relations omitted from the main text for brevity.

\section{Theoretical Framework}
\label{sec:framework}

In this section, we present the theoretical framework for investigating the effects of nucleon SRCs on the EOS of ANM. We first introduce the EOS and its expansion in isospin asymmetry, together with the relevant nuclear matter parameters  in Subsection \ref{subsec:EOS_def}. We then describe the SRC-modified nucleon momentum distribution, characterized by a depleted Fermi sea and a HMT in Subsection \ref{subsec:SRC_HMT}. Finally, we incorporate this momentum distribution into the nonlinear Walecka model to calculate the energy density, pressure, and nucleon chemical potentials, including the rearrangement contributions arising from the isospin dependence of SRCs in Subsection \ref{subsec:Walecka_model}.

\subsection{Definitions of the EOS of ANM}
\label{subsec:EOS_def}

Cold infinite ANM is characterized by the total baryon density $\rho=\rho_{\rm{n}}+\rho_{\rm{p}}$ and isospin asymmetry $\delta=(\rho_{\rm{n}}-\rho_{\rm{p}})/\rho$, where $\rho_{\rm{n}}$ and $\rho_{\rm{p}}$ are the neutron and proton densities, respectively. The proton fraction is $x_{\rm{p}}=\rho_{\rm{p}}/\rho$. The energy per nucleon $E(\rho,\delta)$ is related to the nucleonic energy density $\varepsilon_{\rm{N}}(\rho,\delta)$ by\,\cite{Cai2016}
\begin{equation}
E(\rho,\delta) = \frac{\varepsilon_{\rm{N}}(\rho,\delta)}{\rho} - M_{\rm N},
\label{eq:E_def}
\end{equation}
where $M_{\rm N}\approx939\,\mathrm{MeV}$ is the bare nucleon rest mass. Throughout this work, we use the natural units in which $\hbar=c=G=1$.

Assuming charge symmetry of the nuclear interactions, the EOS can be expanded in even powers of $\delta$ around symmetric nuclear matter (SNM) as\,\cite{Li2008}:
\begin{equation}
E(\rho,\delta) \approx E_0(\rho) + E_{\mathrm{sym}}(\rho)\delta^2 + \mathcal{O}(\delta^4),
\label{eq:E_delta_expansion}
\end{equation}
where $E_0(\rho)\equiv E(\rho,0)$ is the EOS of SNM. The symmetry energy $E_{\mathrm{sym}}(\rho)$ is defined by
\begin{align}
E_{\mathrm{sym}}(\rho) =& \left.\frac{1}{2}\frac{\partial^2 E(\rho,\delta)}{\partial \delta^2}\right|_{\delta=0}.
\label{eq:Esym_def}
\end{align}

Near the saturation density $\rho_0$ of SNM, the energies $E_0(\rho)$ and $E_{\mathrm{sym}}(\rho)$ can be expanded in terms of the dimensionless density deviation $\chi\equiv(\rho-\rho_0)/3\rho_0$\,\cite{Li2008}:
\begin{align}
E_0(\rho) &\approx E_0(\rho_0) + \frac{1}{2}K_0\chi^2 + \frac{1}{6}J_0\chi^3 + \mathcal{O}(\chi^4), \label{eq:E0_Taylor} \\
E_{\mathrm{sym}}(\rho) &\approx E_{\mathrm{sym}}(\rho_0) + L\chi + \frac{1}{2}K_{\mathrm{sym}}\chi^2 + \frac{1}{6}J_{\mathrm{sym}}\chi^3 + \mathcal{O}(\chi^4), \label{eq:Esym_Taylor}
\end{align}
The linear term in $E_0(\rho)$ vanishes because of the saturation condition $P_0(\rho_0)=0$. The incompressibility coefficient $K_0$ and the skewness coefficient $J_0$ of SNM are defined as
\begin{equation}
K_0 = \left.9\rho_0^2\frac{\d^2 E_0(\rho)}{\d\rho^2}\right|_{\rho=\rho_0}, ~~ J_0 = \left.27\rho_0^3\frac{\d^3 E_0(\rho)}{\d\rho^3}\right|_{\rho=\rho_0},
\label{eq:K0_J0_def}
\end{equation}
while the slope parameter $L$, the curvature coefficient $K_{\rm{sym}}$, and the skewness coefficient $J_{\rm{sym}}$ of the symmetry energy are\,\cite{LiCaiXieZhang2021}
\begin{align}
L &= \left.3\rho_0\frac{\d E_{\rm{sym}}(\rho)}{\d\rho}\right|_{\rho=\rho_0}, \label{eq:def-L}\\
K_{\rm{sym}} &= \left.9\rho_0^2\frac{\d^2 E_{\rm{sym}}(\rho)}{\d\rho^2}\right|_{\rho=\rho_0}, \label{eq:def-Ksym}\\
J_{\rm{sym}} &= \left.27\rho_0^3\frac{\d^3 E_{\rm{sym}}(\rho)}{\d\rho^3}\right|_{\rho=\rho_0}.\label{eq:def-Jsym}
\end{align}

The nucleonic pressure follows from the relation
\begin{equation}
P_{\rm{N}}(\rho,\delta)
=\rho^2\frac{\partial E(\rho,\delta)}{\partial\rho}
=\rho\frac{\partial\varepsilon_{\rm{N}}}{\partial\rho}
-\varepsilon_{\rm{N}},
\end{equation}
where the derivatives are taken at fixed $\delta$.
For SNM, this reduces to
$P_0(\rho)=\rho^2\d E_0(\rho)/\d\rho$.

\subsection{Single Nucleon Momentum Distribution with SRC Effects}
\label{subsec:SRC_HMT}

To incorporate the effects of SRCs in ANM, we adopt a phenomenological single nucleon momentum distribution function $n_{\v{k}}^J(\rho,\delta)$ ($J=\rm{n},\rm{p}$), characterized by a depleted Fermi sea and a $|\v{k}|^{-4}$ high-momentum tail (HMT)\,\cite{CaiLi22PRCFFG,CaiLiChen2016,CaiLi2016PLB}:
\begin{equation}
n_{\v{k}}^J(\rho,\delta) = \begin{cases}
\Delta_J, & 0 < |\v{k}| \le k_{\rm{F}}^J, \\[2pt]
\displaystyle C_J \left({k_{\rm{F}}^J}/{|\v{k}|}\right)^4, & k_{\rm{F}}^J < |\v{k}| \le \phi_J k_{\rm{F}}^J, \\[2pt]
0, & |\v{k}| > \phi_J k_{\rm{F}}^J,
\end{cases}
\label{eq:n_k_SRC}
\end{equation}
where $k_{\rm{F}}^J=(3\pi^2\rho_J)^{1/3}=k_{\rm{F}}(1+\tau_3^J\delta)^{1/3}$ is the Fermi momentum of nucleon species $J$, with $k_{\rm{F}}=(3\pi^2\rho/2)^{1/3}$, $\tau_3^{\rm{n}}=+1$, and $\tau_3^{\rm{p}}=-1$. Here, $\Delta_J$ denotes the occupation probability below the Fermi momentum, $C_J$ characterizes the strength of the HMT, and $\phi_J>1$ specifies its dimensionless momentum cutoff.

Motivated by analyses of inclusive and exclusive electron-scattering experiments and microscopic Brueckner-Hartree-Fock and self-consistent Green's function calculations\,\cite{Subedi2008,Hen2014,Rios2009,Yin2013,Rios2014}, we parametrize the isospin dependence of $C_J$ and $\phi_J$ as
\begin{equation}
C_J(\delta) = C_0\left(1 + \tau_3^J C_1 \delta\right), ~~\phi_J(\delta) = \phi_0\left(1 + \tau_3^J \phi_1 \delta\right).
\label{eq:C_phi_isospin}
\end{equation}
The normalization condition for each nucleon species,
\begin{equation}
\rho_J = \frac{2}{(2\pi)^3}\int_0^\infty\d\v{k}  n_{\v{k}}^J(\rho,\delta) = \frac{k_{\rm{F}}^{J,3}}{3\pi^2},
\label{eq:n_k_norm}
\end{equation}
then determines the Fermi-sea occupation factor as
\begin{equation}
\Delta_J(\delta) = 1 - 3C_J(\delta)\left(1 - {\phi_J^{-1}(\delta)}\right).
\label{eq:Delta_tau}
\end{equation}
In SNM with $\delta=0$, $C_{\rm{n}}=C_{\rm{p}}=C_0$, $\phi_{\rm{n}}=\phi_{\rm{p}}=\phi_0$, and $\Delta_{\rm{n}}=\Delta_{\rm{p}}=\Delta_0\equiv1-3C_0(1-1/\phi_0)$.

Following the experimental and theoretical constraints established in Ref.\,\cite{Cai2015}, we adopt $C_0\approx0.161$, $C_1\approx-0.25$, $\phi_0\approx2.38$, and $\phi_1\approx-0.56$. These parameters yield an HMT fraction of $x_{\rm{SNM}}^{\rm{HMT}}=3C_0(1-1/\phi_0)\approx28\%$ in SNM and $x_{\rm{PNM}}^{\rm{HMT}}=3C_0(1+C_1)[1-1/\phi_0(1+\phi_1)]\approx1.6\%$ in PNM.
The standard free Fermi gas (FFG) momentum distribution is recovered by setting $C_0=0$, $C_1=0$, $\phi_0=1$ and $\phi_1=0$, which gives $\Delta_J=1$.

\subsection{Nonlinear Walecka Model with SRC-modified $n_{\v{k}}^J(\rho,\delta)$}
\label{subsec:Walecka_model}

We employ covariant density functional theory based on the nonlinear Walecka model, including the isoscalar-scalar $\sigma$, isoscalar-vector $\omega_\mu$, isovector-scalar $\vec{\delta}$, and isovector-vector $\vec{\rho}_\mu$ mesons, together with nonlinear $\sigma$ and $\omega$ self-interactions and the $\omega$-$\rho$ cross-coupling\,\cite{Mueller1996,Horowitz2001,ToddRutel2005,Kubis1997,Liu2002,Qin25}. The interacting Lagrangian density is given by
\begin{align}
\mathcal{L} =& \overline{\psi}\left[\gamma_\mu\left(i\partial^\mu - g_\omega\omega^\mu + g_\rho\vec{\rho}^{\mu}\cdot\vec{\tau}\right) - \left(M_{\rm{N}} - g_\sigma\sigma + g_\delta\vec{\delta}\cdot\vec{\tau}\right)\right]\psi \notag\\
&+ \frac{1}{2}\left(\partial_\mu\sigma\partial^\mu\sigma - m_\sigma^2\sigma^2\right) - \frac{1}{3}b_\sigma M_{\rm{N}} g_\sigma^3\sigma^3 - \frac{1}{4}c_\sigma g_\sigma^4\sigma^4 \notag\\
& - \frac{1}{4}\omega_{\mu\nu}\omega^{\mu\nu} + \frac{1}{2}m_\omega^2\omega_\mu\omega^\mu + \frac{1}{4}c_\omega g_\omega^4\left(\omega_\mu\omega^\mu\right)^2 \notag\\
&+ \frac{1}{2}\left(\partial_\mu\vec{\delta}\cdot\partial^\mu\vec{\delta} - m_\delta^2\vec{\delta}^{2}\right) - \frac{1}{4}\vec{\rho}_{\mu\nu}\cdot\vec{\rho}^{\mu\nu} + \frac{1}{2}m_\rho^2\vec{\rho}_\mu\cdot\vec{\rho}^{\mu} \notag\\
& + \frac{1}{2}\Lambda_{\rm{V}}g_\omega^2 g_\rho^2\left(\omega_\mu\omega^\mu\right)\left(\vec{\rho}_\nu\cdot\vec{\rho}^{\nu}\right),
\label{eq:Lagrangian}
\end{align}
where $\omega_{\mu\nu} \equiv \partial_\mu\omega_\nu - \partial_\nu\omega_\mu$ and $\vec{\rho}_{\mu\nu} \equiv \partial_\mu\vec{\rho}_\nu - \partial_\nu\vec{\rho}_\mu$ are the vector meson field tensors. Throughout this work, we fix the meson masses at $m_\sigma \approx 500.0\,\rm{MeV}$, $m_\omega \approx 782.5\,\rm{MeV}$, $m_\rho \approx 763.0\,\rm{MeV}$, and $m_\delta \approx 980.0\,\rm{MeV}$.

Previous studies have proposed additional meson interactions, such as the $\sigma^2\vec{\delta}^{2}$ coupling, as a possible mechanism for reconciling massive NSs with the tidal deformability constraint from GW170817\,\cite{Zabari2019,Li2022}. Such extensions modify the density dependence of the EOS through additional interaction channels and associated adjustable parameters. Other studies have also incorporated SRC-HMT and explored how variations in nuclear matter parameters, such as the symmetry energy slope $L$, incompressibility $K_0$, and nucleon effective mass, affect NS masses and tidal deformabilities\,\cite{Gautam2025,Hong2023,Souza2020,Lourenco2022a,Lourenco2022b,Lourenco2026,Hong2025,Lu2021SRC,Lu2022SRC,LiRen2019,HongRen2022,ShangDong2020,WangDong2024,LiuDong2024}. These investigations demonstrate that SRC-HMT can play an important role in satisfying astrophysical constraints when combined with suitable choices of nuclear matter parameters. In this work, we focus on a different aspect of this mechanism: How SRC-HMT modifies the symmetry energy curvature $K_{\rm{sym}}$ and whether the resulting EOSs can accommodate a supersoft symmetry energy at high densities while remaining consistent with massive NSs and GW170817. We retain the four meson degrees of freedom, including the isovector-scalar $\delta$ meson, while avoiding additional interaction terms introduced specifically to satisfy astrophysical constraints. The inclusion of the $\delta$ meson makes the analytical derivation of $K_{\rm{sym}}$ more involved, but also enables a more complete treatment of the symmetry energy curvature within the adopted framework. By keeping the interaction structure relatively simple, we can more clearly identify how SRC-HMT affects $K_{\rm{sym}}$ and explore its implications for the high density behavior of $E_{\rm{sym}}(\rho)$ and properties of NSs.

In the relativistic mean field (RMF) approximation for static, uniform nuclear matter, the meson fields are replaced by their classical ground state expectation values: $\sigma \to \overline{\sigma}$, $\omega_\mu \to \overline{\omega}_0\delta_{\mu 0}$, $\delta^{(a)} \to \overline{\delta}^{(3)}\delta_{a3}$, and $\rho_\mu^{(a)} \to \overline{\rho}_0^{(3)}\delta_{\mu 0}\delta_{a3}$. For SNM, we introduce the shorthand notations $f_0 \equiv \overline{\sigma}(\rho,\delta=0)$ and $w_0 \equiv \overline{\omega}_0(\rho,\delta=0)$\,\cite{Cai2016}. The Dirac effective masses of protons and neutrons split in asymmetric matter according to
\begin{equation}
M_{\rm{p}}^* = M_{\rm{N}} - g_\sigma\overline{\sigma} - g_\delta\overline{\delta}^{(3)}, ~~ M_{\rm{n}}^* = M_{\rm{N}} - g_\sigma\overline{\sigma} + g_\delta\overline{\delta}^{(3)},
\label{eq:Mstar_pn}
\end{equation}
or, equivalently, $M_J^* = M_{\rm{N}} - g_\sigma\overline{\sigma} + \tau_3^J g_\delta\overline{\delta}^{(3)}$ for $J=\rm{n},\rm{p}$. In SNM, both effective masses reduce to 
$
M_0^* = M_{\rm{N}} - g_\sigma\overline{\sigma} = M_{\rm{N}} - g_\sigma f_0$.

The Euler-Lagrange equations yield the coupled nonlinear self-consistent field equations:
\begin{align}
&m_\sigma^2\overline{\sigma} + b_\sigma M_{\rm{N}} g_\sigma^3\overline{\sigma}^2 + c_\sigma g_\sigma^4\overline{\sigma}^3  = g_\sigma\rho^{\rm{s}}, \label{eq:EOM_sigma} \\
&m_\delta^2\overline{\delta}^{(3)} = g_\delta(\rho_{\rm{p}}^{\rm{s}} - \rho_{\rm{n}}^{\rm{s}}), \label{eq:EOM_delta} \\
&m_\omega^2\overline{\omega}_0 + c_\omega g_\omega^4\overline{\omega}_0^3 + \Lambda_{\rm{V}}g_\omega^2 g_\rho^2\overline{\rho}_0^{(3),2}\overline{\omega}_0  = g_\omega\rho, \label{eq:EOM_omega} \\
&m_\rho^2\overline{\rho}_0^{(3)} + \Lambda_{\rm{V}}g_\omega^2 g_\rho^2\overline{\omega}_0^2\overline{\rho}_0^{(3)} = g_\rho(\rho_{\rm{p}} - \rho_{\rm{n}}), \label{eq:EOM_rho}
\end{align}
where $\rho^{\rm s}=\rho_{\rm{n}}^{\rm s}+\rho_{\rm p}^{\rm s}$.
With the SRC-HMT momentum distribution (\ref{eq:n_k_SRC}), the scalar density $\rho_J^{\rm{s}}$ can be evaluated analytically as\,\cite{Cai2016}
\begin{align}
\rho_J^{\rm{s}} =& \frac{2}{(2\pi)^3}\int_0^{\phi_J k_{\rm{F}}^J}\d\v{k} n_{\v{k}}^J\frac{M_J^*}{\sqrt{\v{k}^2 + M_J^{*,2}}} \notag\\
=& \frac{M_J^*\Delta_J}{2\pi^2}\left[k_{\rm{F}}^J E_{\rm{F}}^{J,*} - M_J^{*,2}\operatorname{arcsinh}\left(\frac{k_{\rm{F}}^J}{M_J^*}\right)\right] \notag\\
&+ \frac{C_Jk_{\rm{F}}^{J,4}}{\pi^2 M_J^*}\left[\sqrt{1 + \left(\frac{M_J^*}{k_{\rm{F}}^J}\right)^2} - \sqrt{1 + \left(\frac{M_J^*}{\phi_J k_{\rm{F}}^J}\right)^2}\right],
\label{eq:rho_s_analytic}
\end{align}
where $E_{\rm{F}}^{J,*} = [{k_{\rm{F}}^{J,2} + M_J^{*,2}}]^{1/2}$.

The nucleonic energy density $\varepsilon_{\rm{N}}$ and pressure $P_{\rm{N}}$ of ANM are obtained from the energy momentum tensor $\mathcal{T}^{\mu\nu}$ as
\begin{align}
\varepsilon_{\rm{N}}
=&\sum_{J=\rm{p,n}}\varepsilon_J^{\rm{kin}}
+\frac{1}{2}m_\sigma^2\overline{\sigma}^2
+\frac{1}{3}b_\sigma M_{\rm{N}}g_\sigma^3\overline{\sigma}^3
+\frac{1}{4}c_\sigma g_\sigma^4\overline{\sigma}^4
\notag\\
&+\frac{1}{2}m_\delta^2\overline{\delta}^{(3),2}
+\frac{1}{2}m_\omega^2\overline{\omega}_0^2
+\frac{3}{4}c_\omega g_\omega^4\overline{\omega}_0^4
\notag\\
&+\frac{1}{2}m_\rho^2\overline{\rho}_0^{(3),2}
+\frac{3}{2}\Lambda_{\rm{V}}g_\omega^2g_\rho^2
\overline{\omega}_0^2\overline{\rho}_0^{(3),2},
\label{eq:eps_N}\\
P_{\rm{N}} =& \sum_{J=\rm{p,n}}P^{\rm{kin}}_J - \frac{1}{2}m_\sigma^2\overline{\sigma}^2 - \frac{1}{3}b_\sigma M_{\rm{N}} g_\sigma^3\overline{\sigma}^3 - \frac{1}{4}c_\sigma g_\sigma^4\overline{\sigma}^4 \notag\\
&- \frac{1}{2}m_\delta^2\overline{\delta}^{(3),2} + \frac{1}{2}m_\omega^2\overline{\omega}_0^2 + \frac{1}{4}c_\omega g_\omega^4\overline{\omega}_0^4 \notag\\
&+ \frac{1}{2}m_\rho^2\overline{\rho}_0^{(3),2} + \frac{1}{2}\Lambda_{\rm{V}}g_\omega^2 g_\rho^2\overline{\omega}_0^2\overline{\rho}_0^{(3),2}, \label{eq:P_N}
\end{align}
therefore \begin{equation}
\varepsilon_{\rm{N}}+P_{\rm{N}}
=\sum_{J=\rm{p,n}}
\left(\varepsilon_J^{\rm{kin}}+P_J^{\rm{kin}}\right)
+g_\omega\overline{\omega}_0\rho
+g_\rho\overline{\rho}_0^{(3)}
(\rho_{\rm{p}}-\rho_{\rm{n}}),
\end{equation}
the kinetic energy density $\varepsilon_J^{\rm{kin}}$ and kinetic pressure $P^{\rm{kin}}_J$, including SRC-HMT effects, are given explicitly by\,\cite{Cai2016}
\begin{align}
\varepsilon_J^{\rm{kin}} =& \frac{2}{(2\pi)^3}\int_0^{\phi_J k_{\rm{F}}^J}\d \v{k} n_{\v{k}}^J\sqrt{\v{k}^2 + M_J^{*,2}} \notag\\
=& \frac{\Delta_J}{\pi^2}h_<^{(1)}(k_{\rm{F}}^J,M_J^*) \notag\\
&+ \frac{C_Jk_{\rm{F}}^{J,4}}{\pi^2}\left[h_>^{(1)}(\phi_J k_{\rm{F}}^J,M_J^*) - h_>^{(1)}(k_{\rm{F}}^J,M_J^*)\right], \label{eq:eps_kin} 
\end{align}
and
\begin{align}
P^{\rm{kin}}_J =& \frac{1}{3}\frac{2}{(2\pi)^3}\int_0^{\phi_J k_{\rm{F}}^J} \d \v{k}n_{\v{k}}^J\frac{\v{k}^2}{\sqrt{\v{k}^2 + M_J^{*,2}}} \notag\\
=& \frac{\Delta_J}{3\pi^2}h_<^{(2)}(k_{\rm{F}}^J,M_J^*) \notag\\
&+ \frac{C_Jk_{\rm{F}}^{J,4}}{3\pi^2}\left[h_>^{(2)}(\phi_J k_{\rm{F}}^J,M_J^*) - h_>^{(2)}(k_{\rm{F}}^J,M_J^*)\right], \label{eq:P_kin}
\end{align}
with the auxiliary analytical functions
\begin{align}
h_<^{(1)}(k,m^*)
=& \frac{m^{*,4}}{8}\left(\frac{k}{m^*}\right)
\sqrt{1+\left(\frac{k}{m^*}\right)^2}\cdot
\left[2\left(\frac{k}{m^*}\right)^2+1\right]\notag\\
&-\frac{m^{*,4}}{8}
\operatorname{arcsinh}\left(\frac{k}{m^*}\right),\\
h_>^{(1)}(k,m^*)
=& \operatorname{arcsinh}\left(\frac{k}{m^*}\right)
-\left(\frac{k}{m^*}\right)^{-1}
\sqrt{1+\left(\frac{k}{m^*}\right)^2}, \\
h_<^{(2)}(k,m^*)
=& \frac{m^{*,4}}{8}\left(\frac{k}{m^*}\right)
\sqrt{1+\left(\frac{k}{m^*}\right)^2}\cdot
\left[2\left(\frac{k}{m^*}\right)^2-3\right] \notag\\
&+\frac{3m^{*,4}}{8}
\operatorname{arcsinh}\left(\frac{k}{m^*}\right), \\
h_>^{(2)}(k,m^*)
=& \operatorname{arcsinh}\left(\frac{k}{m^*}\right).
\label{eq:fABCD}
\end{align}

Finally, to ensure thermodynamic consistency in $\beta$-stable NS matter, the nucleon chemical potentials $\mu_J = \partial \varepsilon_{\rm{N}}/\partial \rho_J$ are derived directly from Eq.\,(\ref{eq:eps_N}). Since the HMT parameters $C_J(\delta)$ and $\phi_J(\delta)$ depend explicitly on $\delta$, differentiation of $\varepsilon_J^{\rm{kin}}$ with respect to the neutron and proton densities produces both a direct single particle HMT contribution and an isospin rearrangement contribution\,\cite{Lourenco2026}. The resulting chemical potential difference can be written as
\begin{equation}
\mu_{\rm{n}} - \mu_{\rm{p}} = \mu_{\rm{n}}^{\rm{(D)}} - \mu_{\rm{p}}^{\rm{(D)}}+ \Delta\mu^{\rm{(R)}}_{\rm{SRC}},
\label{eq:mu_diff_total}
\end{equation}
where the direct chemical potential of species $J$ is
\begin{align}
\mu_J^{\rm{(D)}} =& \Delta_J E_{\rm{F}}^{J,*} + 3C_J\left(E_{\rm{F}}^{J,*} - \frac{F_{\rm{F}}^{J,*}}{\phi_J}\right) \notag\\
&+ 4C_J k_{\rm{F}}^J\ln\left(\frac{\phi_J k_{\rm{F}}^J + F_{\rm{F}}^{J,*}}{k_{\rm{F}}^J + E_{\rm{F}}^{J,*}}\right)
+ g_\omega\overline{\omega}_0 - \tau_3^J g_\rho\overline{\rho}_0^{(3)},
\label{eq:mu_dir}
\end{align}
with 
\begin{equation}
F_{\rm{F}}^{J,*} = \sqrt{\left(\phi_J k_{\rm{F}}^J\right)^2 + M_J^{*,2}}.
\end{equation}
The SRC isospin rearrangement term is
\begin{align}
\Delta\mu^{\rm{(R)}}_{\rm{SRC}}
=&\frac{2}{\rho}\left[C_0 C_1
\sum_{J=\rm{n,p}}\tau_3^J
\frac{\partial\varepsilon_{J}^{\rm{kin}}}{\partial C_{J}}
+\phi_0\phi_1\sum_{J=\rm{n,p}}\tau_3^J
\frac{\partial\varepsilon_{J}^{\rm{kin}}}{\partial\phi_{J}}\right]\notag\\
=&\frac{2}{\rho}\left[
C_0 C_1\left(
\frac{\partial\varepsilon_{\rm n}^{\rm{kin}}}{\partial C_{\rm n}}
-\frac{\partial\varepsilon_{\rm p}^{\rm{kin}}}{\partial C_{\rm p}}
\right)+\phi_0\phi_1\left(
\frac{\partial\varepsilon_{\rm n}^{\rm{kin}}}{\partial\phi_{\rm n}}
-\frac{\partial\varepsilon_{\rm p}^{\rm{kin}}}{\partial\phi_{\rm p}}
\right)
\right].
\label{eq:mu_rearr_SRC}
\end{align}
See Appendix \ref{app:app} for the detailed expression for $\Delta\mu^{\rm{(R)}}_{\rm{SRC}}$.

\section{Analytical Expressions for the Curvature Coefficient of Symmetry Energy in the Nonlinear Walecka Model}
\label{sec:Ksym_analytical}

The density dependence of the nuclear symmetry energy is characterized by its slope and curvature coefficients. In this section, we derive analytical expressions for the curvature coefficient $K_{\rm{sym}}(\rho)$ within the nonlinear Walecka model, incorporating nucleon SRCs and the isovector-scalar $\delta$ meson. We separate the kinetic, $\rho$-meson, and $\delta$-meson contributions while accounting for the density dependence of the self-consistent mean fields. Accordingly, the curvature coefficient is decomposed as $
K_{\rm{sym}}(\rho)
=K_{\rm{sym}}^{\rm{kin}}(\rho)
+K_{\rm{sym}}^\rho(\rho)
+K_{\rm{sym}}^\delta(\rho)$.
In Subsection \ref{sub-derivation}, we give these contributions and present the corresponding expressions for the symmetry energy $E_{\rm{sym}}(\rho)$ and its slope parameter $L(\rho)$. In Subsection \ref{sub-FFG}, we examine the FFG limit of the nucleon momentum distribution as a consistency check, retaining the density dependence of the mean fields. Finally, in Subsection \ref{sub-argument}, we use the non-relativistic limit to provide a qualitative explanation for how SRC-HMT can substantially reduce $K_{\rm{sym}}$.

\begin{widetext}

\subsection{Expression for the Curvature $K_{\textmd{sym}}$}\label{sub-derivation}

The SRC-HMT parameters $C_0$, $C_1$, $\phi_0$,
and $\phi_1$ are assumed to be density independent\,\cite{Cai2015,Cai2016}.
Primes on the mean fields denote total derivatives
with respect to $\rho$.
The kinetic contribution follows from differentiating the
SRC-modified kinetic symmetry energy twice with respect to
density. Since the effective mass changes with density, the
result contains both $f_0'\equiv \d f_0/\d\rho $ and $f_0''\equiv \d^2f_0/\d\rho^2$, in addition to the
explicit dependence on the Fermi momentum. The resulting
expression is
\begin{align}
K_{\rm{sym}}^{\rm{kin}}(\rho)
=&\frac{1}{12k_{\rm F}^2\phi_0}
\Biggl\{
\frac{36C_0\phi_0 k_{\rm F}}
{(1+\theta^2)^{3/2}}
\left[
C_1\left(1-\frac{1}{\phi_0}\right)
+\frac{\phi_1}{\phi_0}
\right]
\Biggl[
k_{\rm F}^2\theta(1+2\theta^2)
-\theta k_{\rm F}^2
\left(
\frac{6\rho g_\sigma f_0'}{M_0^*}
+\frac{9\rho^2g_\sigma^2f_0'^2}{M_0^{*,2}}
\right)
+9g_\sigma k_{\rm F}\rho^2
(1+\theta^2)f_0''
\Biggr]
\notag\\
&-\frac{2k_{\rm F}\phi_0\Delta_0\theta}
{(1+\theta^2)^{5/2}}
\Biggl[
k_{\rm F}^2(2+7\theta^2+2\theta^4)
+(\theta^2-2)k_{\rm F}^2
\left(
\frac{6\rho g_\sigma f_0'}{M_0^*}
+\frac{9\rho^2g_\sigma^2f_0'^2}{M_0^{*,2}}
\right)
-9g_\sigma k_{\rm F}\rho^2
\theta(1+\theta^2)f_0''
\Biggr]
\notag\\
&+4C_0(4+3C_1)k_{\rm F}\phi_0
\Biggl\{
\frac{1}{(1+\theta^2)^{3/2}}
\Biggl[
k_{\rm F}^2\theta(1+2\theta^2)
-\theta k_{\rm F}^2
\left(
\frac{6\rho g_\sigma f_0'}{M_0^*}
+\frac{9\rho^2g_\sigma^2f_0'^2}{M_0^{*,2}}
\right)
+9g_\sigma k_{\rm F}\rho^2
(1+\theta^2)f_0''
\Biggr]
\notag\\
&\hspace*{2cm}-
\frac{1+3\phi_1}
{(1+\phi_0^2\theta^2)^{3/2}}
\Biggl[
\phi_0 k_{\rm F}^2\theta
(1+2\phi_0^2\theta^2)
-\phi_0\theta k_{\rm F}^2
\left(
\frac{6\rho g_\sigma f_0'}{M_0^*}
+\frac{9\rho^2g_\sigma^2f_0'^2}{M_0^{*,2}}
\right)
+\frac{9g_\sigma k_{\rm F}\rho^2}{\phi_0}
(1+\phi_0^2\theta^2)f_0''
\Biggr]
\Biggr\}
\notag\\
&+\frac{2C_0k_{\rm F}\phi_0^2}
{(1+\phi_0^2\theta^2)^{5/2}}
\Biggl\{
k_{\rm F}^2\theta
\Bigl[
2(1-3\phi_1)
+(5-24\phi_1-9\phi_1^2)\phi_0^2\theta^2
+6(1+3\phi_1^2)\phi_0^4\theta^4
\Bigr]
\notag\\
&\hspace*{2cm}-\theta k_{\rm F}^2
\left(
\frac{6\rho g_\sigma f_0'}{M_0^*}
+\frac{9\rho^2g_\sigma^2f_0'^2}{M_0^{*,2}}
\right)
\Bigl[
2(1-3\phi_1)
+(5+12\phi_1+27\phi_1^2)\phi_0^2\theta^2
\Bigr]
\notag\\
&\hspace*{2cm}+
\frac{9g_\sigma k_{\rm F}\rho^2f_0''}{\phi_0^2}
(1+\phi_0^2\theta^2)
\Bigl[
2(2+3\phi_1+9\phi_1^2)
+(5+12\phi_1+27\phi_1^2)\phi_0^2\theta^2
\Bigr]
\Biggr\}
\notag\\
&-\frac{2C_0k_{\rm F}\phi_0}
{(1+\theta^2)^{5/2}}
\Biggl[
k_{\rm F}^2\theta(2+5\theta^2+6\theta^4)
-\theta(2+5\theta^2)k_{\rm F}^2
\left(
\frac{6\rho g_\sigma f_0'}{M_0^*}
+\frac{9\rho^2g_\sigma^2f_0'^2}{M_0^{*,2}}
\right)
+9g_\sigma k_{\rm F}\rho^2
(1+\theta^2)(4+5\theta^2)f_0''
\Biggr]
\notag\\
&+\frac{27C_0k_{\rm F}\phi_1(C_1-\phi_1)}
{\theta^3\sqrt{1+\theta^2}}
\Biggl[
k_{\rm F}^2(2\theta^4-3\theta^2-9)
-2(\theta^2+3)k_{\rm F}^2
\left(
\frac{6\rho g_\sigma f_0'}{M_0^*}
+\frac{9\rho^2g_\sigma^2f_0'^2}{M_0^{*,2}}
\right)
+18g_\sigma k_{\rm F}\rho^2
\theta(1+\theta^2)f_0''
\notag\\
&\hspace*{2cm}+
9\frac{\sqrt{1+\theta^2}}{\theta}
\operatorname{arcsinh}\theta
\Biggl[
k_{\rm F}^2
+\frac{2k_{\rm F}^2}{3}
\left(
\frac{6\rho g_\sigma f_0'}{M_0^*}
+\frac{9\rho^2g_\sigma^2f_0'^2}{M_0^{*,2}}
\right)
-2g_\sigma k_{\rm F}\rho^2
\theta f_0''
\Biggr]
\Biggr]
\notag\\
&+8C_0(1+6C_1)k_{\rm F}\phi_0
\Biggl\{
-2k_{\rm F}^2V(\theta)
+\frac{\phi_0}{\sqrt{1+\phi_0^2\theta^2}}
\Biggl[
k_{\rm F}^2\theta
+\theta k_{\rm F}^2
\left(
\frac{6\rho g_\sigma f_0'}{M_0^*}
+\frac{9\rho^2g_\sigma^2f_0'^2}{M_0^{*,2}}
\right)
+\frac{9g_\sigma k_{\rm F}\rho^2}{\phi_0^2}
(1+\phi_0^2\theta^2)f_0''
\Biggr]
\notag\\
&\hspace*{2cm}-
\frac{1}{\sqrt{1+\theta^2}}
\Biggl[
k_{\rm F}^2\theta
+\theta k_{\rm F}^2
\left(
\frac{6\rho g_\sigma f_0'}{M_0^*}
+\frac{9\rho^2g_\sigma^2f_0'^2}{M_0^{*,2}}
\right)
+9g_\sigma k_{\rm F}\rho^2
(1+\theta^2)f_0''
\Biggr]
\Biggr\}
\Biggr\},
\label{eq:Ksym_kin_sum}
\end{align}
Here, $
    \theta={k_{\rm F}}/{M_0^\ast}$, 
as well as $
    E_{\rm F}^*=\sqrt{k_{\rm F}^2+M_0^{*,2}}$ and $
F_{\rm F}^*=\sqrt{p_{\rm F}^2+M_0^{*,2}}$, 
with $p_{\rm F}=\phi_0k_{\rm F}$. Moreover, 
\begin{equation}
V(\theta)=\operatorname{arcsinh}(\phi_0\theta)
-\operatorname{arcsinh}\theta.\end{equation} The derivatives of the SNM scalar field entering the
curvature coefficient are obtained by differentiating its
self-consistency equation with respect to density. Keeping
track of both the explicit density dependence of the scalar
density and its dependence on the effective mass gives
\begin{align}
f_0'
=&\frac{g_\sigma}{R_\sigma}
\frac{\partial\rho^{\rm s}}{\partial\rho},
~~
f_0''
=\frac{1}{R_\sigma}
\left[
g_\sigma
\frac{\partial^2\rho^{\rm s}}{\partial\rho^2}
-2g_\sigma^2f_0'
\frac{\partial^2\rho^{\rm s}}
{\partial\rho\partial M_0^*}+
\left(
g_\sigma^3
\frac{\partial^2\rho^{\rm s}}
{\partial M_0^{*,2}}
-\ddot{W}_\sigma
\right)f_0'^2
\right],
\label{eq:d2sigma0_drho2}
\end{align}
where $W_\sigma(f_0) \equiv m_\sigma^2 f_0 + b_\sigma M_{\mathrm{N}} g_\sigma^3 f_0^2 + c_\sigma g_\sigma^4 f_0^3$, as well as
\begin{align}
\rho^{\rm s}
=&\rho^{\rm s}_{\rm(1)}
+\rho^{\rm s}_{\rm(2)},
~~\text{with}~~\rho^{\rm s}_{\rm(1)}
=\frac{\Delta_0M_0^{*,3}}{\pi^2}
\left(
\theta\sqrt{1+\theta^2}
-\operatorname{arcsinh}\theta
\right),~~
\rho^{\rm s}_{\rm(2)}
=\frac{3\rho C_0}{M_0^*}
\left(
E_{\rm F}^*
-\frac{F_{\rm F}^*}{\phi_0}
\right),
\\
R_\sigma
=&\dot{W}_\sigma
+g_\sigma^2
\frac{\partial\rho^{\rm s}}{\partial M_0^*},~~
\dot{W}_\sigma\equiv\frac{\d W_\sigma}{\d f_0}
=m_\sigma^2
+2b_\sigma M_{\rm N}g_\sigma^3f_0
+3c_\sigma g_\sigma^4f_0^2,
~~
\ddot{W}_\sigma\equiv\frac{\d^2W_\sigma}{\d f_0^2}
=2b_\sigma M_{\rm N}g_\sigma^3
+6c_\sigma g_\sigma^4f_0.
\end{align}
Here, the partial derivatives with respect to
$\rho$ and $M_0^*$ are taken with the other
variable held fixed. This distinction is important because
the total density derivatives of the scalar field also include
the implicit density dependence of $M_0^*$; some detailed relations on the derivative of the scalar density are given in Appendix \ref{app:C}.
The $\rho$-meson contribution is obtained from the density
dependence of the isovector-vector interaction. In particular,
the $\omega$-$\rho$ coupling enters through the effective
meson denominators and the density dependence of $w_0$.
The resulting contribution is
\begin{align}
K_{\rm{sym}}^\rho(\rho)
=&-\frac{9\Lambda_{\rm V}g_\omega^3g_\rho^4
w_0\rho^2}{Q_\omega Q_\rho^2}
\left[
2+\frac{g_\omega\rho}{w_0Q_\omega}
\left(
1-\frac{6c_\omega g_\omega^4w_0^2}
{Q_\omega}
\right)-
\frac{4\Lambda_{\rm V}g_\rho^2g_\omega^3
w_0\rho}{Q_\omega Q_\rho}
\right],
\label{eq:Ksym_rho_analytic}
\end{align}
where\,\cite{Cai2012PLB,Cai2016} \begin{equation}
Q_\rho=m_\rho^2+
\Lambda_{\rm V}g_\rho^2g_\omega^2w_0^2,~~
Q_\omega=m_\omega^2+
3c_\omega g_\omega^4w_0^2.\end{equation}

The $\delta$-meson contribution requires additional care
because the isovector-scalar response depends on the scalar
density and the effective mass. It is convenient to retain the
numerator and denominator of the corresponding symmetry-energy
contribution as two functions, $B$ and $D$. Differentiating
this ratio twice yields
\begin{align}
K_{\rm{sym}}^\delta(\rho)
=&9\rho^2 E_{\rm{sym}}^\delta(\rho)
\left[
2\frac{B''}{B}
-\frac{D''}{D}
+2\left(\frac{B'}{B}-\frac{D'}{D}\right)^2
-\frac{4B'}{\rho B}
+\frac{2D'}{\rho D}
+\frac{2}{\rho^2}
\right], ~~\text{with}~~E_{\rm{sym}}^\delta(\rho)
=-\frac{g_\delta^2}{2m_\delta^2\rho}
\frac{B^2}{D}.
\label{eq:Ksym_delta_exact}
\end{align}
The numerator $B$ collects the scalar density and Fermi surface
terms, including their SRC corrections, whereas $D$ accounts
for the response of the isovector-scalar field. Explicitly,
\begin{align}
B
=&-\frac{3C_0}{\Delta_0}
\left[
C_1\left(1-\frac{1}{\phi_0}\right)
+\frac{\phi_1}{\phi_0}
\right]\rho^{\rm s}_{\rm(1)}
+\frac{4+3C_1}{3}\rho^{\rm s}_{\rm(2)}
+\rho M_0^*
\left[
\frac{\Delta_0-C_0}{E_{\rm F}^*}
+\frac{C_0(1+3\phi_1)}
{\phi_0F_{\rm F}^*}
\right],
\label{eq:B_delta_def}\\
D
=&1+\frac{3g_\delta^2}{m_\delta^2}
\Biggl[
\frac{\rho^{\rm s}_{\rm(1)}}{M_0^*}
-\frac{\Delta_0\rho}{E_{\rm F}^*}
+
\frac{\rho C_0k_{\rm F}^2}{M_0^{*,2}}
\left(
\frac{\phi_0}{F_{\rm F}^*}
-\frac{1}{E_{\rm F}^*}
\right)
\Biggr].
\label{eq:D_delta_def}
\end{align}

To evaluate the curvature coefficient, we differentiate $B$
and $D$ along the self-consistent SNM solution. The first and
second density derivatives of $B$ are
\begin{align}
B'
=&-\frac{3C_0}{\Delta_0}
\left[
C_1\left(1-\frac{1}{\phi_0}\right)
+\frac{\phi_1}{\phi_0}
\right](\rho^{\rm s}_{\rm(1)})'
+\frac{4+3C_1}{3}(\rho^{\rm s}_{\rm(2)})'
\notag\\
&+\left(M_0^*+\rho M_0^{*\prime}\right)
\left[
\frac{\Delta_0-C_0}{E_{\rm F}^*}
+\frac{C_0(1+3\phi_1)}
{\phi_0F_{\rm F}^*}
\right]
-\rho M_0^*
\left[
(\Delta_0-C_0)
\frac{E_{\rm F}^{*\prime}}{E_{\rm F}^{*,2}}
+\frac{C_0(1+3\phi_1)}{\phi_0}
\frac{F_{\rm F}^{*\prime}}{F_{\rm F}^{*,2}}
\right],
\label{eq:B_delta_prime}
\\
B''
=&-\frac{3C_0}{\Delta_0}
\left[
C_1\left(1-\frac{1}{\phi_0}\right)
+\frac{\phi_1}{\phi_0}
\right](\rho^{\rm s}_{\rm(1)})''
+\frac{4+3C_1}{3}(\rho^{\rm s}_{\rm(2)})''
+\left(2M_0^{*\prime}
+\rho M_0^{*\prime\prime}\right)
\left[
\frac{\Delta_0-C_0}{E_{\rm F}^*}
+\frac{C_0(1+3\phi_1)}
{\phi_0F_{\rm F}^*}
\right]
\notag\\
&-2\left(M_0^*+\rho M_0^{*\prime}\right)
\left[
(\Delta_0-C_0)
\frac{E_{\rm F}^{*\prime}}{E_{\rm F}^{*,2}}
+\frac{C_0(1+3\phi_1)}{\phi_0}
\frac{F_{\rm F}^{*\prime}}{F_{\rm F}^{*,2}}
\right]
\notag\\
&+\rho M_0^*
\Biggl\{
(\Delta_0-C_0)
\left[
\frac{2(E_{\rm F}^{*\prime})^2}{E_{\rm F}^{*,3}}
-\frac{E_{\rm F}^{*\prime\prime}}{E_{\rm F}^{*,2}}
\right]+
\frac{C_0(1+3\phi_1)}{\phi_0}
\left[
\frac{2(F_{\rm F}^{*\prime})^2}{F_{\rm F}^{*,3}}
-\frac{F_{\rm F}^{*\prime\prime}}{F_{\rm F}^{*,2}}
\right]
\Biggr\}.
\label{eq:B_delta_second}
\end{align}
The corresponding derivatives of $D$ follow from the same procedure, accounting for the density dependence of the scalar density, the effective mass, and the single-nucleon energies evaluated at the Fermi momentum and the upper cutoff of the SRC-HMT:
\begin{align}
D'
=&\frac{3g_\delta^2}{m_\delta^2}
\Biggl\{
\frac{(\rho^{\rm s}_{\rm(1)})'}{M_0^*}
-\frac{\rho^{\rm s}_{\rm(1)}M_0^{*\prime}}{M_0^{*,2}}
-\Delta_0
\left(
\frac{1}{E_{\rm F}^*}
-\frac{\rho E_{\rm F}^{*\prime}}{E_{\rm F}^{*,2}}
\right)
+
C_0\left[
\frac{k_{\rm F}^2}{M_0^{*,2}}
+\rho\left(
\frac{2k_{\rm F}k_{\rm F}'}{M_0^{*,2}}
-\frac{2k_{\rm F}^2M_0^{*\prime}}{M_0^{*,3}}
\right)
\right]
\left(
\frac{\phi_0}{F_{\rm F}^*}
-\frac{1}{E_{\rm F}^*}
\right)
+
\frac{\rho C_0k_{\rm F}^2}{M_0^{*,2}}
\left(
-\frac{\phi_0F_{\rm F}^{*\prime}}{F_{\rm F}^{*,2}}
+\frac{E_{\rm F}^{*\prime}}{E_{\rm F}^{*,2}}
\right)
\Biggr\},
\label{eq:D_delta_prime}
\\
D''
=&\frac{3g_\delta^2}{m_\delta^2}
\Biggl\{
\frac{(\rho^{\rm s}_{\rm(1)})''}{M_0^*}
-\frac{2(\rho^{\rm s}_{\rm(1)})'M_0^{*\prime}}{M_0^{*,2}}
-\frac{\rho^{\rm s}_{\rm(1)}M_0^{*\prime\prime}}{M_0^{*,2}}
+\frac{2\rho^{\rm s}_{\rm(1)}(M_0^{*\prime})^2}{M_0^{*,3}}
+
\Delta_0
\left[
\frac{2E_{\rm F}^{*\prime}}{E_{\rm F}^{*,2}}
+\frac{\rho E_{\rm F}^{*\prime\prime}}{E_{\rm F}^{*,2}}
-\frac{2\rho(E_{\rm F}^{*\prime})^2}{E_{\rm F}^{*,3}}
\right]
\notag\\
&\hspace{1cm}+
C_0\Biggl[
2\left(
\frac{2k_{\rm F}k_{\rm F}'}{M_0^{*,2}}
-\frac{2k_{\rm F}^2M_0^{*\prime}}{M_0^{*,3}}
\right)
+
\rho\left(
\frac{2[(k_{\rm F}')^2+k_{\rm F}k_{\rm F}'']}{M_0^{*,2}}
-\frac{8k_{\rm F}k_{\rm F}'M_0^{*\prime}}{M_0^{*,3}}
-\frac{2k_{\rm F}^2M_0^{*\prime\prime}}{M_0^{*,3}}
+\frac{6k_{\rm F}^2(M_0^{*\prime})^2}{M_0^{*,4}}
\right)
\Biggr]
\left(
\frac{\phi_0}{F_{\rm F}^*}
-\frac{1}{E_{\rm F}^*}
\right)
\notag\\
&\hspace{1cm}+
2C_0\left[
\frac{k_{\rm F}^2}{M_0^{*,2}}
+\rho\left(
\frac{2k_{\rm F}k_{\rm F}'}{M_0^{*,2}}
-\frac{2k_{\rm F}^2M_0^{*\prime}}{M_0^{*,3}}
\right)
\right]
\left[
-\phi_0\frac{F_{\rm F}^{*\prime}}{F_{\rm F}^{*,2}}
+\frac{E_{\rm F}^{*\prime}}{E_{\rm F}^{*,2}}
\right]
+
\frac{\rho C_0k_{\rm F}^2}{M_0^{*,2}}
\left[
\phi_0\left(
\frac{2(F_{\rm F}^{*\prime})^2}{F_{\rm F}^{*,3}}
-\frac{F_{\rm F}^{*\prime\prime}}{F_{\rm F}^{*,2}}
\right)
-\frac{2(E_{\rm F}^{*\prime})^2}{E_{\rm F}^{*,3}}
+\frac{E_{\rm F}^{*\prime\prime}}{E_{\rm F}^{*,2}}
\right]
\Biggr\}.
\label{eq:D_delta_second}
\end{align}

The density dependence of the effective mass appearing in
these expressions is determined by the same scalar mean field
used in the kinetic contribution. Specifically, the density
derivatives of the Dirac effective mass
$M_0^*=M_{\rm N}-g_\sigma f_0$ are given by
$M_0^{*\prime}=-g_\sigma\partial f_0/\partial\rho$
and $M_0^{*\prime\prime}=-g_\sigma\partial^2 f_0/\partial\rho^2$,
where $f_0$ is the self-consistent isoscalar-scalar mean field
in SNM. Similarly, the first and second density derivatives
of the Fermi momentum are
$k_{\rm F}'=k_{\rm F}/3\rho$ and
$k_{\rm F}''=-2k_{\rm F}/9\rho^2$,
respectively.
For the upper cutoff momentum of the HMT,
$p_{\rm F}=\phi_0 k_{\rm F}$, the corresponding derivatives are
$p_{\rm F}'=p_{\rm F}/3\rho$ and
$p_{\rm F}''=-2p_{\rm F}/9\rho^2$.

Having obtained the curvature coefficient, we now give the
corresponding lower-order quantities in the same decomposition.
For completeness, the symmetry energy and its slope parameter
are written as
$
E_{\rm{sym}}(\rho)
=E_{\rm{sym}}^{\rm{kin}}(\rho)
+E_{\rm{sym}}^\rho(\rho)
+E_{\rm{sym}}^\delta(\rho)$ and $
L(\rho)
=L^{\rm{kin}}(\rho)
+L^\rho(\rho)
+L^\delta(\rho)$.
The corresponding expression for the kinetic sector
is\,\cite{Cai2016}
\begin{align}
E_{\rm{sym}}^{\rm{kin}}(\rho)
=&\frac{k_{\rm F}^2}{6E_{\rm F}^*}\Delta_0
-3E_{\rm F}^*C_0
\left[
C_1\left(1-\frac{1}{\phi_0}\right)
+\frac{\phi_1}{\phi_0}
\right]\notag\\
&-\frac{9M_0^*C_0\phi_1(C_1-\phi_1)}
{8\phi_0\theta^3}\left[
2\theta(1+\theta^2)^{3/2}
-\theta\sqrt{1+\theta^2}
-\operatorname{arcsinh}\theta
\right]\notag\\
&+\frac{2k_{\rm F}C_0(6C_1+1)}{3}
\left[
V(\theta)
-\sqrt{1+\frac{1}{\phi_0^2\theta^2}}
+\sqrt{1+\frac{1}{\theta^2}}
\right]\notag\\
&+\frac{k_{\rm F}C_0}{6}
\left[
(1+3\phi_1)^2\frac{p_{\rm F}}{F_{\rm F}^*}
-2(2+3\phi_1+9\phi_1^2)
\frac{F_{\rm F}^*}{p_{\rm F}}
-\frac{k_{\rm F}}{E_{\rm F}^*}
+\frac{4E_{\rm F}^*}{k_{\rm F}}
\right]\notag\\
&+\frac{C_0(4+3C_1)}{3}
\left[
\frac{(1+3\phi_1)F_{\rm F}^*}{\phi_0}
-E_{\rm F}^*
\right].
\label{eq:Esym_kin_sum}
\end{align}
Taking one density derivative of the kinetic symmetry energy,
including the implicit density dependence of $M_0^*$, gives\,\cite{Cai2016}
\begin{align}
L^{\rm{kin}}(\rho)
=&\frac{\Delta_0k_{\rm F}^2}{6E_{\rm F}^{*,3}}
\left(
E_{\rm F}^{*,2}+M_0^{*,2}
+3\rho g_\sigma M_0^*f_0'
\right)
-\frac{3C_0}{E_{\rm F}^*}
\left[
C_1\left(1-\frac{1}{\phi_0}\right)
+\frac{\phi_1}{\phi_0}
\right]
\left(
k_{\rm F}^2-3\rho g_\sigma M_0^*f_0'
\right)
\notag\\
&-\frac{9C_0\phi_1(C_1-\phi_1)M_0^*}
{8\phi_0\theta^3}
\left\{
\left(
3+\frac{12\rho g_\sigma f_0'}{M_0^*}
\right)\operatorname{arcsinh}\theta
+\frac{\theta}{\sqrt{1+\theta^2}}
\left[
2\theta^4-\theta^2-3
-\frac{12\rho g_\sigma f_0'}{M_0^*}
(1+\theta^2)
\right]
\right\}
\notag\\
&+\frac{2k_{\rm F}C_0(6C_1+1)}{3}
\left[
V(\theta)
-\sqrt{1+\frac{1}{\phi_0^2\theta^2}}
+\sqrt{1+\frac{1}{\theta^2}}
\right]
+\frac{2C_0(6C_1+1)}{3}
\left(
1+\frac{3\rho g_\sigma f_0'}{M_0^*}
\right)
\left(
\frac{F_{\rm F}^*}{\phi_0}
-E_{\rm F}^*
\right)
\notag\\
&+\frac{C_0(4+3C_1)}{3}
\left[
\frac{(1+3\phi_1)p_{\rm F}k_{\rm F}}
{F_{\rm F}^*}
-\frac{k_{\rm F}^2}{E_{\rm F}^*}
-3\rho g_\sigma M_0^*f_0'
\left(
\frac{1+3\phi_1}{\phi_0F_{\rm F}^*}
-\frac{1}{E_{\rm F}^*}
\right)
\right]
\notag\\
&+\frac{k_{\rm F}C_0}{6}
\left[
(1+3\phi_1)^2
\frac{p_{\rm F}}{F_{\rm F}^*}
-2(2+3\phi_1+9\phi_1^2)
\frac{F_{\rm F}^*}{p_{\rm F}}
-\frac{k_{\rm F}}{E_{\rm F}^*}
+\frac{4E_{\rm F}^*}{k_{\rm F}}
\right]
\notag\\
&+\frac{k_{\rm F}C_0M_0^{*,2}}{6}
\left(
1+\frac{3\rho g_\sigma f_0'}{M_0^*}
\right)
\left[
(1+3\phi_1)^2
\frac{p_{\rm F}}{F_{\rm F}^{*,3}}
+\frac{2(2+3\phi_1+9\phi_1^2)}
{p_{\rm F}F_{\rm F}^*}
-\frac{k_{\rm F}}{E_{\rm F}^{*,3}}
-\frac{4}{k_{\rm F}E_{\rm F}^*}
\right].
\label{eq:Lsym_kin_full}
\end{align}

\end{widetext}

The $\rho$-meson contributions to the symmetry energy and
its slope involve the same effective denominators $Q_\rho$
and $Q_\omega$ introduced above\,\cite{Cai2012PLB}:
\begin{align}
E_{\rm{sym}}^\rho(\rho)
=&\frac{g_\rho^2\rho}{2Q_\rho},~~
L^\rho(\rho)
=\frac{3g_\rho^2\rho}{2Q_\rho}
-\frac{3\Lambda_{\rm V}g_\omega^3g_\rho^4
w_0\rho^2}{Q_\omega Q_\rho^2}.
\label{eq:L_rho_analytic}
\end{align}
Finally, differentiating the $\delta$-meson symmetry energy
contribution once gives
\begin{equation}
L^\delta(\rho)
=3E_{\rm{sym}}^\delta(\rho)
\left(
2\rho\frac{B'}{B}
-\rho\frac{D'}{D}
-1
\right).
\label{eq:L_delta}
\end{equation}
Here, factors $B$ and $D$ are defined in
Eqs.\,(\ref{eq:B_delta_def}) and
(\ref{eq:D_delta_def}), respectively, while
their first density derivatives $B'$ and $D'$
are given in Eqs.\,(\ref{eq:B_delta_prime})
and (\ref{eq:D_delta_prime}), respectively.

An important feature of these results is that the kinetic symmetry energy, its slope, and its curvature generally exhibit different and nontrivial density dependences, even when the nucleon momentum distribution takes the FFG form. In this limit, the mean field interactions remain present, and the Dirac effective mass retains its density dependence. Consequently, the slope and curvature involve the first and second density derivatives of the effective mass, respectively, making them sensitive not only to its magnitude but also to how rapidly it changes with density. By comparison, in the noninteracting ideal FFG with a constant nucleon mass, these three quantities share the same simple power law density dependence only in the non-relativistic and ultra-relativistic limits. The density dependence of the scalar mean field therefore introduces additional complexity into the kinetic symmetry energy and its derivatives, which cannot be captured by simply replacing the nucleon mass with an effective mass in the ideal FFG expressions.
Subsection \ref{sub-argument} gives an explicit demonstration.

These differences have important implications for the total symmetry energy at high densities. The kinetic, $\rho$-meson, and $\delta$-meson contributions generally respond differently to increasing density, and their relative importance can change substantially. In particular, the curvature is sensitive to changes in the balance among these contributions, even when their combined effect on the symmetry energy itself is relatively modest. Their interplay can therefore produce a variety of high density behaviors, including softening, stiffening, and potentially non-monotonic behavior. SRC-HMT further modifies this balance by changing the kinetic contribution and its density derivatives, without necessarily requiring changes to the interaction parameters. These results illustrate how the interplay between nucleon momentum distributions and self consistent mean fields can generate diverse density dependences of the symmetry energy within the nonlinear Walecka model.

\subsection{Expressions under the FFG Momentum Distribution}\label{sub-FFG}

As a consistency check, we consider the FFG limit of the
nucleon momentum distribution by setting $C_0=0$ and
$\phi_0=1$, for which $\Delta_0=1$ and all SRC
contributions vanish. This limit refers only to the
momentum distribution and does not remove the mean field
interactions of the nonlinear Walecka model. In particular,
the Dirac effective mass $M_0^*$ remains density dependent.
In this limit, the kinetic symmetry energy, its slope,
and its curvature reduce to
\begin{align}
\widehat{E}_{\rm{sym}}^{\rm{kin}}(\rho)
=&\frac{k_{\rm F}^2}{6E_{\rm F}^*},
\\
\widehat{L}^{\rm{kin}}(\rho)
=&\frac{k_{\rm F}^2}{6E_{\rm F}^{*,3}}
\left(
E_{\rm F}^{*,2}+M_0^{*,2}
+3\rho g_\sigma M_0^*f_0'
\right),
\\
\widehat{K}_{\rm{sym}}^{\rm{kin}}(\rho)
=&-\frac{k_{\rm F}^2}{6E_{\rm F}^{*,5}}
\Biggl[
2M_0^{*,4}+7k_{\rm F}^2M_0^{*,2}
+2k_{\rm F}^4
\notag\\
&+M_0^{*,2}\left(k_{\rm F}^2-2M_0^{*,2}\right)
\left(
\frac{6\rho g_\sigma f_0'}{M_0^*}
+\frac{9\rho^2g_\sigma^2f_0'^2}{M_0^{*,2}}
\right)
\notag\\
&-9\rho^2g_\sigma M_0^*E_{\rm F}^{*,2}f_0''
\Biggr].\label{def-Ksymhat}
\end{align}
These expressions retain the density dependence of the
Dirac effective mass through the self-consistent scalar
field and provide a direct check of the general kinetic
results.

The $\rho$-meson contributions retain their original
forms. For the $\delta$-meson contribution, the functions
$B$ and $D$ become\,\cite{Liu2002}
\begin{align}
\widehat{B}
=&\frac{\rho M_0^*}{E_{\rm F}^*},\\
\widehat{D}
=&1+\frac{g_\delta^2}{m_\delta^2}
\frac{\partial\rho^{\rm s}}{\partial M_0^*}
=1+\frac{3g_\delta^2}{m_\delta^2}
\left(
\frac{{\rho}^{\rm s}}{M_0^\ast}
-\frac{\rho}{E_{\rm F}^\ast}
\right)\notag\\
=&1+\frac{g_\delta^2M_0^{*,2}}{\pi^2m_\delta^2}
\left[
\frac{\theta(3+\theta^2)}{\sqrt{1+\theta^2}}
-3\operatorname{arcsinh}\theta
\right],
\end{align}
where the partial derivative is evaluated at fixed
$k_{\rm F}$ using the FFG momentum distribution.
The corresponding symmetry energy\,\cite{Liu2002}, slope\,\cite{Cai2012PLB}, and curvature
are
\begin{align}
\widehat{E}_{\rm{sym}}^\delta(\rho)
=&-\frac{g_\delta^2\rho}{2m_\delta^2\widehat{D}}
\left(\frac{M_0^*}{E_{\rm F}^*}\right)^2,\label{eq:Esym_delta_FFG}
\\
\widehat{L}^\delta(\rho)
=&3\widehat{E}_{\rm{sym}}^\delta(\rho)
\left[
1-\frac{2k_{\rm F}^2}{E_{\rm F}^{*,2}}
\left(
\frac{1}{3}+\frac{\rho g_\sigma f_0'}{M_0^*}
\right)
-\rho\frac{\widehat{D}'}{\widehat{D}}
\right],
\\
\widehat{K}_{\rm{sym}}^\delta(\rho)
=&9\rho^2\widehat{E}_{\rm{sym}}^\delta(\rho)
\Biggl\{
-\frac{2k_{\rm F}^2}{E_{\rm F}^{*,2}}
\Biggl[
\frac{2M_0^{*,2}}{E_{\rm F}^{*,2}}
\left(
\frac{1}{3\rho}+\frac{g_\sigma f_0'}{M_0^*}
\right)^2
-\frac{1}{3\rho^2}
\notag\\
&+\frac{g_\sigma f_0''}{M_0^*}
+\frac{g_\sigma^2f_0'^2}{M_0^{*,2}}
\Biggr]
-\frac{\widehat{D}''}{\widehat{D}}
+\left(\frac{\widehat{D}'}{\widehat{D}}\right)^2
\notag\\
&+\left[
\frac{2k_{\rm F}^2}{E_{\rm F}^{*,2}}
\left(
\frac{1}{3\rho}+\frac{g_\sigma f_0'}{M_0^*}
\right)
+\frac{\widehat{D}'}{\widehat{D}}
\right]^2
\notag\\
&-\frac{4k_{\rm F}^2}{\rho E_{\rm F}^{*,2}}
\left(
\frac{1}{3\rho}+\frac{g_\sigma f_0'}{M_0^*}
\right)
-\frac{2}{\rho}\frac{\widehat{D}'}{\widehat{D}}
\Biggr\}.
\end{align}
The expressions for $\widehat{D}'$ and $\widehat{D}''$ together with $\widehat{B}'$ and $\widehat{B}''$ can be obtained through Eqs.\,(\ref{eq:B_delta_prime}) to (\ref{eq:D_delta_second}), and we give them in Appendix \ref{app:B}.
These FFG limits provide consistency checks of the
kinetic, $\rho$-meson, and $\delta$-meson contributions.
They also illustrate that, even without SRCs, the
density dependence of the mean fields remains important
for the slope and curvature of the symmetry energy.
To the best of our knowledge, this is the first time
that an explicit analytical expression for $K_{\rm{sym}}$
has been derived within the nonlinear Walecka model
for the FFG nucleon momentum distribution.

\subsection{How SRC-HMT Can Reduce $K_{\textmd{sym}}$}\label{sub-argument}

We use these expressions to make an analytical argument to explain
how SRC-HMT can reduce the curvature $K_{\rm{sym}}$.
To illustrate this effect, we consider the non-relativistic limit
$\theta=k_{\rm F}/M_0^*\ll1$, while retaining the density dependence of
the Dirac effective mass $M_0^*$. For simplicity, we set
$b_\sigma=c_\sigma=0$, and neglect the $\delta$ meson and the
$\rho$-$\omega$ cross coupling. The $\rho$ meson is retained, but its
contribution to $K_{\rm{sym}}$ vanishes because
$E_{\rm{sym}}^\rho$ is proportional to $\rho$ in the absence
of cross couplings.

We first consider the FFG momentum distribution.
Expanding the corresponding expression for
$\widehat{K}_{\rm{sym}}$ of Eq.\,(\ref{def-Ksymhat}) in the non-relativistic limit gives
\begin{equation}
\widehat{K}_{\rm{sym}}\approx
\frac{k_{\rm F}^2}{6M_0^*}
\left(
-2+\frac{12\rho g_\sigma f_0'}{M_0^*}
+\frac{18\rho^2g_\sigma^2f_0'^2}{M_0^{*,2}}
+\frac{9\rho^2g_\sigma f_0''}{M_0^*}
\right).
\end{equation}
The first term comes from the density dependence of
$k_{\rm F}\sim\rho^{1/3}$, while the remaining terms arise
from the density dependence of the scalar mean field.
In the present limit, the scalar density satisfies
$\rho^{\rm{s}}\approx\rho$, so the scalar field equation reduces to
$m_\sigma^2f_0\approx g_\sigma\rho$.
Consequently,
$\rho g_\sigma f_0'\approx M_{\rm N}-M_0^*$
and $f_0''\approx0$ at leading order.
Substituting these relations, we obtain
\begin{equation}\label{Ksym-a1}
\widehat{K}_{\rm{sym}}\approx
\frac{k_{\rm F}^2}{6M_0^*}
\left[
18\left(\frac{M_{\rm N}}{M_0^*}\right)^2
-24\left(\frac{M_{\rm N}}{M_0^*}\right)+4
\right].
\end{equation}
The sign of $\widehat{K}_{\rm{sym}}$ is therefore controlled
by the effective mass. It is positive for
$M_0^*/M_{\rm N}\lesssim0.879$, which includes the typical
value $M_0^*/M_{\rm N}\approx0.6\sim0.7$ near saturation density.

We next consider the SRC-HMT momentum distribution.
Expanding the corresponding kinetic symmetry energy to the
same order, with the SRC-HMT parameters held constant, gives
\begin{equation}\label{Ksym-a2}
K_{\rm{sym}}\approx
\Upsilon_{\rm{sym}}\widehat{K}_{\rm{sym}},
\end{equation}
where $\widehat{K}_{\rm{sym}}$ is defined in Eq.\,(\ref{Ksym-a1}), and the SRC-HMT factor $\Upsilon_{\rm{sym}}$ is given by\,\cite{Cai2015}
\begin{align}
\Upsilon_{\rm{sym}}={}&
1+C_0(1+3C_1)
\left(5\phi_0-8+\frac{3}{\phi_0}\right)\notag\\
&+3C_0\phi_1
\left(1+\frac{3C_1}{5}\right)
\left(5\phi_0-\frac{3}{\phi_0}\right)
+\frac{27C_0\phi_1^2}{5\phi_0}.
\end{align}
For the empirical SRC-HMT parameters adopted in this work,
$\Upsilon_{\rm{sym}}\approx-1.12$\,\cite{Cai2015}. Thus, for the same effective mass,
the FFG and SRC-HMT momentum distributions give opposite
signs for the curvature. In particular,
at $M_0^*/M_{\rm N}\approx0.6\sim0.7$, the FFG result is positive,
whereas the SRC-HMT result is negative.

For an ideal Fermi gas without interactions, $M_0^*\to M_{\rm N}$,  and Eqs.\,(\ref{Ksym-a1}) and (\ref{Ksym-a2}) reduce to $-k_{\rm F}^2/3M_{\rm N}$ and $-\Upsilon_{\rm{sym}} k_{\rm F}^2/3M_{\rm N}$, respectively. The former is negative, whereas the latter is positive for the empirical SRC-HMT parameters adopted here. These results are consistent with the familiar non-relativistic expressions for the kinetic symmetry energy.

This analysis provides a qualitative understanding of how SRC-HMT affects $K_{\rm{sym}}$. The density dependence of the scalar mean field plays an important role in determining the curvature, while SRC-HMT can modify the kinetic contribution and even reverse its sign under typical conditions. Moreover, the full relativistic calculation shows that the $\rho$-meson contribution in Eq.\,(\ref{eq:Ksym_rho_analytic}) is negative (see Appendix \ref{app:D}) and substantially reduces $K_{\rm{sym}}$, an effect that may be enhanced by SRC-HMT.

\section{SRC-HMT Effects on $K_{\textmd{sym}}$ and Density Dependence of Symmetry Energy}
\label{sec:Esym_results}

In this section, we investigate how SRC-HMT affects the density dependence of the nuclear symmetry energy and the allowed range of its curvature $K_{\rm{sym}}$ within the nonlinear RMF framework. In Subsection \ref{subsec:fitting-scheme}, we first calibrate the HMT and FFG models using the same empirical nuclear matter properties at saturation and examine their consistency with existing constraints on the SNM EOS. In Subsection \ref{subsec:Esym_SRC-HMT}, we then compare the symmetry energies predicted by the two frameworks and determine the ranges of $K_{\rm{sym}}$ compatible with the tidal deformability constraint from GW170817 and the NS maximum mass requirement. Particular attention is given to the role of SRC-HMT in softening the symmetry energy while slightly stiffening the SNM EOS, providing a physical mechanism for reconciling these astrophysical constraints. We demonstrate that SRC-HMT permits substantially more negative values of $K_{\rm{sym}}$ and significantly enlarges the physically allowed parameter space compared with the FFG framework.

\subsection{Model Calibration and Nuclear Matter Constraints}
\label{subsec:fitting-scheme}

We calibrate the nonlinear RMF models in both the HMT and FFG frameworks using empirical nuclear matter properties at saturation. For a consistent comparison, we adopt the same saturation density $\rho_0\approx0.15\,\rm{fm}^{-3}$, binding energy per nucleon $E_0(\rho_0)\approx-16.0\,\rm{MeV}$, Dirac effective mass ratio $M_0^*/M_{\rm N}\approx0.65$, and incompressibility coefficient $K_0\approx230\,\rm{MeV}$ in both frameworks. These values are consistent with empirical constraints from nuclear structure and nuclear matter studies\,\cite{LiCaiXieZhang2021}. In the isovector sector, we fix the symmetry energy at saturation to $E_{\rm{sym}}(\rho_0)\approx32.0\,\rm{MeV}$ and investigate its slope $L$ and curvature $K_{\rm{sym}}$. The remaining isoscalar parameter is determined through the SNM skewness coefficient $J_0$, which governs the behavior of EOS of SNM at high densities\,\cite{Cai2017NST}. To incorporate the NS maximum mass constraint, we choose $J_0$ such that the parameter set with the minimum achievable $K_{\rm{sym}}$ at $L=40\,\rm{MeV}$ supports $M_{\rm{NS}}^{\max}\approx2.01M_\odot$\,\cite{Antoniadis2013}. This gives $J_0\approx-528\,\rm{MeV}$ for HMT and $J_0\approx-455\,\rm{MeV}$ for FFG, corresponding to $K_{\rm{sym}}\approx-339\,\rm{MeV}$ and $-120\,\rm{MeV}$, respectively. The resulting isoscalar couplings and SRC parameters are summarized in TAB.\,\ref{tab:isoscalar_params}.

\begin{figure}[h!]
    \centering
    \includegraphics[width=0.95\columnwidth]{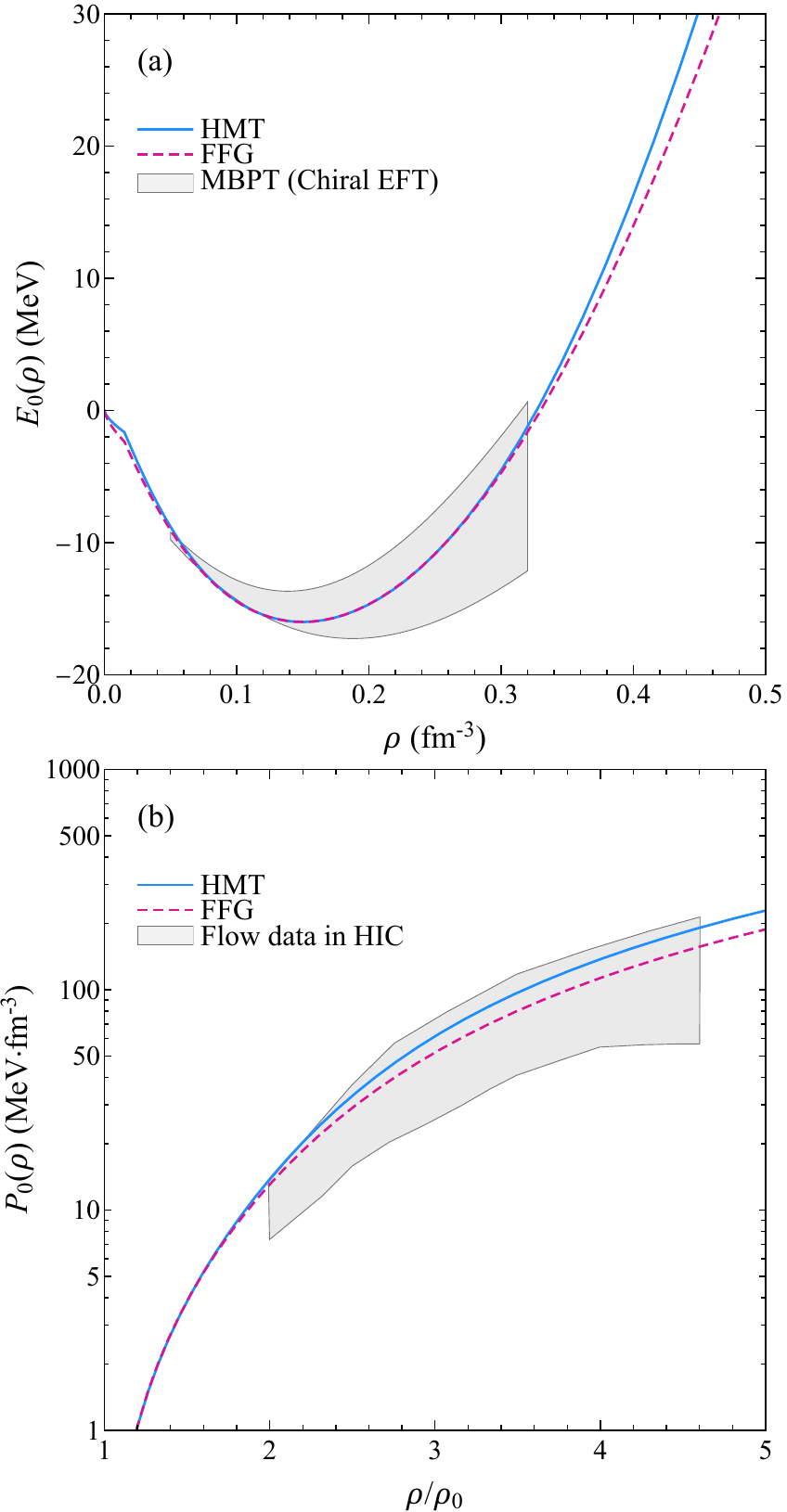}
\caption{(Color online). Density dependence of the energy per nucleon $E_0(\rho)$ (a) and pressure $P_0(\rho)$ (b) in SNM for the HMT (solid lightblue lines) and FFG (dashed magenta lines) frameworks, using the parameter sets listed in TAB.\,\ref{tab:isoscalar_params}. Both frameworks reproduce the same empirical saturation properties but have different skewness coefficients, $J_0\approx-528\,\mathrm{MeV}$ for HMT and $J_0\approx-455\,\mathrm{MeV}$ for FFG, determined by the NS maximum mass requirement $M_{\rm{NS}}^{\max}\approx2.01M_\odot$. The shaded regions represent the microscopic MBPT calculations based on chiral effective field theory\,\cite{Drischler2026} in (a) and the collective flow constraints from heavy ion collisions\,\cite{Danielewicz2002} in panel (b).}
    \label{fig:SNM_E0_P0}
\end{figure}

FIG.\,\ref{fig:SNM_E0_P0} compares the density dependence of the energy per nucleon $E_0(\rho)$ and pressure $P_0(\rho)$ in SNM for the HMT and FFG frameworks, using the parameter sets listed in TAB.\,\ref{tab:isoscalar_params}. As shown in FIG.\,\ref{fig:SNM_E0_P0}(a), the two energy curves are nearly indistinguishable around saturation density and remain very close up to $\rho/\rho_0\approx2$. Both are consistent with the microscopic MBPT calculations based on chiral effective field theory\,\cite{Drischler2026}. At higher densities, the HMT EOS becomes slightly stiffer than the FFG EOS. This difference reflects the combined effects of high momentum nucleons and the smaller $\omega$-meson self coupling $c_\omega$ in the HMT framework, which reshapes the EOS at these densities.
As shown in FIG.\,\ref{fig:SNM_E0_P0}(b), the pressures predicted by the two frameworks are also very similar near saturation but gradually separate as density increases, with the HMT framework giving a higher pressure. Both predictions remain consistent with the collective flow constraints from heavy-ion collisions (HICs)\,\cite{Danielewicz2002} over the density range considered. These results show that the HMT and FFG frameworks can reproduce the same empirical saturation properties while maintaining consistency with independent constraints on the SNM EOS. The relatively small differences between their SNM predictions provide a useful basis for examining how SRCs influence the symmetry energy and its curvature.

\renewcommand*{\tablename}{\small TAB.}
\begin{table}[h!]
    \centering
    \renewcommand{\arraystretch}{1.1}
    \setlength{\tabcolsep}{2pt}
    \begin{tabular}{ccc}
        \hline\hline
        Quantity & HMT $n_{\v{k}}^J$ of Eq.\,(\ref{eq:n_k_SRC}) & FFG $n_{\v{k}}^J$ \\
        \hline
        $\rho_0$ ($\mathrm{fm}^{-3}$)
            & 0.15 & 0.15 \\
        $E_0(\rho_0)$ (MeV)
            & $-16.0$ & $-16.0$ \\
        $K_0$ (MeV)
            & 230 & 230 \\
        $M_0^*(\rho_0)/M_{\rm N}$
            & 0.65 & 0.65 \\
        $E_{\rm{sym}}(\rho_0)$ (MeV)
            & 32.0 & 32.0 \\
        $J_0$ (MeV)
            & $-528$ & $-455$ \\
        $L$ (MeV)
            & 40.0 & 40.0 \\
        $K_{\rm{sym}}$ (MeV)
            & $-339$ & $-120$ \\
        \hline
        $g_\sigma$
            & 9.9087 & 9.6486 \\
        $g_\omega$
            & 11.0679 & 12.1540 \\
        $b_\sigma$
            & 0.003627 & 0.002320 \\
        $c_\sigma$
            & $-0.004019$ & $-0.000608$ \\
        $c_\omega$
            & 0.003373 & 0.004093 \\
        $g_\rho$
            & 8.1249 & 8.3274 \\
        $g_\delta$
            & 1.9589 & 7.9080 \\
        $\Lambda_{\rm V}$
            & 0.05686 & 0.05757 \\
        \hline
        $C_0$
            & 0.161 & 0 \\
        $C_1$
            & $-0.25$ & / \\
        $\phi_0$
            & 2.38 & 1 \\
        $\phi_1$
            & $-0.56$ & / \\
        \hline\hline
    \end{tabular}
    \caption{
    Nuclear matter saturation properties, RMF coupling
    constants, and SRC parameters for the HMT and FFG
    frameworks. Both frameworks reproduce the same
    empirical saturation properties. The
    skewness coefficients $J_0$  are chosen such that,
    at $L=40\,\mathrm{MeV}$, the parameter sets with
    the minimum achievable symmetry energy curvature
    $K_{\rm{sym}}$ satisfy the NS maximum
    mass requirement $M_{\rm{NS}}^{\max}\approx
    2.01M_\odot$.
    }
    \label{tab:isoscalar_params}
\end{table}

\subsection{SRC-HMT Broadens the Allowed Symmetry Energy Curvature Range Consistent with Astrophysical Constraints}\label{subsec:Esym_SRC-HMT}

\begin{figure}[h!]
    \centering
    \includegraphics[width=0.95\columnwidth]{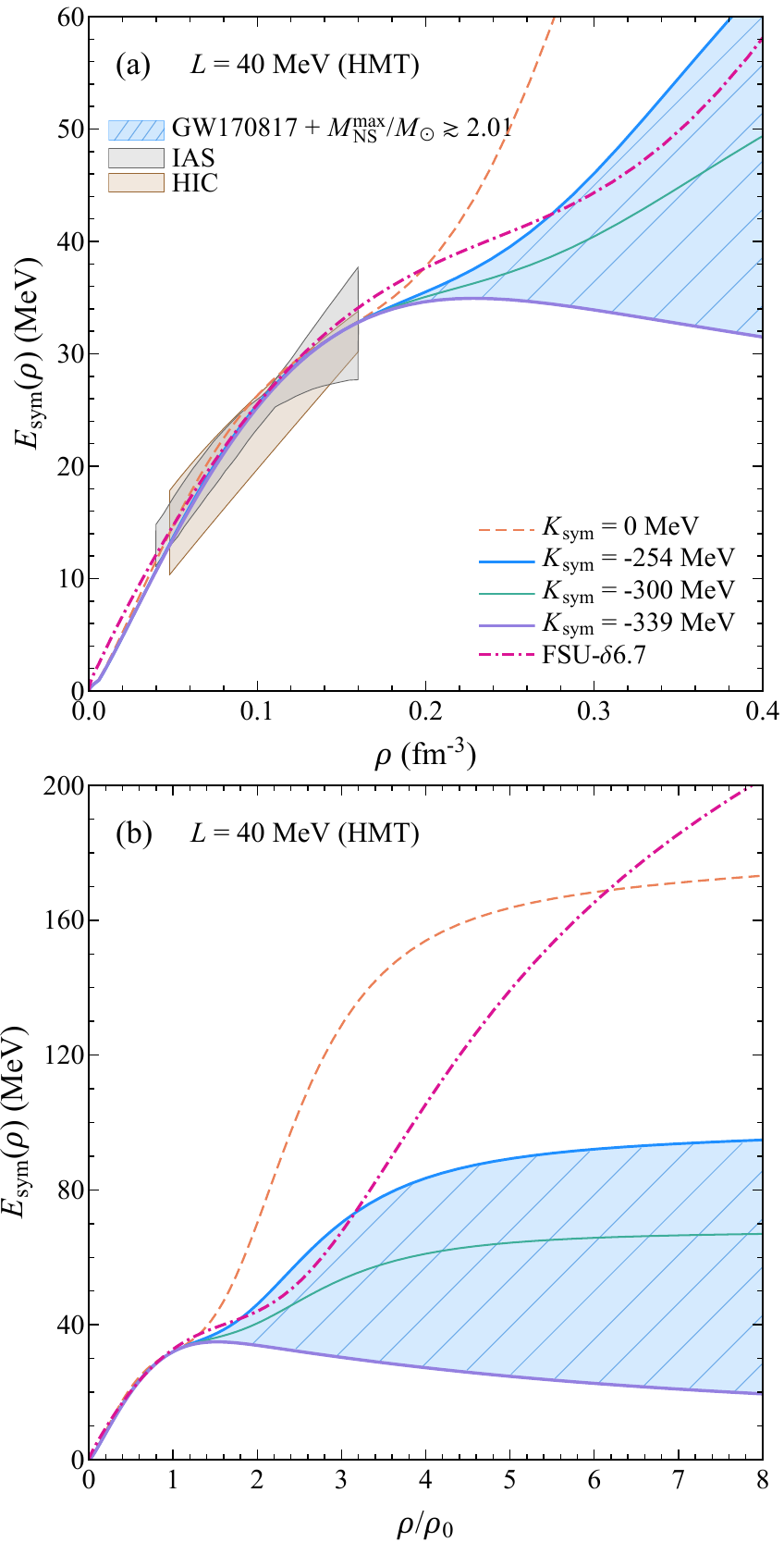}
    \caption{(Color online).
Density dependence of nuclear symmetry energy $E_{\rm{sym}}(\rho)$ with SRC-HMT at $L=40\,\mathrm{MeV}$ for (a) $0\leq\rho\leq0.4\,\mathrm{fm}^{-3}$ and (b) $0\leq\rho\leq8\rho_0$. The curves correspond to $K_{\rm{sym}}=0$\,MeV, $-254$\,MeV, $-300$\,MeV, and $-339\,\mathrm{MeV}$, with FSU-$\delta6.7$\,\cite{Li2022} shown for comparison. The blue shaded region satisfies the GW170817 tidal deformability constraint\,\cite{Abbott2017,Abbott2018} and the NS maximum mass requirement $M_{\rm{NS}}^{\max}/M_\odot\gtrsim2.01$\,\cite{Antoniadis2013}. The gray and brown regions in (a) indicate IAS\,\cite{Danielewicz2014}  and HIC\,\cite{Tsang2009,Tsang2012} constraints, respectively. Sufficiently negative $K_{\rm{sym}}$ allows supersoft symmetry energy at high densities (solid violet) while remaining consistent with both astrophysical constraints.}
\label{fig:Esym_HMT}
\end{figure}

Having established the model parameters and examined the EOS of SNM, we now investigate how SRC-HMT modifies the density dependence of the nuclear symmetry energy, particularly its curvature $K_{\rm{sym}}$ and the possibility of supersoft behavior at high densities. FIG.\,\ref{fig:Esym_HMT} shows $E_{\rm{sym}}(\rho)$ in the HMT framework at $L=40\,\mathrm{MeV}$ over two density ranges: (a) $0\leq\rho\leq0.4\,\mathrm{fm}^{-3}$ and (b) $0\leq\rho\leq8\rho_0$. The curves correspond to $K_{\rm{sym}}=0$\,MeV, $-254$\,MeV, $-300$\,MeV, and $-339\,\mathrm{MeV}$, together with the FSU-$\delta6.7$ model for comparison; the latter has an interaction term in the form of $\sigma^2\vec{\delta}^2$\,\cite{Li2022}. Since the RMF framework can readily produce positive values of $K_{\rm{sym}}$, the $K_{\rm{sym}}=0$\,MeV curve is included only as a reference. All curves reproduce the same symmetry energy and slope at saturation density but diverge progressively at higher densities. In particular, increasingly negative $K_{\rm{sym}}$ leads to stronger softening and can eventually produce a supersoft symmetry energy that decreases with increasing density.

The blue shaded region in FIG.\,\ref{fig:Esym_HMT} represents the range of symmetry energies consistent with the GW170817 tidal deformability constraint and the NS maximum mass requirement. Its upper boundary corresponds to $K_{\rm{sym}}\approx-254\,\mathrm{MeV}$, for which $\Lambda_{1.4}\approx580$, which is the adopted upper limit from GW170817\,\cite{Abbott2018}. The lower boundary corresponds to $K_{\rm{sym}}\approx-339\,\mathrm{MeV}$ (solid violet), giving the softest symmetry energy achievable within the present RMF model at $L=40\,\mathrm{MeV}$, while satisfying astrophysical constraints and keeping the other nuclear matter parameters fixed as in TAB.\,\ref{tab:isoscalar_params}.
Thus, the allowed range is approximately $-339\,\mathrm{MeV}\lesssim K_{\rm{sym}}\lesssim-254\,\mathrm{MeV}$. 
The isobaric analog state (IAS)\,\cite{Danielewicz2014} and heavy ion collision (HIC)\,\cite{Tsang2009,Tsang2012} constraints shown in FIG.\,\ref{fig:Esym_HMT}(a) provide complementary information near the saturation density, where the predicted symmetry energies remain broadly consistent with experimental constraints.

The high density behavior of $E_{\rm{sym}}(\rho)$ is particularly interesting. As shown in panel (b) of FIG.\,\ref{fig:Esym_HMT}, the FSU-$\delta6.7$ model\,\cite{Li2022}, incorporating an effective $\sigma^2\vec{\delta}^2$ self-interaction term, predicts a rapidly increasing symmetry energy, reaching approximately $140$-$165\,\mathrm{MeV}$ at $\rho\approx5$-$6\rho_0$. In comparison, the HMT model with $K_{\rm{sym}}\approx-300\,\mathrm{MeV}$ (green solid) exhibits a nearly flat density dependence at high densities. Interestingly, the two models predict similar symmetry energies below approximately $2\rho_0$, but markedly different trends at higher densities. With a more negative $K_{\rm{sym}}\approx-339\,\mathrm{MeV}$, the HMT model exhibits a supersoft symmetry energy that decreases at high densities, reaching only about $25$-$30\,\mathrm{MeV}$ over the same density range. This pronounced softening remains compatible with astrophysical constraints: The corresponding EOS supports a maximum NS mass of $M_{\rm{NS}}^{\max}\approx2.01M_\odot$ while satisfying the GW170817 tidal deformability constraint\,\cite{Abbott2017,Abbott2018}. As $K_{\rm{sym}}$ increases, the maximum mass also increases, so the mass requirement remains satisfied. These results demonstrate that SRC-HMT can accommodate a supersoft symmetry energy without requiring a correspondingly soft overall EOS. Although the numerical boundaries depend on the adopted nuclear matter parameters, SRC-HMT substantially extends the range of symmetry energy curvatures and high density behaviors compatible with nuclear and astrophysical constraints. The implications for NS masses, radii, and tidal deformabilities are examined in the next section.

\renewcommand*{\tablename}{\small TAB.}
\begin{table}[h!]
    \centering
    \renewcommand{\arraystretch}{1.1}
    \setlength{\tabcolsep}{6pt}
    \begin{tabular}{cccc}
        \hline\hline
        $K_{\rm{sym}}$ (MeV) &
        $J_{\rm{sym}}$ (MeV) &
        $\rho/\rho_0\leftrightarrow L(\rho)=0$ &
        $L_{\min}$ (MeV) \\
        \hline
        $-254$ & 2688 & /    & 34.9 \\
        $-300$ & 1991 & /    & 24.8 \\
        $-339$ & 1202 & 1.5 & $-14.5$ \\
        \hline\hline
    \end{tabular}
    \caption{
    Symmetry energy curvature $K_{\rm{sym}}$, skewness
    $J_{\rm{sym}}$, the density
    at which $L(\rho)$ vanishes, and its minimum
    value $L_{\min}$ within the truncated expansion,
    for the selected SRC-HMT models at $L=40$\,MeV.
    Slashes ``/'' indicate no zero within this approximation
    at $\rho\geq\rho_0$.
    }
    \label{tab:Jsym_HMT}
\end{table}

\begin{figure}[h!]
    \centering
    \includegraphics[width=0.95\columnwidth]{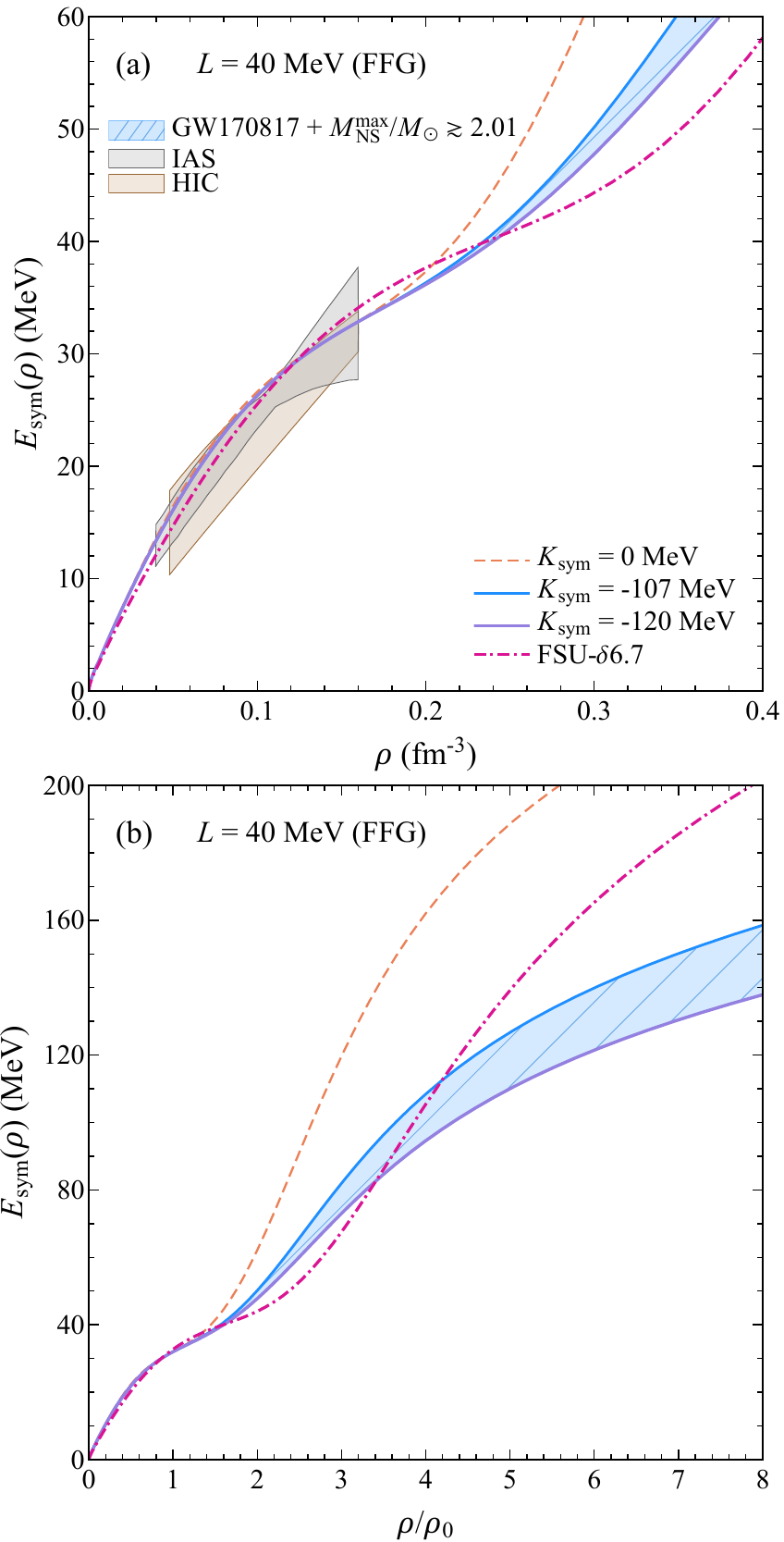}
\caption{(Color online). Same as FIG.\,\ref{fig:Esym_HMT}, but for the FFG framework at $L=40\,\mathrm{MeV}$. The curves correspond to $K_{\rm{sym}}=0$\,MeV, $-107$\,MeV, and $-120\,\mathrm{MeV}$. The blue shaded region represents the symmetry energies consistent with the GW170817 tidal deformability constraint and the NS maximum mass requirement, bounded by $K_{\rm{sym}}\approx-107\,\mathrm{MeV}$ and $-120\,\mathrm{MeV}$.}
\label{fig:Esym_FFG}
\end{figure}

To better understand the supersoft behavior in
FIG.\,\ref{fig:Esym_HMT}, we examine the symmetry
energy skewness $J_{\rm{sym}}$, with the corresponding
values listed in TAB.\,\ref{tab:Jsym_HMT}.
At fixed slope $L=40$\,MeV, $J_{\rm{sym}}$ decreases
from about $2688$\,MeV to $1202$\,MeV as
$K_{\rm{sym}}$ becomes more negative, while remaining
positive. Expanding the density dependent slope and
curvature around saturation to the order involving
$J_{\rm{sym}}$, we obtain
\begin{align}
&L(\rho)
\approx({\rho}/{\rho_0})
\left(L+K_{\rm{sym}}\chi
+2^{-1} J_{\rm{sym}}\chi^2\right),\\
&K_{\rm{sym}}(\rho)
\approx({\rho}/{\rho_0})^2
\left(K_{\rm{sym}}+J_{\rm{sym}}\chi\right).
\end{align}
The density at which $L(\rho)$ vanishes within
the truncated expansion is listed in
TAB.\,\ref{tab:Jsym_HMT}.
Only the case with $K_{\rm{sym}}\approx-339$\,MeV
develops a negative $L(\rho)$, beginning at approximately
$1.5\rho_0$, where the symmetry energy reaches a maximum value 
of about $34.9$\,MeV.
This is consistent with the onset of supersoft behavior
shown in FIG.\,\ref{fig:Esym_HMT}.
The minimum of $L(\rho)$ is approximately $-14.5$\,MeV,
while $L(\rho)$ remains positive for the other two cases.
Interestingly, $K_{\rm{sym}}(\rho)$ for the case $K_{\rm{sym}}\approx-339\,\rm{MeV}$ becomes positive above
approximately $1.9\rho_0$, even though the symmetry energy
continues to decrease.
This illustrates that a decreasing symmetry energy does
not necessarily require negative curvature, since the
slope and curvature characterize different aspects of
its density dependence.
Moreover, the full EOS calculation confirms that the
$E_{\rm{sym}}(\rho)$ continues to decrease with density
for $K_{\rm{sym}}\approx-339$\,MeV, exhibiting the supersoft behavior.

For comparison, FIG.\,\ref{fig:Esym_FFG} shows the corresponding symmetry energies in the FFG framework at the same slope parameter $L=40\,\mathrm{MeV}$. The blue shaded region is bounded by $K_{\rm{sym}}\approx-107\,\mathrm{MeV}$, corresponding to $\Lambda_{1.4}\approx580$, and $K_{\rm{sym}}\approx-120\,\mathrm{MeV}$, the minimum achievable curvature in this framework under the adopted nuclear matter parameters and NS maximum mass requirement. Compared with the HMT results in FIG.\,\ref{fig:Esym_HMT}, the allowed curvature range is considerably narrower and shifted toward less negative values. Moreover, the symmetry energy increases monotonically with density for all the FFG parameter sets shown, including those with the most negative allowed curvature. Thus, within the present FFG framework, the symmetry energy cannot exhibit supersoft behavior or even substantial softening at high densities; instead, it increases rapidly and becomes increasingly stiff beyond approximately $2\rho_0$. This comparison demonstrates that SRC-HMT substantially extends the range of symmetry energy behaviors compatible with the same nuclear matter properties and astrophysical constraints.

\begin{figure}[h!]
    \centering
    \includegraphics[width=0.95\columnwidth]{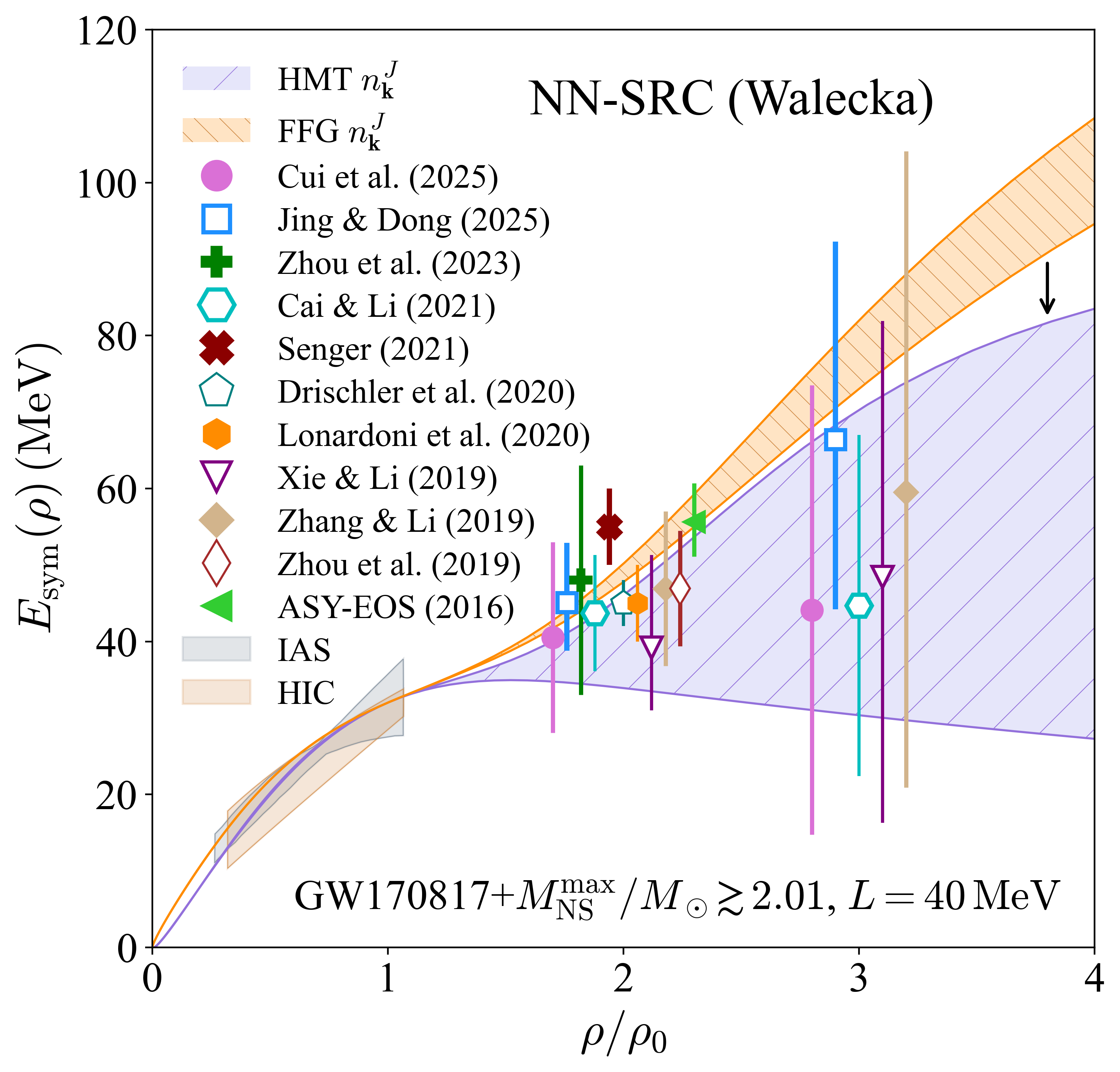}
\caption{(Color online). Density dependence of
$E_{\rm{sym}}(\rho)$ for the HMT and FFG momentum distributions
over $0\le\rho/\rho_0\le4$, represented by the lavender and tan
bands, respectively.
Both bands are consistent with the constraints from GW170817
and $M_{\rm{NS}}^{\rm{max}}/M_\odot\gtrsim2.01$,
with $L=40$\,MeV and the remaining parameters fixed
at the values listed in TAB.\,\ref{tab:isoscalar_params}.
Also shown are partial constraints on $E_{\rm{sym}}(\rho)$ at
$2\rho_0$ and $3\rho_0$ from previous
studies\,\cite{Cui2025NST,JingDong2025,CaiLi2021,Drischler2020,
Lonardoni2020,XieLi2019,ZhangLi2019,Zhou2019,Russotto2016,Zhou2023,Senger2021}.
For clarity, constraints at the same density are slightly
displaced horizontally.}
\label{fig:Esym_comp}
\end{figure}

Comparing the allowed bands in the lower panels of FIG.\,\ref{fig:Esym_HMT} and FIG.\,\ref{fig:Esym_FFG} confirms the well established tendency of SRCs to soften the symmetry energy at supra-saturation densities, at a fixed slope parameter $L$\,\cite{Cai2016,CaiLi2022Gogny}. In particular, the HMT distribution permits substantially lower symmetry energies as the density increases, highlighting the important role of nucleon momentum distributions in shaping the high density $E_{\rm{sym}}(\rho)$.

To further assess our results, FIG.\,\ref{fig:Esym_comp} compares the predicted symmetry energy bands with partial existing constraints at $2\rho_0$ and $3\rho_0$ from various theoretical analyses, heavy-ion collision experiments, and NS observations\,\cite{Cui2025NST,JingDong2025,CaiLi2021,Drischler2020,Lonardoni2020,XieLi2019,ZhangLi2019,Zhou2019,Russotto2016,Zhou2023,Senger2021}. These studies employ different approaches, including microscopic nuclear many-body calculations, transport model analyses of heavy-ion collisions, Bayesian analyses of NS observations, and phenomenological reconstructions of the density dependence of the symmetry energy. The comparison is made at $L=40$\,MeV, with both the FFG and HMT bands satisfying the constraints from GW170817 and massive NSs. Around $2\rho_0$, most existing estimates favor a symmetry energy of approximately $40$-$50$\,MeV\,\cite{LiCaiXieZhang2021,Li2026Sym}, with uncertainties that generally overlap both predicted bands. This suggests that the available constraints at intermediate densities do not strongly distinguish between the FFG and HMT momentum distributions. At $3\rho_0$, however, the uncertainties become considerably larger, reflecting the limited knowledge of nuclear interactions at high densities and the model dependence involved in extracting the symmetry energy from NS observations. In particular, NS observables constrain the total EOS of neutron-rich matter rather than the symmetry energy alone, allowing different density dependences of $E_{\rm{sym}}(\rho)$ to remain compatible with similar observations. As shown in the figure, the HMT band extends to substantially lower symmetry energies than the FFG band, particularly beyond $2\rho_0$, while remaining compatible with the available constraints. Thus, current nuclear and astrophysical information does not exclude a supersoft symmetry energy induced by SRCs, even when the EOS supports massive NSs and satisfies the GW170817 constraint. Finally, we emphasize that the comparison is illustrated for $L=40$\,MeV. Other choices of $L$ may shift the predicted bands and modify their overlap with existing constraints, although the general tendency of SRCs to soften the symmetry energy at supra-saturation densities is expected to persist.

\begin{figure}[h!]
    \centering
    \includegraphics[width=0.95\columnwidth]{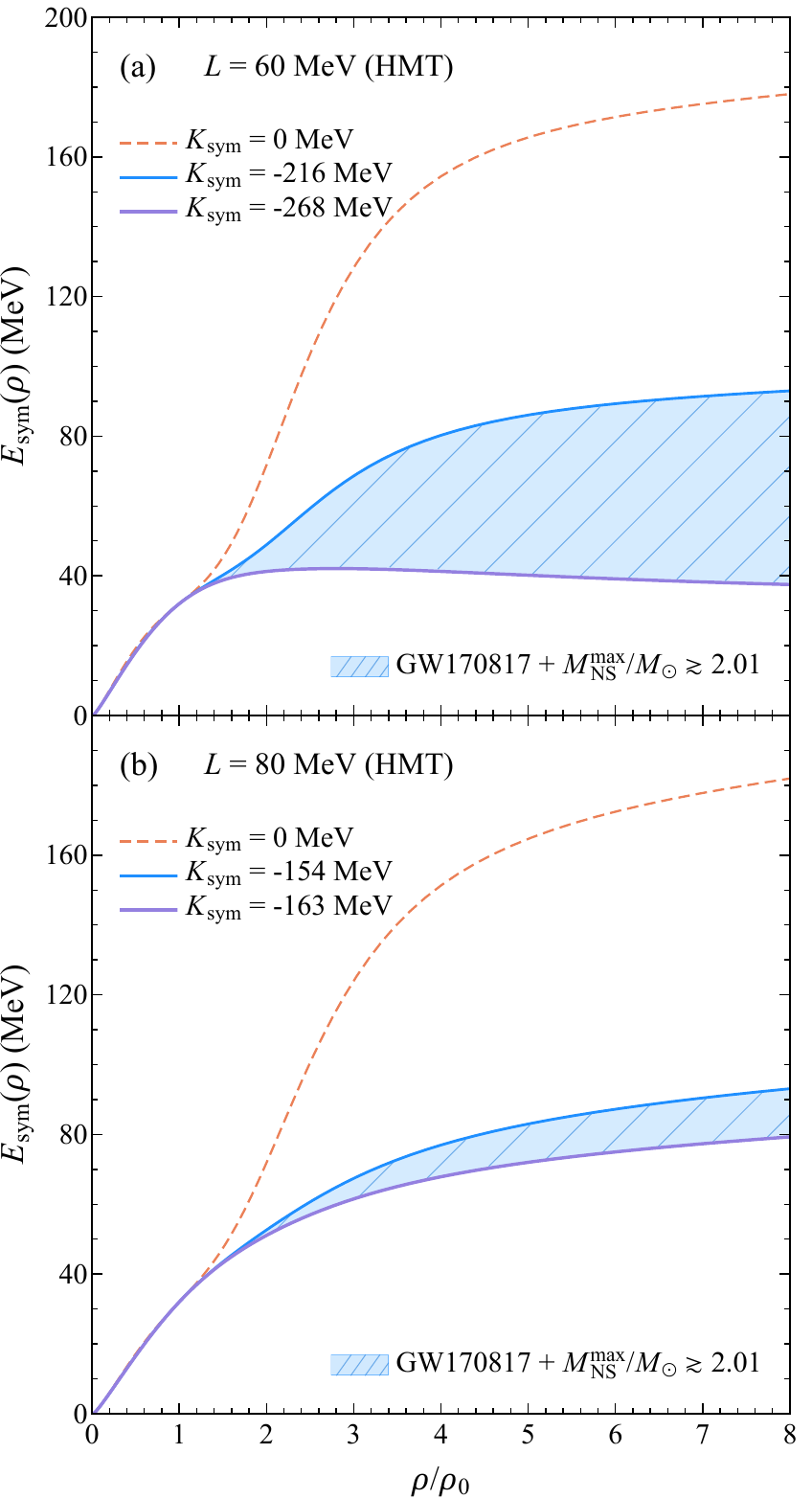}
\caption{(Color online). Density dependence of $E_{\rm{sym}}(\rho)$ in the HMT framework at (a) $L=60\,\mathrm{MeV}$ and (b) $L=80\,\mathrm{MeV}$. The curves correspond to $K_{\rm{sym}}=0$\,MeV, $-216$\,MeV, and $-268\,\mathrm{MeV}$ in (a), and $K_{\rm{sym}}=0$\,MeV, $-154$\,MeV, and $-163\,\mathrm{MeV}$ in (b). The blue shaded regions represent the ranges of symmetry energies consistent with the tidal deformability constraint from GW170817 and the NS maximum mass requirement $M_{\rm{NS}}^{\max}/M_\odot\gtrsim2.01$.}
\label{fig:Esym_HMT_L6080}
\end{figure}

We next examine how the allowed range of $K_{\rm{sym}}$ changes with increasing symmetry energy slope $L$. FIG.\,\ref{fig:Esym_HMT_L6080} shows the corresponding results within the HMT framework for $L=60$ and $80$\,MeV. At $L=60\,\mathrm{MeV}$, the allowed curvature ranges from approximately $K_{\rm{sym}}\approx-268$\,MeV to $-216\,\mathrm{MeV}$, while at $L=80\,\mathrm{MeV}$, it ranges from approximately $-163$\,MeV to $-154\,\mathrm{MeV}$. In both cases, the upper boundaries are determined by the adopted GW170817 constraint $\Lambda_{1.4}\lesssim580$, whereas the lower boundaries correspond to the minimum achievable $K_{\rm{sym}}$ within the present model. As $L$ increases, the allowed curvature shifts toward less negative values and its range becomes narrower. At $L=60\,\mathrm{MeV}$, the symmetry energy can become nearly flat or even decrease slightly at high densities, whereas at $L=80\,\mathrm{MeV}$, it continues to increase but exhibits a substantially softened density dependence. Thus, although a larger $L$ increasingly restricts the possibility of supersoft symmetry energy, SRC-HMT still accommodates relatively soft high density symmetry energies consistent with both the GW170817 tidal deformability and NS maximum mass constraints.

This comparison is particularly significant in understanding the role of SRC-HMT in extending the parameter space compatible with astrophysical observations. A larger $L$ generally favors higher pressure above saturation density, tending to increase NS radii and tidal deformabilities, while sufficiently negative $K_{\rm{sym}}$ can compensate for this effect. However, the FFG framework has a much more restricted range of achievable negative curvatures. As summarized in TAB.\,\ref{tab:Ksym_allowed}, at $L=40\,\mathrm{MeV}$, the allowed range of $K_{\rm{sym}}$ spans approximately $85\,\mathrm{MeV}$ in the HMT framework, compared with only $13\,\mathrm{MeV}$ in the FFG framework, a difference of more than a factor of six. As $L$ increases to $60$\,MeV and $80\,\mathrm{MeV}$, the HMT framework still admits allowed curvature ranges of approximately $52$\,MeV and $9\,\mathrm{MeV}$, respectively, whereas no FFG parameter set simultaneously satisfies the adopted tidal deformability and NS maximum mass constraints. Thus, SRC-HMT substantially extends the viable parameter space in both $L$ and $K_{\rm{sym}}$, allowing greater flexibility in the high density behavior of the symmetry energy.

\renewcommand*{\tablename}{\small TAB.}
\begin{table}[h!]
    \centering
    \renewcommand{\arraystretch}{1.1}
    \setlength{\tabcolsep}{5pt}
    \begin{tabular}{ccc}
        \hline\hline
        $L$ (MeV) & Allowed $K_{\rm{sym}}$ range (MeV)
        & $\Delta K_{\rm{sym}}$ (MeV) \\
        \hline
        \multicolumn{3}{c}{HMT} \\
        \hline
        40 & $-339\lesssim K_{\rm{sym}}\lesssim-254$
           & 85 \\
        60 & $-268\lesssim K_{\rm{sym}}\lesssim-216$
           & 52 \\
        80 & $-163\lesssim K_{\rm{sym}}\lesssim-154$
           & 9 \\
        \hline
        \multicolumn{3}{c}{FFG} \\
        \hline
        40 & $-120\lesssim K_{\rm{sym}}\lesssim-107$
           & 13 \\
        60 & / & / \\
        80 & / & / \\
        \hline\hline
    \end{tabular}
    \caption{
    Allowed ranges of the symmetry energy curvature
    $K_{\rm{sym}}$ and their widths
    $\Delta K_{\rm{sym}}$ in the HMT and FFG
    frameworks at different values of $L$.
    The ranges satisfy the GW170817 constraint
    $\Lambda_{1.4}\lesssim580$ and the NS maximum
    mass requirement
    $M_{\rm{NS}}^{\max}/M_\odot\gtrsim2.01$.
    ``/'' indicates that no parameter set
    satisfies both constraints.
    }
    \label{tab:Ksym_allowed}
\end{table}

More importantly, these results demonstrate that the existence of massive NSs does not, by itself, rule out a supersoft symmetry energy at high densities, even when the relatively small tidal deformability inferred from GW170817\,\cite{Abbott2017,Abbott2018}  is taken into account. The key point is that the stiffness of beta-stable matter depends on the overall EOS rather than the symmetry energy alone. In particular, SRC-HMT slightly stiffens the SNM EOS while substantially softening the symmetry energy, allowing these effects to partially compensate in dense NS matter. Moreover, a softer symmetry energy generally favors a smaller proton fraction $x_{\rm p}$ as shown in FIG.\,\ref{fig:xp}; since the energy cost of neutron-proton asymmetry is reduced, this corresponds to a larger isospin asymmetry $\delta$. Within the quadratic approximation, the symmetry energy contributes $E_{\rm{sym}}(\rho)\delta^2$ to the energy per nucleon. Thus, a softer symmetry energy with a larger $\delta$ and a stiffer symmetry energy with a smaller $\delta$ may yield comparable contributions to the energy of beta-stable matter. Although this compensation does not necessarily imply similar pressures, it helps explain why a supersoft symmetry energy need not produce an excessively soft NS EOS. Together with the slightly stiffer SNM EOS, this composition effect helps explain how SRC-HMT can accommodate both massive NSs and GW170817. These findings illustrate how SRC-HMT influences macroscopic NS properties through modifications of the nucleon momentum distribution, without requiring additional interaction terms in the present framework. Other mechanisms, including meson interactions\,\cite{Li2022}, can also accommodate these astrophysical constraints, while SRC-HMT offers a complementary perspective based on nucleon correlations.

Another important implication concerns the interpretation of the symmetry energy curvature $K_{\rm{sym}}$. Conventional model analyses and empirical estimates favor values around $K_{\rm{sym}}\approx-100\pm100\,\mathrm{MeV}$\,\cite{LiCaiXieZhang2021,Li2026Sym}, although substantial model dependence remains\,\cite{Mondal2017,BiswasRosswog2025,JingDong2025,Seif2025Ksym,Carvalho2024,GrundlerLi2026}. Our results indicate that considerably more negative curvatures should not be excluded solely on the basis of these conventional expectations. In particular, the strong softening associated with negative $K_{\rm{sym}}$ can remain compatible with nuclear and astrophysical constraints when SRC-HMT is included, even for EOSs supporting massive NSs. This suggests that constraints on $K_{\rm{sym}}$ are closely connected to assumptions about the nucleon momentum distribution, and that a broader range of microscopic nuclear dynamics should be considered when inferring the high density symmetry energy from NS observations.

Finally, these conclusions are obtained within the nonlinear RMF framework based on the Walecka model. Although SRC-HMT provides a physically motivated mechanism for modifying the symmetry energy, the quantitative ranges of $K_{\rm{sym}}$, the emergence of supersoft behavior, and the balance between the isoscalar and isovector contributions may depend on the adopted nuclear interaction model. Investigating whether similar effects persist in other theoretical frameworks will be important for assessing their generality.

\section{Effects of SRC-HMT on NS Properties}
\label{sec:NS_properties}

Having shown that SRC-HMT can accommodate a supersoft symmetry energy at high densities while satisfying astrophysical constraints, we now investigate how SRC-HMT and the symmetry energy curvature $K_{\rm{sym}}$ affect the properties of NSs. Using the calibrated RMF parameter sets, we first examine the proton fraction in beta-equilibrated NS matter in Subsection \ref{subsec:proton_fraction}, followed by the core-crust transition density and pressure in Subsection \ref{subsec:core_crust}. We then study the NS mass-radius relations and tidal deformabilities in Subsections \ref{subsec:mass_radius} and \ref{subsec:tidal_deformability}, respectively. Finally, we explore the density dependence of the EOS-parameter $\phi\equiv P/\varepsilon$ and the squared speed of sound $s^2$ in Subsection \ref{subsec:sound}. Throughout these analyses, we compare the HMT and FFG frameworks under the astrophysical constraints from GW170817 and massive NS observations, focusing on how SRC-HMT and $K_{\rm{sym}}$ influence NS properties. For simplicity, we focus on the case of $L=40\,\mathrm{MeV}$, as the analysis for other values of $L$ follows similarly.

\subsection{Proton Fraction $x_{\textmd{p}}=\rho_{\textmd{p}}/\rho$}
\label{subsec:proton_fraction}

The proton fraction $x_{\rm{p}}$ in cold, neutrino-free NS matter is determined by beta equilibrium and charge neutrality. For the processes $\rm{n}\leftrightarrow\rm{p}+\rm{e}^{-}+\overline{\nu}_{\rm{e}}$, the equilibrium condition reads\,\cite{Xu2009}
\begin{equation}\label{elechem1}
\mu_{\rm{n}}-\mu_{\rm{p}}
=\mu_{\rm{e}}
=2\frac{\partial E(\rho,\delta)}{\partial\delta},
\end{equation}
where $\delta=1-2x_{\rm{p}}$ and $\mu_J$ denotes the chemical potential of particle species $J$. The neutron-proton chemical potential difference is governed by the isospin dependence of the nuclear EOS and therefore determines the equilibrium proton fraction.

Treating leptons as relativistic free Fermi gases, their chemical potentials are given by $
\mu_\ell=
[{m_\ell^2+(3\pi^2\rho x_\ell)^{2/3}}]^{1/2},
$ with $\ell=\rm{e},\mu$.
Once the electron chemical potential exceeds the muon rest mass, muons appear in NS matter. The beta equilibrium and charge neutrality conditions then become $
\mu_{\rm{n}}-\mu_{\rm{p}}
=\mu_{\rm{e}}=\mu_\mu$ and $
x_{\rm{p}}=x_{\rm{e}}+x_\mu$.
These equations determine the equilibrium proton fraction and lepton abundances at each baryon density.

The total energy density and pressure of beta-equilibrated npe$\mu$ matter are expressed as $
\varepsilon(\rho,\delta)
=\rho[M_{\rm N}+E(\rho,\delta)]
+\varepsilon_{\rm{e}}+\varepsilon_\mu$,
and $
P=P_{\rm{N}}+P_{\rm{e}}+P_\mu$,
respectively, where the baryonic and leptonic pressures satisfy
$P_{\rm{N}}=\rho_{\rm{n}}\mu_{\rm{n}}
+\rho_{\rm{p}}\mu_{\rm{p}}-\varepsilon_{\rm{N}}$
and
$P_\ell=\rho_\ell\mu_\ell-\varepsilon_\ell$.
Along the beta equilibrium trajectory, the total pressure also satisfies the thermodynamic relation
$P=\rho^2\d(\varepsilon/\rho)/\d\rho$.

\begin{figure}[h!]
    \centering
    \includegraphics[width=0.95\columnwidth]{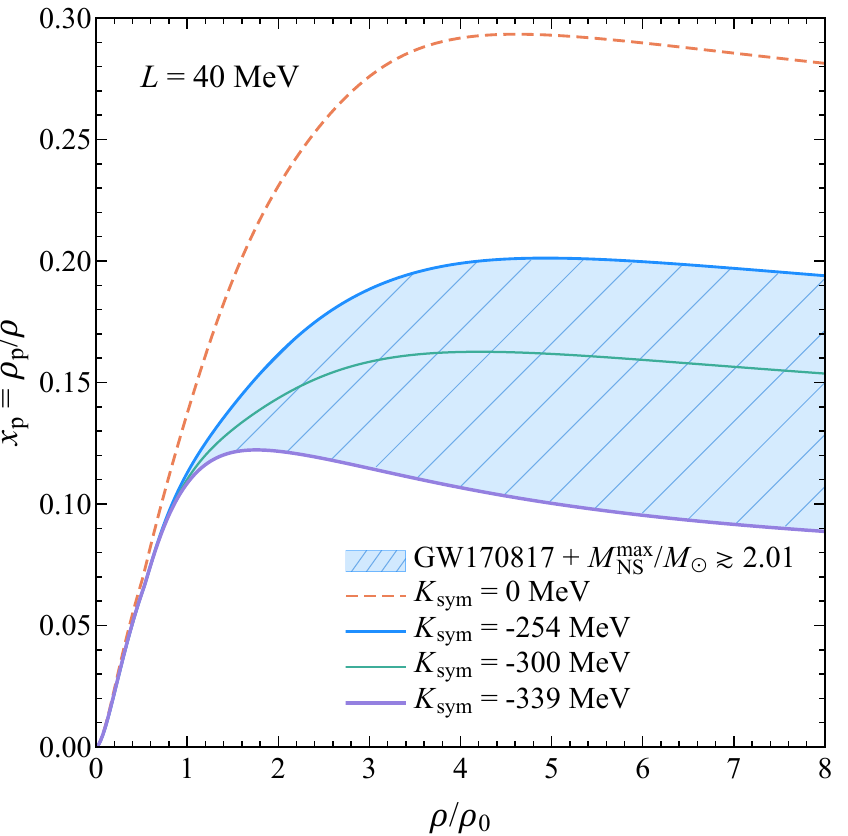}
\caption{(Color online). Density dependence of the proton fraction $x_{\rm{p}}$ in beta-equilibrated NS matter with SRC-HMT at $L=40\,\mathrm{MeV}$. The curves correspond to different symmetry energy curvatures, $K_{\rm{sym}}=0$\,MeV, $-254$\,MeV, $-300$\,MeV, and $-339\,\mathrm{MeV}$. The blue shaded region represents the range of proton fractions consistent with the tidal deformability constraint from GW170817 and the NS maximum mass requirement $M_{\rm{NS}}^{\max}/M_\odot\gtrsim2.01$. More negative values of $K_{\rm{sym}}$ generally lead to smaller proton fractions at supra-saturation densities, reflecting the softening of the symmetry energy induced by SRC-HMT.}\label{fig:xp}
\end{figure}

Since SRC-HMT substantially modifies the density dependence of the symmetry energy, it can also affect the proton fraction $x_{\rm{p}}$ in beta-equilibrated NS matter. In particular, a softer symmetry energy generally reduces the neutron-proton chemical potential difference and favors a smaller proton fraction. FIG.\,\ref{fig:xp} shows the density dependence of $x_{\rm{p}}$ at $L=40\,\mathrm{MeV}$ for different values of $K_{\rm{sym}}$ in the HMT framework. At low densities, the proton fraction increases rapidly, with relatively small differences among the predictions. At higher densities, however, the curves gradually separate, with more negative values of $K_{\rm{sym}}$ leading to substantially smaller proton fractions. For $K_{\rm{sym}}\approx-254\,\mathrm{MeV}$, $x_{\rm{p}}$ reaches approximately $0.20$ around $5\rho_0$ before decreasing slightly. In contrast, for $K_{\rm{sym}}\approx-339\,\mathrm{MeV}$, it reaches a maximum of approximately $0.12$ near $2\rho_0$ and then decreases to about $0.09$ at $8\rho_0$. This non-monotonic behavior reflects the softening of the symmetry energy at high densities, which suppresses the proton abundance.

The blue shaded region indicates the range of proton fractions consistent with the GW170817 tidal deformability constraint and the NS maximum mass requirement. Remarkably, at $5\rho_0$, the allowed proton fraction varies from approximately $0.10$ to $0.20$, corresponding to a factor of two difference between the softest and stiffest allowed symmetry energies. This substantial variation demonstrates that considerable uncertainty in the proton fraction remains within the SRC-HMT framework, even after imposing astrophysical constraints. Such uncertainty may have important implications for NS cooling, since the $x_{\rm p}$ determines whether efficient neutrino emission processes, such as the nucleonic direct Urca process, are kinematically allowed.

\subsection{Core-crust Transition}
\label{subsec:core_crust}

The core-crust transition separates the uniform liquid core from the inhomogeneous inner crust of NSs. Its transition density $\rho_{\rm{t}}$ and pressure $P_{\rm{t}}$ depend sensitively on the nuclear EOS at sub-saturation densities, particularly the density dependence of the symmetry energy. To investigate the effects of SRC-HMT on the core-crust transition, we employ the thermodynamical method, which requires the intrinsic stability conditions\,\cite{Xu2009}
\begin{equation}\label{ther2}
-\left(\frac{\partial P}{\partial v}\right)_{\mu_{\rm{np}}}>0,
~~
-\left(\frac{\partial\mu_{\rm{np}}}{\partial q_{\rm{c}}}\right)_v>0,
\end{equation}
where $v$ and $q_{\rm{c}}$ denote the volume and electric charge per baryon, respectively, and $\mu_{\rm{np}}=\mu_{\rm{n}}-\mu_{\rm{p}}$. For beta-equilibrated npe$\mu$ matter, the lepton pressures depend only on the common lepton chemical potential, which equals $\mu_{\rm{np}}$. Consequently, the mechanical stability condition at fixed $\mu_{\rm{np}}$ can be expressed in terms of the nucleonic EOS as\,\cite{Xu2009,Moustakidis2010,Ducoin2011,GonzalezBoquera2017,Cai2012}
\begin{align}\label{Vther}
K_\mu
=&2\rho\frac{\partial E(\rho,x_{\rm{p}})}{\partial\rho}
+\rho^2\frac{\partial^2E(\rho,x_{\rm{p}})}{\partial\rho^2}
\notag\\
&-\frac{\rho^2
(\partial^2E(\rho,x_{\rm{p}})/
\partial\rho\partial x_{\rm{p}})^2}
{\partial^2E(\rho,x_{\rm{p}})/\partial x_{\rm{p}}^2}
>0.
\end{align}
Here, the partial derivatives are evaluated at fixed values of the other independent variable, while the equilibrium composition is determined self consistently from beta equilibrium and charge neutrality. Uniform matter is mechanically stable when $K_\mu>0$, whereas $K_\mu<0$ signals the spinodal instability. The transition density is identified by $K_\mu(\rho_{\rm{t}})=0$ along the beta-equilibrium trajectory, provided the chemical stability condition is satisfied. The corresponding transition pressure is calculated from the total pressure at $\rho_{\rm{t}}$.

\begin{figure}[h!]
    \centering
    \includegraphics[width=0.95\columnwidth]{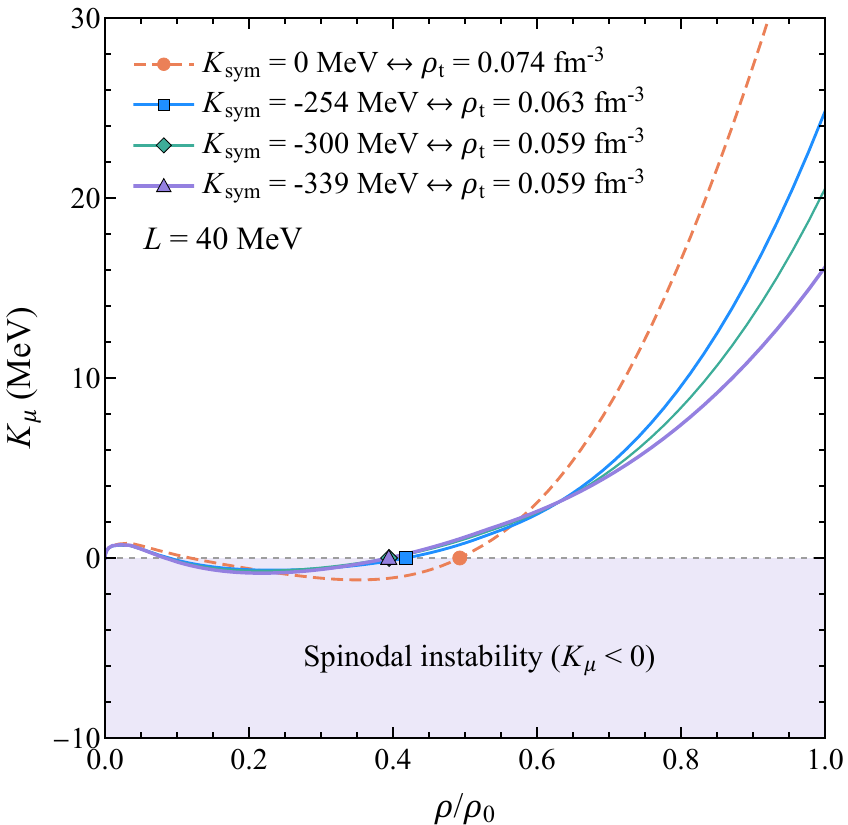}
    \caption{(Color online). Density dependence of the thermodynamical stability function $K_\mu$ in beta-equilibrated NS matter with SRC-HMT at $L=40\,\mathrm{MeV}$ for different values of $K_{\rm{sym}}$. The shaded region indicates the spinodal instability ($K_\mu<0$), while the symbols mark the core-crust transition densities $\rho_{\rm{t}}$, determined by $K_\mu(\rho_{\rm{t}})=0$.}
    \label{fig:Kmu}
\end{figure}

FIG.\,\ref{fig:Kmu} illustrates the density dependence of $K_\mu$ in the HMT framework at $L=40\,\mathrm{MeV}$ for different values of $K_{\rm{sym}}$. At low densities, $K_\mu$ becomes negative, indicating that uniform matter is unstable against density and composition fluctuations. As density increases, $K_\mu$ crosses zero and becomes positive, marking the onset of the stable uniform liquid core. The transition density depends appreciably on $K_{\rm{sym}}$: As $K_{\rm{sym}}$ decreases from $0$\,MeV to $-254\,\mathrm{MeV}$, $\rho_{\rm{t}}$ decreases from approximately $0.074$ to $0.063\,\mathrm{fm}^{-3}$. A further decrease to $K_{\rm{sym}}\approx-300\,\mathrm{MeV}$ lowers $\rho_{\rm{t}}$ to about $0.059\,\mathrm{fm}^{-3}$, while the result remains nearly unchanged at $K_{\rm{sym}}\approx-339\,\mathrm{MeV}$. Interestingly, the stability functions exhibit substantial differences at higher densities, even when their transition densities are nearly identical. This indicates that the core-crust transition is governed by the behavior of the EOS near the stability boundary rather than by the magnitude of $K_\mu$ at higher densities.

\renewcommand*{\tablename}{\small TAB.}
\begin{table}[h!]
    \centering
    \renewcommand{\arraystretch}{1.1}
    \setlength{\tabcolsep}{4pt}
    \begin{tabular}{ccccc}
        \hline\hline
        $K_{\rm{sym}}$ (MeV)
        & $\rho_{\rm{t}}$ ($\mathrm{fm}^{-3}$)
        & $\Delta\rho_{\rm{t}}$ (\%)
        & $P_{\rm{t}}$ ($\mathrm{MeV}/\mathrm{fm}^{3}$)
        & $\Delta P_{\rm{t}}$ (\%) \\
        \hline
        \multicolumn{5}{c}{HMT ($L=40\,\mathrm{MeV}$)} \\
        \hline
        0      & 0.074 & $+25.4$ & 0.732 & $+64.9$ \\
        $-254$ & 0.063 & $+6.8$  & 0.509 & $+14.6$ \\
        $-300$ & 0.059 & 0.0     & 0.450 & $+1.4$ \\
        $-339$ & 0.059 & 0.0     & 0.444 & 0.0 \\
        \hline
        \multicolumn{5}{c}{HMT ($L=60\,\mathrm{MeV}$)} \\
        \hline
        0      & 0.046 & $+9.5$ & 0.261 & $+38.1$ \\
        $-216$ & 0.041 & $-2.4$ & 0.184 & $-2.6$ \\
        $-268$ & 0.042 & 0.0    & 0.189 & 0.0 \\
        \hline
        \multicolumn{5}{c}{HMT ($L=80\,\mathrm{MeV}$)} \\
        \hline
        0      & 0.037 & $-2.6$ & 0.125 & $-0.8$ \\
        $-154$ & 0.038 & 0.0    & 0.124 & $-1.6$ \\
        $-163$ & 0.038 & 0.0    & 0.126 & 0.0 \\
        \hline
        \multicolumn{5}{c}{FFG ($L=40\,\mathrm{MeV}$)} \\
        \hline
        0      & 0.100 & $-2.0$ & 0.970 & $+12.0$ \\
        $-107$ & 0.102 & 0.0    & 0.894 & $+3.2$ \\
        $-120$ & 0.102 & 0.0    & 0.866 & 0.0 \\
        \hline\hline
    \end{tabular}
    \caption{
    Core-crust transition density $\rho_{\rm{t}}$
    and pressure $P_{\rm{t}}$ predicted by the HMT
    and FFG frameworks for different values of
    $L$ and $K_{\rm{sym}}$.
    The relative changes $\Delta\rho_{\rm{t}}$
    and $\Delta P_{\rm{t}}$ are calculated with
    respect to the softest symmetry energy
    (most negative $K_{\rm{sym}}$) at the same $L$
    within each framework.
    }
    \label{tab:core_crust}
\end{table}

TAB.\,\ref{tab:core_crust} summarizes the predicted core-crust transition densities and pressures in the HMT and FFG frameworks. A striking result is that SRC-HMT substantially lowers both quantities. For $L=40\,\mathrm{MeV}$, the HMT parameter sets satisfying the astrophysical constraints predict $\rho_{\rm{t}}\approx0.059$-$0.063\,\mathrm{fm}^{-3}$ and $P_{\rm{t}}\approx0.444$-$0.509\,\mathrm{MeV}/\mathrm{fm}^{3}$, compared with $\rho_{\rm{t}}\approx0.102\,\mathrm{fm}^{-3}$ and $P_{\rm{t}}\approx0.866$-$0.894\,\mathrm{MeV}/\mathrm{fm}^{3}$ in the FFG framework. Thus, the HMT predictions are approximately $40\%$ lower in transition density and $45\%$ lower in transition pressure than the FFG predictions within their respective allowed parameter ranges. Even for the same $L=40\,\mathrm{MeV}$ and $K_{\rm{sym}}=0$\,MeV, SRC-HMT reduces $\rho_{\rm{t}}$ from $0.100\,\rm{fm}^{-3}$ to $0.074\,\mathrm{fm}^{-3}$ and $P_{\rm{t}}$ from $0.970\,\mathrm{MeV}/\mathrm{fm}^{3}$ to $0.732\,\mathrm{MeV}/\mathrm{fm}^{3}$, corresponding to reductions of approximately $26\%$ and $25\%$, respectively. These results demonstrate that SRC-HMT can significantly modify the stability of neutron-rich matter at sub-saturation densities, beyond the effects associated with varying the symmetry energy curvature alone.
The transition properties also exhibit different sensitivities to $L$ and $K_{\rm{sym}}$. Within the HMT framework, increasing $L$ from $40$ to $80\,\mathrm{MeV}$ generally lowers both $\rho_{\rm{t}}$ and $P_{\rm{t}}$. At $L=40\,\mathrm{MeV}$, the astrophysically allowed $K_{\rm{sym}}$ range produces variations of approximately $7\%$ in $\rho_{\rm{t}}$ and $15\%$ in $P_{\rm{t}}$, whereas the corresponding variations in the FFG framework are negligible for $\rho_{\rm{t}}$ and about $3\%$ for $P_{\rm{t}}$. For HMT at $L=60\,\mathrm{MeV}$, both quantities vary by only a few percent, while at $L=80\,\mathrm{MeV}$ they remain nearly unchanged across the allowed $K_{\rm{sym}}$ range. Notably, the differences between HMT and FFG at $L=40\,\mathrm{MeV}$ are much larger than the variations within either framework, indicating that the treatment of nucleon momentum distributions is a major source of uncertainty in the predicted core-crust transition.

These findings may have important implications for NS crusts, since the core-crust transition density and pressure affect the crust thickness and the density range over which nuclear pasta phases may exist\,\cite{Gearheart2011,Sotani2017,Newton2022Crust}. Changes in the crustal structure, particularly the possible formation and distribution of nuclear pasta, can influence neutrino scattering and transport, with potential consequences for the thermal evolution of NSs\,\cite{Horowitz2004Pasta,Horowitz2015Pasta}.

\subsection{NS Mass-radius Relations}
\label{subsec:mass_radius}

We investigate the NS mass-radius (M-R) relations by solving the Tolman--Oppenheimer--Volkoff (TOV) equations\,\cite{Tolman1939,Oppenheimer1939,Misner1973},
\begin{align}
\frac{\d }{\d r}P
=&-\frac{[\varepsilon(r)+P(r)][M(r)+4\pi r^3P(r)]}
{r[r-2M(r)]},\\
\frac{\d }{\d r}M
=&4\pi r^2\varepsilon(r),
\end{align}
where $P(r)$, $\varepsilon(r)$, and $M(r)$ denote the pressure, energy density, and enclosed gravitational mass at radius $r$, respectively. For a given EOS $P(\varepsilon)$, these equations are integrated from the stellar center, with $M(0)=0$ and $P(0)=P_{\rm c}$, to the surface defined by $P(R)=0$, yielding the NS mass $M_{\rm{NS}}\equiv M(R)$ and radius $R$.
The NS core EOS is constructed from beta-equilibrated npe$\mu$ matter, as described in Subsection \ref{subsec:proton_fraction}. For the outer crust, we adopt the Baym-Pethick-Sutherland and Feynman-Metropolis-Teller EOSs\,\cite{BPS1971}. The inner crust is described by a polytropic form $P=a+b\varepsilon^{4/3}$\,\cite{Shi2026,Carriere2003}, where the constants $a$ and $b$ are determined by matching continuously to the outer crust and core EOSs. The core-crust transition density is calculated self-consistently using the thermodynamical method introduced above.

\begin{figure}[h!]
    \centering
    \includegraphics[width=0.95\columnwidth]{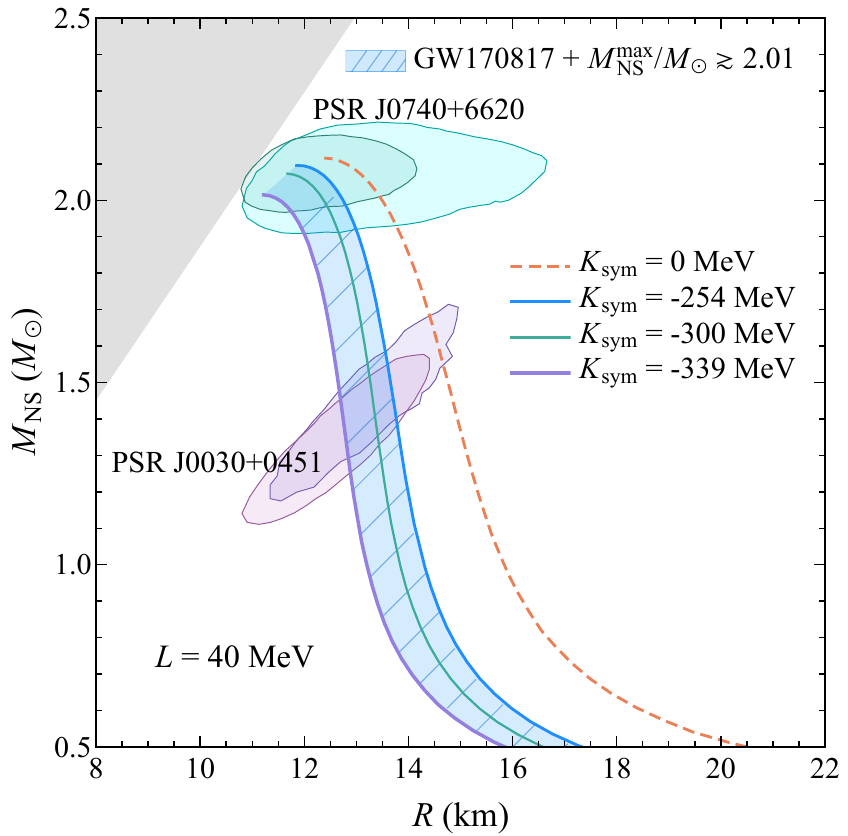}
    \caption{(Color online). M-R relations of NSs in the HMT framework at $L=40\,\mathrm{MeV}$ for different values of $K_{\rm{sym}}$, together with the observational constraints from PSR J0030+0451\,\cite{Miller2019} and PSR J0740+6620\,\cite{Riley21}. The blue shaded region represents the M-R relations consistent with the GW170817 tidal deformability constraint and the NS maximum mass requirement $M_{\rm{NS}}^{\max}/M_\odot\gtrsim2.01$. The gray dashed line denotes the effective causality boundary $R/\mathrm{km}\gtrsim4.73M_{\rm{NS}}/M_\odot+1.14$\,\cite{CaiLiZhang2023ApJ,CaiLiZhang2023PRD} derived within the IPAD-TOV framework, with the gray shaded region disfavored by this constraint.}
    \label{fig:MR}
\end{figure}

\renewcommand*{\tablename}{\small TAB.}
\begin{table}[h!]
    \centering
    \renewcommand{\arraystretch}{1.1}
    \setlength{\tabcolsep}{5pt}
    \begin{tabular}{cccc}
        \hline\hline
        $K_{\rm{sym}}$ (MeV)
        & $R_{1.4}$ (km)
        & $M_{\rm{NS}}^{\max}/M_\odot$
        & $\rho_{\rm{c}}/\rho_0$ \\
        \hline
        \multicolumn{4}{c}{HMT ($L=40\,\mathrm{MeV}$)} \\
        \hline
        0        & 14.94 & 2.12 & 5.97 \\
        $-254$ & 13.77 & 2.10 & 6.38 \\
        $-300$   & 13.36 & 2.07 & 6.42 \\
        $-339$   & 12.75 & 2.01 & 6.92 \\
        \hline
        \multicolumn{4}{c}{HMT ($L=60\,\mathrm{MeV}$)} \\
        \hline
        0        & 14.51 & 2.12 & 5.96 \\
        $-216$ & 13.44 & 2.09 & 6.37 \\
        $-268$   & 12.87 & 2.03 & 6.88 \\
        \hline
        \multicolumn{4}{c}{HMT ($L=80\,\mathrm{MeV}$)} \\
        \hline
        0      & 14.22 & 2.13 & 5.94 \\
        $-154$ & 13.37 & 2.09 & 6.35 \\
        $-163$ & 13.25 & 2.07 & 6.37 \\
        \hline
        \multicolumn{4}{c}{FFG ($L=40\,\mathrm{MeV}$)} \\
        \hline
        0      & 14.35 & 2.03 & 6.09 \\
        $-107$ & 13.54 & 2.02 & 6.47 \\
        $-120$ & 13.32 & 2.01 & 6.48 \\
        \hline\hline
    \end{tabular}
    \caption{
    Radius $R_{1.4}$ of a $1.4M_\odot$ NS,
    maximum NS mass $M_{\rm{NS}}^{\max}$,
    and central density $\rho_{\rm{c}}$
    of the maximum mass configuration
    predicted by the HMT and FFG frameworks
    for different values of $L$ and $K_{\rm{sym}}$.
    }
    \label{tab:MR_properties}
\end{table}

FIG.\,\ref{fig:MR} shows the resulting M-R relations in the HMT framework at $L=40\,\mathrm{MeV}$, together with the observational constraints from PSR J0030+0451\,\cite{Miller2019} and PSR J0740+6620\,\cite{Riley21}. The gray dashed line denotes the effective causality boundary $R/\mathrm{km}\gtrsim4.73M_{\rm{NS}}/M_\odot+1.14$\,\cite{CaiLiZhang2023ApJ,CaiLiZhang2023PRD} derived within the IPAD-TOV framework, with the gray shaded region disfavored by this constraint. The blue shaded region represents the M-R relations satisfying both the GW170817 tidal deformability constraint and the NS maximum mass requirement.

As $K_{\rm{sym}}$ becomes more negative, the NS M-R curves shift toward smaller radii, reflecting the softening of the symmetry energy and the resulting reduction in the pressure of neutron-rich matter. The corresponding NS properties are summarized in TAB.\,\ref{tab:MR_properties}. Within the astrophysically allowed range at $L=40\,\mathrm{MeV}$, $R_{1.4}$ decreases from $13.77$\,km to $12.75$\,km, corresponding to a difference of approximately $1.02$\,km, or $7.4\%$. Remarkably, this approximately $1$\,km variation remains even after imposing the astrophysical constraints and is comparable to the radius precision targeted by future gravitational wave observations with next generation detectors\,\cite{Huxford2024,Walker2024,Khadkikar2025}. Such measurements could therefore provide valuable constraints on $K_{\rm{sym}}$ and the effects of SRC-HMT on dense nuclear matter.

The maximum NS mass also decreases as $K_{\rm{sym}}$ becomes more negative. Within the same allowed range, $M_{\rm{NS}}^{\max}$ decreases from $2.10M_\odot$ to $2.01M_\odot$, corresponding to a reduction of approximately $4.3\%$. Meanwhile, the central density of the maximum mass configuration increases from $6.38\rho_0$ to $6.92\rho_0$, or by approximately $8.5\%$. Similar trends are observed for $L=60$\,MeV and $80\,\mathrm{MeV}$. These results demonstrate that a softer symmetry energy generally leads to more compact NS configurations, with smaller radii, lower maximum masses, and higher central densities.

A comparison between the HMT and FFG frameworks further reveals the importance of SRC-HMT for NS structure. At the same $L=40\,\mathrm{MeV}$ and $K_{\rm{sym}}=0$\,MeV, the HMT framework predicts a larger canonical radius ($14.94$\,km versus $14.35$\,km) and a higher maximum mass ($2.12M_\odot$ versus $2.03M_\odot$), corresponding to increases of approximately $4.1\%$ and $4.4\%$, respectively. These differences reflect the modifications of the underlying EOS associated with the nucleon momentum distributions. More importantly, after imposing the GW170817 tidal deformability and NS maximum mass constraints, the HMT framework permits a much wider range of $K_{\rm{sym}}$ than the FFG framework. Consequently, the allowed HMT predictions span $R_{1.4}\approx12.75$-$13.77$\,km, a width of approximately $1.02$\,km, compared with only $0.22$\,km in the FFG framework, where $R_{1.4}\approx13.32$-$13.54$\,km. The allowed radius variation is therefore about $4.6$ times larger in the HMT framework. Similarly, the allowed maximum mass varies by approximately $0.09M_\odot$ in the HMT framework, compared with only $0.01M_\odot$ in the FFG framework. These findings demonstrate that SRC-HMT substantially broadens the range of NS properties compatible with astrophysical observations, allowing relatively small radii while maintaining sufficient pressure to support massive NSs.

\subsection{Tidal Deformabilities}
\label{subsec:tidal_deformability}

In this subsection, we investigate the effects of SRC-HMT on the tidal deformability of NSs. The dimensionless tidal deformability is defined as\,\cite{Hinderer2008}
\begin{equation}
\Lambda=\frac{2}{3}k_2\left(\frac{R}{M_{\rm{NS}}}\right)^5
=\frac{2k_2}{3\xi^5},
\end{equation}
where $k_2$ is the quadrupolar tidal Love number and $\xi=M_{\rm{NS}}/R$ is the NS compactness\,\cite{Cai2025PRD}. The Love number $k_2$ is determined by the stellar structure and the underlying EOS through the linear tidal perturbation equations\,\cite{Hinderer2008}. Owing to its strong dependence on compactness, $\Lambda$ is particularly sensitive to the NS radius\,\cite{Shi2026PRD} and the internal distribution of matter, making it a valuable probe of the symmetry energy and SRC-HMT effects in dense matter.

\begin{figure}[h!]
    \centering
    \includegraphics[width=0.95\columnwidth]{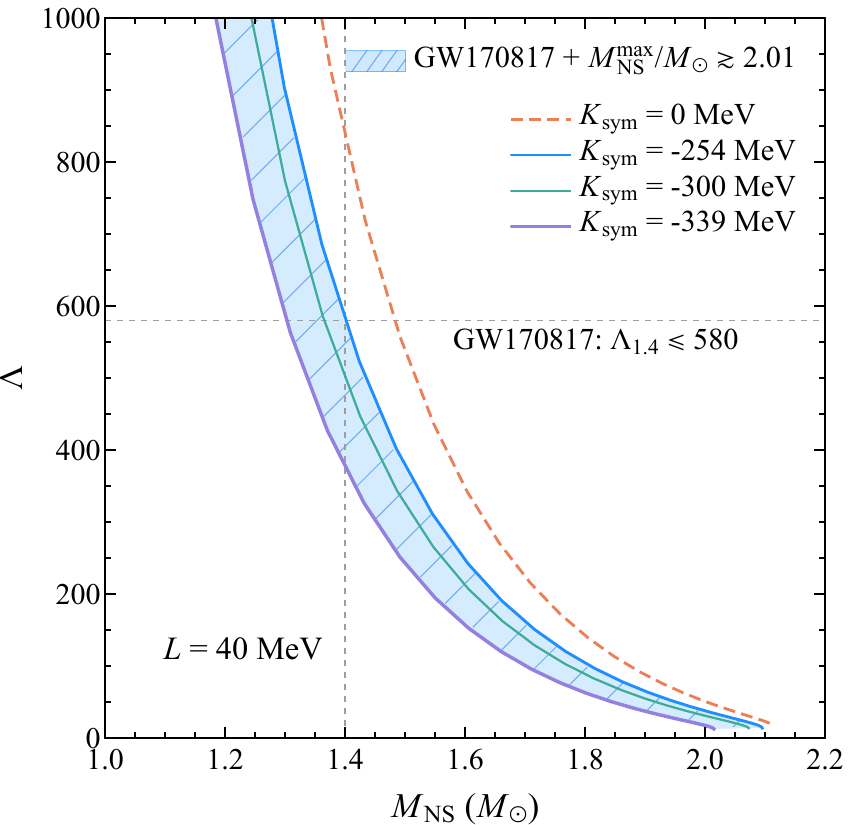}
    \caption{(Color online). Tidal deformability $\Lambda$ as a function of NS mass $M_{\rm{NS}}$ in the HMT framework at $L=40\,\mathrm{MeV}$ for different values of $K_{\rm{sym}}$. The horizontal dashed line indicates the GW170817 constraint $\Lambda_{1.4}\lesssim580$\,\cite{Abbott2018}, while the vertical dashed line marks $M_{\rm{NS}}=1.4M_\odot$. The blue shaded region represents the predictions consistent with the GW170817 tidal deformability constraint and the NS maximum mass requirement $M_{\rm{NS}}^{\max}/M_\odot\gtrsim2.01$. More negative values of $K_{\rm{sym}}$ generally lead to smaller tidal deformabilities at a given NS mass.}
    \label{fig:Lambda}
\end{figure}

FIG.\,\ref{fig:Lambda} shows the tidal deformability $\Lambda$ as a function of NS mass in the HMT framework at $L=40\,\mathrm{MeV}$ for different values of $K_{\rm{sym}}$. As expected, $\Lambda$ decreases rapidly with increasing NS mass, reflecting the increasing compactness of more massive stars. At a given mass, more negative values of $K_{\rm{sym}}$ lead to smaller tidal deformabilities, consistent with the reduction in NS radii discussed in Subsection \ref{subsec:mass_radius}. For a canonical $1.4M_\odot$ NS, $\Lambda_{1.4}$ decreases from approximately $834$ at $K_{\rm{sym}}=0$\,MeV to $375$ at $K_{\rm{sym}}\approx-339\,\mathrm{MeV}$, a reduction of about $55\%$. The GW170817 constraint $\Lambda_{1.4}\lesssim580$ excludes the relatively stiff symmetry energies\,\cite{Abbott2018}, while the NS maximum mass requirement limits how soft the EOS can become. Together, these constraints define the blue shaded region in FIG.\,\ref{fig:Lambda}.
The corresponding values of $\Lambda_{1.4}$ for all parameter sets are summarized in TAB.\,\ref{tab:tidal_deformability}. Within the astrophysically allowed $K_{\rm{sym}}$ range at $L=40\,\mathrm{MeV}$, the HMT framework predicts $\Lambda_{1.4}\approx375$-$580$, corresponding to a variation of approximately $55\%$ relative to the softest symmetry energy. This substantial variation remains despite imposing both the GW170817 and maximum mass constraints. At $L=60\,\mathrm{MeV}$, the allowed range narrows to $\Lambda_{1.4}\approx434$-$580$, corresponding to a relative variation of about $34\%$, while at $L=80\,\mathrm{MeV}$ it further narrows to $\Lambda_{1.4}\approx545$-$580$, or approximately $6\%$. These results demonstrate that the uncertainty in tidal deformability associated with the allowed symmetry energy curvature depends strongly on the slope coefficient $L$, becoming considerably larger at smaller symmetry energy slopes.

\renewcommand*{\tablename}{\small TAB.}
\begin{table}[h!]
    \centering
    \renewcommand{\arraystretch}{1.1}
    \setlength{\tabcolsep}{9pt}
    \begin{tabular}{ccc}
        \hline\hline
        $K_{\rm{sym}}$ (MeV)
        & $\Lambda_{1.4}$
        & $\Delta\Lambda_{1.4}$ (\%) \\
        \hline
        \multicolumn{3}{c}{HMT ($L=40\,\mathrm{MeV}$)} \\
        \hline
        0      & 834 & $+122.4$ \\
        $-254$ & 580 & $+54.7$ \\
        $-300$ & 498 & $+32.8$ \\
        $-339$ & 375 & 0.0 \\
        \hline
        \multicolumn{3}{c}{HMT ($L=60\,\mathrm{MeV}$)} \\
        \hline
        0      & 860 & $+98.2$ \\
        $-216$ & 580 & $+33.6$ \\
        $-268$ & 434 & 0.0 \\
        \hline
        \multicolumn{3}{c}{HMT ($L=80\,\mathrm{MeV}$)} \\
        \hline
        0      & 870 & $+59.6$ \\
        $-154$ & 580 & $+6.4$ \\
        $-163$ & 545 & 0.0 \\
        \hline
        \multicolumn{3}{c}{FFG ($L=40\,\mathrm{MeV}$)} \\
        \hline
        0      & 760 & $+42.1$ \\
        $-107$ & 580 & $+8.4$ \\
        $-120$ & 535 & 0.0 \\
        \hline\hline
    \end{tabular}
    \caption{
    Dimensionless tidal deformability $\Lambda_{1.4}$
    of canonical NSs predicted by the HMT
    and FFG frameworks for different values of
    $L$ and $K_{\rm{sym}}$.
    The relative change $\Delta\Lambda_{1.4}$
    is calculated with respect to the softest
    symmetry energy (most negative $K_{\rm{sym}}$)
    at the same $L$.
    }
    \label{tab:tidal_deformability}
\end{table}

A comparison with the FFG framework further highlights the impact of SRC-HMT. At $L=40\,\mathrm{MeV}$, the astrophysically allowed FFG parameter sets predict only $\Lambda_{1.4}\approx535$-$580$, corresponding to a relative variation of about $8\%$. In contrast, the HMT framework allows a much broader range, $\Lambda_{1.4}\approx375$-$580$. The width of the allowed interval increases from $45$ in FFG to $205$ in HMT, or by a factor of approximately $4.6$. Moreover, the minimum allowed $\Lambda_{1.4}$ decreases from $535$ in FFG to $375$ in HMT, a reduction of approximately $30\%$. Interestingly, at fixed $L=40\,\mathrm{MeV}$ and $K_{\rm{sym}}=0$\,MeV, the HMT framework predicts a larger $\Lambda_{1.4}$ than FFG ($834$ versus $760$), indicating that SRC-HMT does not necessarily reduce the tidal deformability at fixed symmetry energy parameters. Rather, its principal effect is to broaden the range of $K_{\rm{sym}}$ compatible with the astrophysical constraints, thereby allowing substantially smaller tidal deformabilities while maintaining sufficiently large maximum NS masses. Together with the approximately $1$\,km variation in $R_{1.4}$ found in Subsection \ref{subsec:mass_radius}, these results demonstrate the importance of nucleon momentum distributions for interpreting tidal measurements and constraining the EOS of dense nuclear matter.

\subsection{EOS-parameter and Sound Speed Profiles}
\label{subsec:sound}

In this subsection, we investigate the effects of SRC-HMT and the symmetry energy curvature $K_{\rm{sym}}$ on the EOS-parameter $P/\varepsilon$\,\cite{CaiLiMa2026phi,CaiLiMa2026X} and the squared speed of sound $s^2=\d P/\d\varepsilon$ in dense NS matter. The EOS-parameter measures the pressure relative to the energy density and plays an important role in determining the relativistic structure and compactness of NSs. It can also be interpreted as the average squared sound speed from the stellar surface\,\cite{Saes22,Saes24,Marc24}, where $P=0$, to the point under consideration. Moreover, the quantity $\Delta=1/3-P/\varepsilon$ defines the dimensionless trace anomaly\,\cite{Fujimoto2022,Li2026PRL}, which characterizes the breaking of conformal symmetry and provides a useful connection to the high density EOS studied within perturbative QCD. In contrast to $P/\varepsilon$, the squared sound speed $s^2$ characterizes the local stiffness of the EOS and governs the propagation of small pressure perturbations through dense matter\,\cite{CaiLi2024SSS}. It is also subject to the causality requirement $0\leq s^2\leq1$ for a stable, causal EOS. Its density dependence may also reveal non-monotonic behavior, including possible peaks associated with changes in the stiffness of dense matter. Although closely related, $P/\varepsilon$ and $s^2$ describe different aspects of the EOS: the former characterizes the average pressure response, while the latter measures the local response to changes in energy density. Their density dependence therefore provides complementary information about dense nuclear matter, including possible changes in EOS stiffness and the properties of massive NSs\,\cite{CaiLi2025Review}.

\begin{figure}[h!]
    \centering
    \includegraphics[width=0.95\columnwidth]{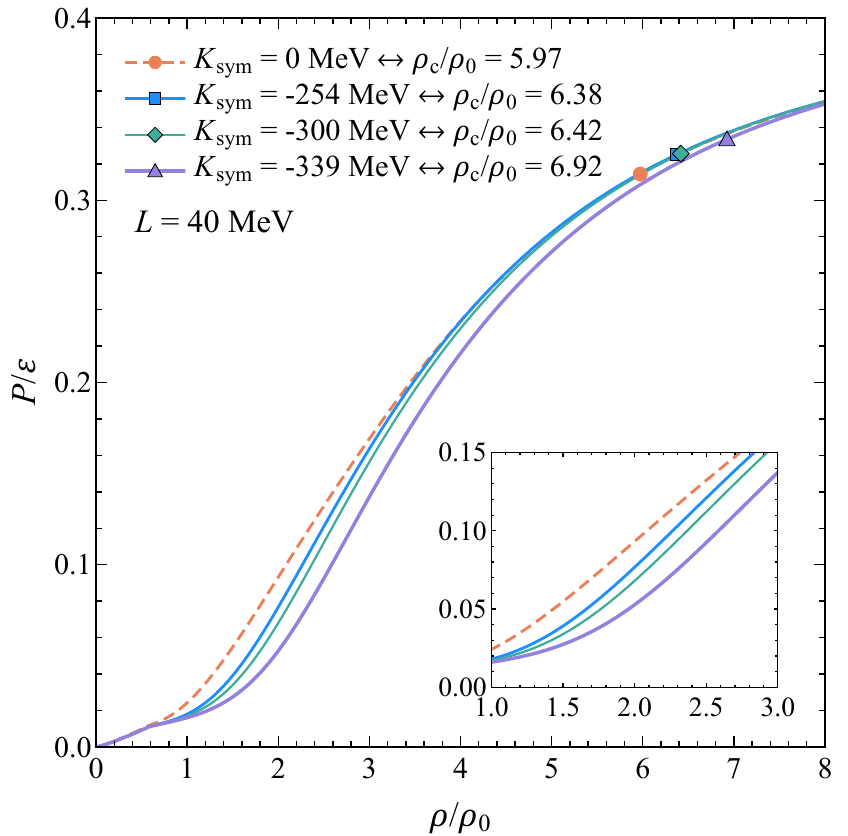}
    \caption{(Color online). The EOS-parameter $P/\varepsilon$ as a function of normalized baryon density $\rho/\rho_0$ in the HMT framework at $L=40\,\mathrm{MeV}$ for different values of $K_{\rm{sym}}$. The symbols indicate the central densities $\rho_{\rm c}$ of the corresponding maximum mass NS configurations, with their values given in the legend. The inset highlights the differences among the predictions at lower densities.}
    \label{fig:phi}
\end{figure}

\begin{figure}[h!]
    \centering
    \includegraphics[width=0.95\columnwidth]{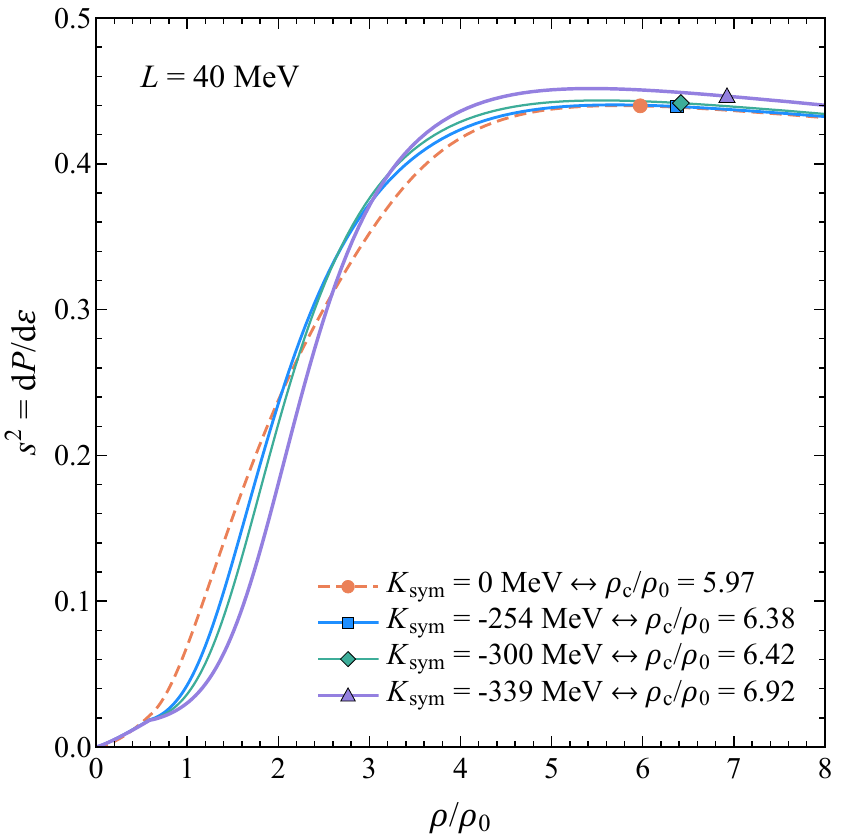}
    \caption{(Color online). Squared speed of sound $s^2=\mathrm{d}P/\mathrm{d}\varepsilon$ as a function of normalized baryon density $\rho/\rho_0$ in the HMT framework at $L=40\,\mathrm{MeV}$ for different values of $K_{\rm{sym}}$. The symbols mark the central densities $\rho_{\rm c}$ of the corresponding maximum mass NS configurations, with their values given in the legend. All curves exhibit a broad maximum at high densities, with more negative values of $K_{\rm{sym}}$ generally leading to a slightly higher peak and a shift toward lower densities.}
    \label{fig:sss}
\end{figure}

In FIG.\,\ref{fig:phi}, we show the density dependence of $P/\varepsilon$ in the HMT framework at $L=40\,\mathrm{MeV}$ for different values of $K_{\rm{sym}}$. The EOS-parameter increases monotonically with density, while more negative values of $K_{\rm{sym}}$ generally lead to smaller $P/\varepsilon$, particularly around $\rho\approx1$-$2\rho_0$. Within the astrophysically allowed range of $K_{\rm{sym}}$, the relative variation in $P/\varepsilon$ can reach approximately $40\%$-$50\%$ around $\rho\approx1.5$-$2\rho_0$, indicating a substantial influence of the symmetry energy curvature on the average stiffness of the EOS. At higher densities, these differences become relatively smaller, and the central EOS-parameters of the maximum mass NS configurations remain close to one another despite their different central densities, suggesting a reduced sensitivity to $K_{\rm{sym}}$ in the most massive NSs.

The corresponding squared sound speed $s^2$ is shown in FIG.\,\ref{fig:sss}. Around $\rho\approx1$-$2\rho_0$, more negative values of $K_{\rm{sym}}$ generally reduce $s^2$. Even within the astrophysically allowed range of $K_{\rm{sym}}$, the relative variation in $s^2$ can reach approximately $20\%$-$50\%$, with the largest differences occurring around $1.5\rho_0$. This substantial variation indicates that the symmetry energy curvature strongly affects the local stiffness of the EOS near nuclear saturation density. At higher densities, all curves develop a broad maximum, with more negative $K_{\rm{sym}}$ generally producing a slightly higher peak at a lower density.

\begin{figure}[h!]
    \centering
    \includegraphics[width=0.95\columnwidth]{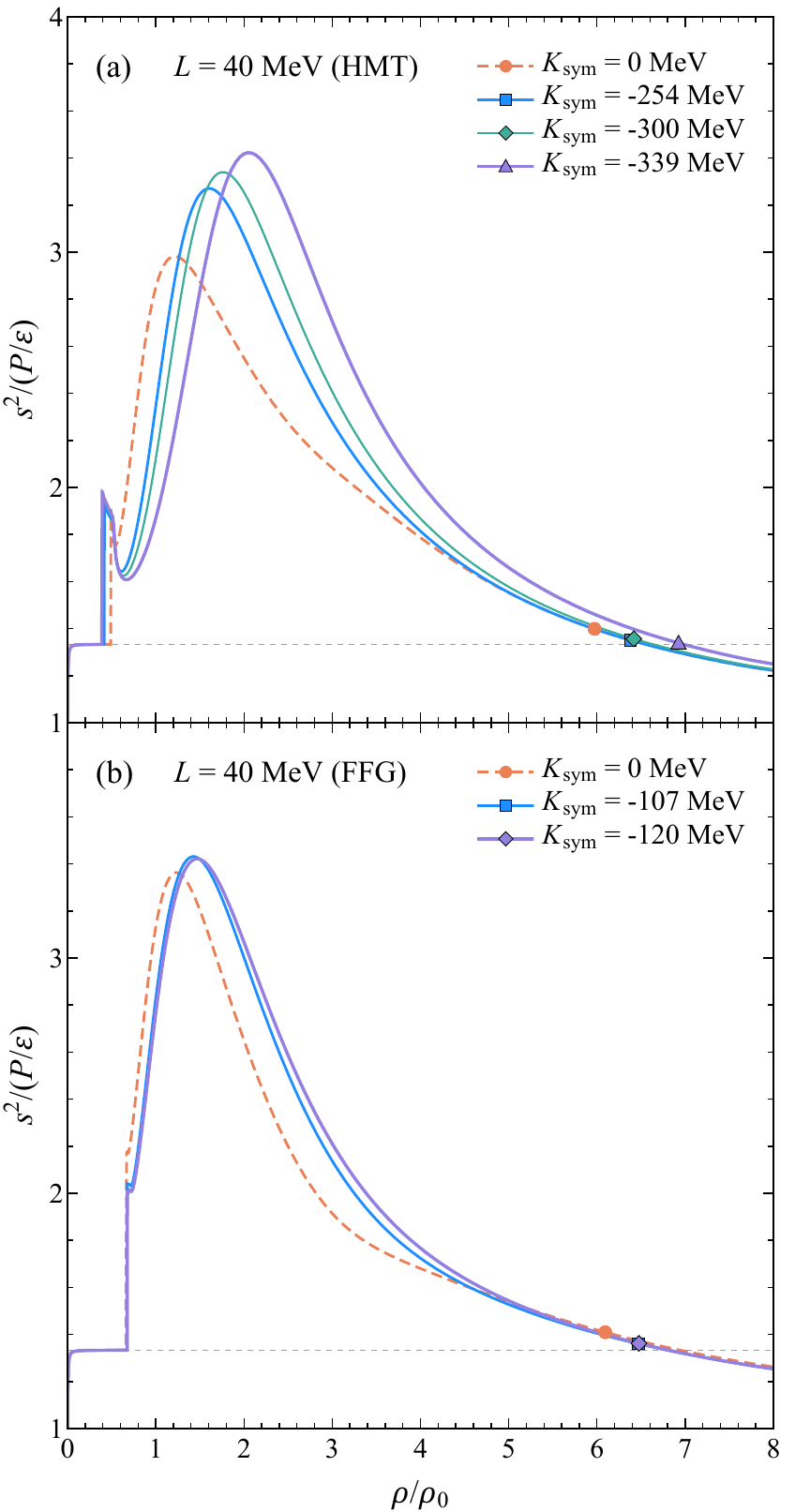}
   \caption{(Color online). Logarithmic EOS slope $\gamma=s^2/(P/\varepsilon)$ versus density for $L=40$\,MeV with (a) HMT and (b) FFG distributions and the indicated $K_{\rm{sym}}$ values. Symbols mark the central densities of maximum-mass NSs; horizontal dashed lines indicate $\gamma=4/3$.}
    \label{fig:gamma}
\end{figure}

To further characterize the density dependence of EOS stiffness,
FIG.\,\ref{fig:gamma} shows the logarithmic slope\,\cite{CaiLi2025Review}
\begin{equation}
\gamma\equiv\frac{\d\ln P}{\d\ln\varepsilon}
=\frac{s^2}{P/\varepsilon}.
\end{equation}
This quantity measures the fractional pressure increase
associated with a fractional increase in energy density and
characterizes the local power law behavior of the EOS.
Unlike $s^2$, which measures the local pressure response,
$\gamma$ also accounts for the ratio $P/\varepsilon$,
providing a complementary measure of EOS stiffness.
Between the core-crust transition and approximately $2\rho_0$,
the HMT EOSs with strongly negative $K_{\rm{sym}}$ exhibit a local
decrease in $\gamma$, indicating slower pressure growth
relative to energy density.
This behavior is consistent with the softening discussed above
and is absent from the FFG results shown here.
The subsequent increase in $\gamma$ accompanies the rapid rise
in $s^2$ around $2$-$3\rho_0$, indicating rapid stiffening of the EOS;
see also FIG.\,\ref{fig:sss}.
The growing departure of $\gamma$ from unity in this region
also indicates a deviation from conformal scaling,
characterized by $\gamma=1$\,\cite{Ann20}.
At higher densities, $\gamma$ decreases to values close to $4/3$
near the central densities of the maximum-mass NSs. The EOS can thus be approximated locally by
$P\sim\varepsilon^{4/3}$, corresponding to a polytropic index
$n\approx3$ under the convention
$P\sim\varepsilon^{1+1/n}$\,\cite{ShapiroTeukolsky1983}.
The proximity of these values across different $K_{\rm{sym}}$ suggests a reduced sensitivity of the high density logarithmic EOS slope to the symmetry energy curvature. Together, these results indicate that SRC-induced modifications of the symmetry energy curvature can influence both EOS softening in the outer core and its subsequent stiffening at higher densities.

\section{Summary and Outlook}
\label{sec:summary}

We investigated the effects of SRC-HMT on the symmetry energy and NS properties within the nonlinear Walecka framework. For the first time, we derived analytical expressions for the symmetry energy curvature $K_{\rm{sym}}$ for both FFG and HMT nucleon momentum distributions, allowing us to examine how SRC-induced modifications affect the density dependence of the symmetry energy and the EOS of dense matter. The model parameters were calibrated to reproduce empirical nuclear matter properties around saturation, while the resulting EOSs were further constrained by astrophysical observations.

Our central finding is that SRC-HMT can accommodate a supersoft symmetry energy at high densities while remaining consistent with massive NS observations and the relatively small tidal deformability inferred from GW170817. In particular, SRC-HMT allows substantially more negative values of $K_{\rm{sym}}$, reaching approximately $-300\,\mathrm{MeV}$ or below, for which the symmetry energy can begin decreasing near $2\rho_0$. Despite this pronounced softening, the EOS can remain sufficiently stiff to support massive NSs without predicting excessively large radii or tidal deformabilities. This compatibility can be partly attributed to the slight stiffening of the SNM EOS by SRC-HMT, which counteracts the softening of the symmetry energy. In addition, a softer symmetry energy reduces the proton fraction and increases the isospin asymmetry of beta-stable matter, partially compensating for the reduced symmetry energy contribution to the EOS. These findings demonstrate that a supersoft symmetry energy does not necessarily imply an excessively soft NS EOS.
Compared with the FFG momentum distribution, SRC-HMT substantially broadens the range of $K_{\rm{sym}}$ values and NS properties compatible with astrophysical observations. Beyond its impact on the high density symmetry energy, SRC-HMT also modifies the proton fraction and core-crust transition. Furthermore, the EOS-parameter $P/\varepsilon$ and squared speed of sound $s^2$ show considerable sensitivity to $K_{\rm{sym}}$, particularly around $\rho\approx1$-$2\rho_0$. Together, these results highlight the role of SRC-HMT in determining the density dependence of the EOS, as well as NS composition and structure.

Our findings suggest several directions for future studies. An important question is whether a supersoft symmetry energy compatible with massive NSs and GW170817 can also emerge in other models. Exploring its implications for heavy ion collisions\,\cite{Xu2026,Senger2024,Vinas2024,Fang2026,Estee2021,Wang2025,Liu2025}, NS mergers\,\cite{Most2021,Jacobi2024,Raithel2021,Bauswein2019}, and supernovae\,\cite{Fischer2014,Hempel2012,Lattimer1991,Steiner2013} can provide additional tests of the symmetry energy under different physical conditions\,\cite{Xiao2009,Wen2009,LiXie2020}. Equally important is quantifying uncertainties associated with SRC parameters, nuclear interactions, and observational constraints. Together with joint statistical analyses of nuclear and astrophysical data, such studies will help clarify the role of SRC in shaping the density dependence of the symmetry energy and the dense matter EOS.

\section*{Acknowledgment} 
This work was supported in part by the National Natural Science Foundation of China under contract Nos. 12688101 and 12547102, and the STCSM under Grant Nos.~23590780100 and 23JC1400200.

\appendix
\renewcommand{\theequation}{\Alph{section}.\arabic{equation}}
\setcounter{equation}{0}

\begin{widetext}

\section{SRC Rearrangement Contribution to the Chemical Potential Difference}
\label{app:app}

In this appendix, we give the SRC rearrangement contribution $\Delta\mu_{\rm{SRC}}^{\rm{(R)}}$ to the neutron-proton chemical potential difference. This contribution arises from the isospin dependence of the nucleon momentum distribution through the SRC-HMT parameters $C_J$ and $\phi_J$, characterizing the HMT amplitude and upper momentum cutoff, respectively. Their density variations generate additional terms in the chemical potentials beyond the explicit density dependence of the kinetic energy. The single nucleon effective energies are defined as $E_J^\ast(k)=[{k^2+M_J^{*,2}}]^{1/2}$, where $J=\rm{n,p}$ and $M_J^\ast$ is the Dirac effective mass. Using the SRC-HMT momentum distributions introduced in the main text, the rearrangement contribution reads
\begin{align}
\Delta\mu_{\rm{SRC}}^{\rm{(R)}}
=\frac{2}{\pi^2\rho}\Bigg\{
C_0C_1\Bigg[&
k_{\rm F}^{\rm n,4}\Bigg(
\operatorname{arcsinh}\left(\frac{\phi_{\rm n}k_{\rm F}^{\rm n}}{M_{\rm n}^\ast}\right)
-\operatorname{arcsinh}\left(\frac{k_{\rm F}^{\rm n}}{M_{\rm n}^\ast}\right)
-\frac{E_{\rm n}^\ast(\phi_{\rm n}k_{\rm F}^{\rm n})}{\phi_{\rm n}k_{\rm F}^{\rm n}}
+\frac{E_{\rm n}^\ast(k_{\rm F}^{\rm n})}{k_{\rm F}^{\rm n}}
\Bigg)
\notag\\
&-\frac{3}{8}\left(1-\frac{1}{\phi_{\rm n}}\right)
\Bigg[
k_{\rm F}^{\rm n}E_{\rm n}^\ast(k_{\rm F}^{\rm n})
\left(2k_{\rm F}^{\rm n,2}+M_{\rm n}^{*,2}\right)
-M_{\rm n}^{\ast,4}\ln\left(
\frac{k_{\rm F}^{\rm n}+E_{\rm n}^\ast(k_{\rm F}^{\rm n})}{M_{\rm n}^\ast}
\right)
\Bigg]
\notag\\
&-k_{\rm F}^{\rm p,4}\Bigg(
\operatorname{arcsinh}\left(\frac{\phi_{\rm p}k_{\rm F}^{\rm p}}{M_{\rm p}^\ast}\right)
-\operatorname{arcsinh}\left(\frac{k_{\rm F}^{\rm p}}{M_{\rm p}^\ast}\right)
-\frac{E_{\rm p}^\ast(\phi_{\rm p}k_{\rm F}^{\rm p})}{\phi_{\rm p}k_{\rm F}^{\rm p}}
+\frac{E_{\rm p}^\ast(k_{\rm F}^{\rm p})}{k_{\rm F}^{\rm p}}
\Bigg)
\notag\\
&+\frac{3}{8}\left(1-\frac{1}{\phi_{\rm p}}\right)
\Bigg[
k_{\rm F}^{\rm p}E_{\rm p}^\ast(k_{\rm F}^{\rm p})
\left(2k_{\rm F}^{\rm p,2}+M_{\rm p}^{*,2}\right)
-M_{\rm p}^{\ast,4}\ln\left(
\frac{k_{\rm F}^{\rm p}+E_{\rm p}^\ast(k_{\rm F}^{\rm p})}{M_{\rm p}^\ast}
\right)
\Bigg]
\Bigg]
\notag\\
+\phi_0\phi_1\Bigg[&
\frac{C_{\rm n}}{\phi_{\rm n}^2}
\Bigg\{
k_{\rm F}^{\rm n,3}E_{\rm n}^\ast(\phi_{\rm n}k_{\rm F}^{\rm n})
-\frac{3}{8}
\Bigg[
k_{\rm F}^{\rm n}E_{\rm n}^\ast(k_{\rm F}^{\rm n})
\left(2k_{\rm F}^{\rm n,2}+M_{\rm n}^{*,2}\right)
-M_{\rm n}^{\ast,4}\ln\left(
\frac{k_{\rm F}^{\rm n}+E_{\rm n}^\ast(k_{\rm F}^{\rm n})}{M_{\rm n}^\ast}
\right)
\Bigg]
\Bigg\}
\notag\\
&-\frac{C_{\rm p}}{\phi_{\rm p}^2}
\Bigg\{
k_{\rm F}^{\rm p,3}E_{\rm p}^\ast(\phi_{\rm p}k_{\rm F}^{\rm p})
-\frac{3}{8}
\Bigg[
k_{\rm F}^{\rm p}E_{\rm p}^\ast(k_{\rm F}^{\rm p})
\left(2k_{\rm F}^{\rm p,2}+M_{\rm p}^{*,2}\right)
-M_{\rm p}^{\ast,4}\ln\left(
\frac{k_{\rm F}^{\rm p}+E_{\rm p}^\ast(k_{\rm F}^{\rm p})}{M_{\rm p}^\ast}
\right)
\Bigg]
\Bigg\}
\Bigg]
\Bigg\}.
\label{eq:mu_rearr_SRC-1}
\end{align}

The two contributions proportional to $C_0C_1$ and
$\phi_0\phi_1$ originate from the isospin dependence of the
SRC-HMT amplitude and cutoff, respectively.
Eq.\,(\ref{eq:mu_rearr_SRC-1}) retains the neutron-proton
effective mass splitting, $M_{\rm n}^\ast\neq M_{\rm p}^\ast$,
and is therefore applicable to the present framework with
an isovector scalar field.
In the absence of this field, the effective masses become equal,
$M_{\rm n}^\ast=M_{\rm p}^\ast$.
Furthermore, the rearrangement contribution vanishes when
$C_0=0$ and $\phi_0=1$ (FFG case), since the SRC-HMT parameters then have no
explicit dependence on the isospin asymmetry.

\section{Scalar Density and Its Derivatives with SRC-HMT}
\label{app:C}

In SNM, where $\theta = k_{\mathrm{F}}/M_0^*$, and $F_{\mathrm{F}}^* = [{(\phi_0 k_{\mathrm{F}})^2 + M_0^{*,2}}]^{1/2}$, the total scalar density $\rho^{\mathrm{s}} = \rho_{\rm n}^{\mathrm{s}} + \rho_{\rm p}^{\mathrm{s}}$ consists of the depleted Fermi sea and HMT contributions,
\begin{equation}
\rho^{\mathrm{s}}(\rho, M_0^*) = \rho_{(1)}^{\mathrm{s}} + \rho_{(2)}^{\mathrm{s}}, ~~ \text{with} ~~
\rho_{(1)}^{\mathrm{s}} = \frac{\Delta_0 M_0^{*,3}}{\pi^2}\left(\theta\sqrt{1+\theta^2} - \operatorname{arcsinh}\theta\right), ~~
\rho_{(2)}^{\mathrm{s}} = \frac{3\rho C_0}{M_0^*}\left(E_{\mathrm{F}}^* - \frac{F_{\mathrm{F}}^*}{\phi_0}\right).
\end{equation}
The density dependence of the scalar mean field follows from the SNM scalar field equation $W_\sigma(f_0) \equiv m_\sigma^2 f_0 + b_\sigma M_{\mathrm{N}} g_\sigma^3 f_0^2 + c_\sigma g_\sigma^4 f_0^3 = g_\sigma \rho^{\mathrm{s}}(\rho, M_0^*)$. Since $M_0^*$ itself varies with $\rho$ through the scalar field, differentiating this equation requires both the explicit density dependence of $\rho^{\mathrm{s}}$ and its implicit dependence through $M_0^*$. Using the chain rule $
{\d}/{\d\rho} = {\partial}/{\partial \rho}|_{M_0^*} - g_\sigma f_0' {\partial}/{\partial M_0^*}|_{\rho}$,
and differentiating twice with respect to $\rho$, we obtain
\begin{align}
f_0' &= \frac{g_\sigma}{R_\sigma}\frac{\partial \rho^{\mathrm{s}}}{\partial \rho},~~
f_0'' = \frac{1}{R_\sigma}\left[g_\sigma \frac{\partial^2 \rho^{\mathrm{s}}}{\partial \rho^2} - 2g_\sigma^2 f_0' \frac{\partial^2 \rho^{\mathrm{s}}}{\partial \rho \partial M_0^*} + \left(g_\sigma^3 \frac{\partial^2 \rho^{\mathrm{s}}}{\partial M_0^{*,2}} - \ddot{W}_\sigma\right)f_0'^2\right], \label{eq:f0pp}
\end{align}
where $f_0' \equiv \d f_0/\d\rho$, $f_0'' \equiv \d^2 f_0/\d\rho^2$, and the auxiliary functions are defined as
\begin{equation}
R_\sigma = \dot{W}_\sigma + g_\sigma^2 \frac{\partial \rho^{\mathrm{s}}}{\partial M_0^*}, ~~
\dot{W}_\sigma \equiv \frac{\d W_\sigma}{\d f_0} = m_\sigma^2 + 2b_\sigma M_{\mathrm{N}} g_\sigma^3 f_0 + 3c_\sigma g_\sigma^4 f_0^2, ~~
\ddot{W}_\sigma \equiv \frac{\d^2 W_\sigma}{\d f_0^2} = 2b_\sigma M_{\mathrm{N}} g_\sigma^3 + 6c_\sigma g_\sigma^4 f_0.
\end{equation}

It remains to evaluate the partial derivatives of $\rho^{\mathrm{s}}$ entering these expressions. We first consider the derivatives with respect to $M_0^*$ at fixed $\rho$, for which $k_{\mathrm{F}}$ is also fixed. The basic relations $\partial\theta/\partial M_0^* = -\theta/M_0^*$, $\partial E_{\mathrm{F}}^*/\partial M_0^* = M_0^*/E_{\mathrm{F}}^*$, and $\partial F_{\mathrm{F}}^*/\partial M_0^* = M_0^*/F_{\mathrm{F}}^*$ then give
\begin{align}
\frac{\partial \rho^{\mathrm{s}}}{\partial M_0^*} &= 3\left[\frac{\rho_{(1)}^{\mathrm{s}}}{M_0^*} - \frac{\Delta_0 \rho}{E_{\mathrm{F}}^*} + \frac{\rho C_0 k_{\mathrm{F}}^2}{M_0^{*,2}}\left(\frac{\phi_0}{F_{\mathrm{F}}^*} - \frac{1}{E_{\mathrm{F}}^*}\right)\right], \\
\frac{\partial^2 \rho^{\mathrm{s}}}{\partial M_0^{*,2}} &= \frac{6\rho_{(1)}^{\mathrm{s}}}{M_0^{*,2}} - \frac{3\Delta_0 \rho (2M_0^{*,2} + 3k_{\mathrm{F}}^2)}{M_0^* E_{\mathrm{F}}^{*,3}} + \frac{3\rho C_0 k_{\mathrm{F}}^2}{M_0^{*,3}}\left[\frac{3M_0^{*,2} + 2k_{\mathrm{F}}^2}{E_{\mathrm{F}}^{*,3}} - \phi_0\frac{3M_0^{*,2} + 2(\phi_0 k_{\mathrm{F}})^2}{F_{\mathrm{F}}^{*,3}}\right].
\end{align}

Instead of evaluating the remaining density derivatives independently, they can be obtained from the mass derivatives above by exploiting the scaling property of the scalar density. For density independent SRC-HMT parameters $\Delta_0$, $C_0$, and $\phi_0$, $\rho^{\mathrm{s}}$ is a homogeneous function of degree three in $(k_{\mathrm{F}},M_0^*)$ and therefore satisfies
$
k_{\mathrm{F}}{\partial \rho^{\mathrm{s}}}/{\partial k_{\mathrm{F}}}|_{M_0^*} + M_0^*{\partial \rho^{\mathrm{s}}}/{\partial M_0^*}|_{k_{\mathrm{F}}} = 3\rho^{\mathrm{s}}$.
Since $k_{\mathrm{F}} = (3\pi^2\rho/2)^{1/3}$ depends only on $\rho$ at fixed $M_0^*$, with $\d k_{\mathrm{F}}/\d\rho = k_{\mathrm{F}}/3\rho$, derivatives with respect to $k_{\mathrm{F}}$ can be converted into derivatives with respect to $\rho$ according to $
3\rho{\partial}/{\partial \rho}|_{M_0^*} =3\rho[{\d k_{\mathrm{F}}}/{\d\rho}]{\partial}/{\partial k_{\mathrm{F}}}|_{M_0^*} = k_{\mathrm{F}}{\partial}/{\partial k_{\mathrm{F}}}|_{M_0^*}$.
The scaling identity can therefore be written directly in terms of $(\rho, M_0^*)$ as
\begin{equation}
3\rho\frac{\partial \rho^{\mathrm{s}}}{\partial \rho} + M_0^*\frac{\partial \rho^{\mathrm{s}}}{\partial M_0^*} = 3\rho^{\mathrm{s}}.
\end{equation}
This relation immediately gives the first density derivative. Further differentiation with respect to $M_0^*$ and $\rho$ then gives the mixed and second density derivatives, respectively:
\begin{align}
\frac{\partial \rho^{\mathrm{s}}}{\partial \rho} &= \frac{1}{3\rho}\left(3\rho^{\mathrm{s}} - M_0^* \frac{\partial \rho^{\mathrm{s}}}{\partial M_0^*}\right), ~~
\frac{\partial^2 \rho^{\mathrm{s}}}{\partial \rho \partial M_0^*} = \frac{1}{3\rho}\left(2\frac{\partial \rho^{\mathrm{s}}}{\partial M_0^*} - M_0^* \frac{\partial^2 \rho^{\mathrm{s}}}{\partial M_0^{*,2}}\right),~~
\frac{\partial^2 \rho^{\mathrm{s}}}{\partial \rho^2} = -\frac{M_0^*}{3\rho}\frac{\partial^2 \rho^{\mathrm{s}}}{\partial \rho \partial M_0^*}.
\end{align}

Together, these relations provide all derivatives of the scalar density required to determine $f_0'$ and $f_0''$. In particular, once the derivatives with respect to $M_0^*$ are known, the density and mixed derivatives follow directly from the scaling identity, avoiding their separate evaluation from the full SRC-HMT scalar density. They thus provide a closed set of relations for determining the density derivatives of the scalar mean field in SNM with SRC-HMT.

\section{Density Derivatives of $\widehat{B}$ and $\widehat{D}$ in the FFG Momentum Distribution}
\label{app:B}

In this appendix, we collect the first and second density derivatives
of $\widehat{B}$ and $\widehat{D}$ for the FFG nucleon momentum
distribution. These expressions follow from
Eqs.\,(\ref{eq:B_delta_prime}) to (\ref{eq:D_delta_second})
by taking $C_0=0$ and $\phi_0=1$, for which $\Delta_0=1$
and the HMT contributions vanish. Primes denote total derivatives
with respect to $\rho$, and all coupling constants are independent
of density.
The results are:
\begin{align}
\widehat{B}'
=&\frac{M_0^*}{E_{\rm F}^*}
+\frac{k_{\rm F}^2}{E_{\rm F}^{*,3}}
\left(
\rho M_0^{*\prime}-\frac{M_0^*}{3}
\right),
~~
\widehat{B}''
=\frac{k_{\rm F}^2}{E_{\rm F}^{*,3}}
\Biggl[
\rho M_0^{*\prime\prime}
+\frac{4}{3}M_0^{*\prime}
-\frac{2M_0^*}{9\rho}
-\frac{3\rho M_0^*}{E_{\rm F}^{*,2}}
\left(
M_0^{*\prime}-\frac{M_0^*}{3\rho}
\right)^2
\Biggr],
\\
\widehat{D}'
=&\frac{g_\delta^2}{\pi^2m_\delta^2}
\left\{
2M_0^*M_0^{*\prime}
\left[
\frac{k_{\rm F}(3M_0^{*,2}+k_{\rm F}^2)}
{M_0^{*,2}E_{\rm F}^*}
-3\operatorname{arcsinh}\left(\frac{k_{\rm F}}{M_0^*}\right)
\right]
+\frac{2k_{\rm F}^5}{M_0^*E_{\rm F}^{*,3}}
\left(
\frac{M_0^*}{3\rho}-M_0^{*\prime}
\right)
\right\},
\\
\widehat{D}''
=&\frac{g_\delta^2}{\pi^2m_\delta^2}
\Biggl\{
2\left[
(M_0^{*\prime})^2+M_0^*M_0^{*\prime\prime}
\right]
\left[
\frac{k_{\rm F}(3M_0^{*,2}+k_{\rm F}^2)}
{M_0^{*,2}E_{\rm F}^*}
-3\operatorname{arcsinh}\left(\frac{k_{\rm F}}{M_0^*}\right)
\right]
+\frac{8k_{\rm F}^5M_0^{*\prime}}
{M_0^*E_{\rm F}^{*,3}}
\left(
\frac{1}{3\rho}-\frac{M_0^{*\prime}}{M_0^*}
\right)
\notag\\
&\hspace{1.5cm}+\frac{2k_{\rm F}^5(4M_0^{*,2}+k_{\rm F}^2)}
{M_0^{*,2}E_{\rm F}^{*5}}
\left(
\frac{M_0^*}{3\rho}-M_0^{*\prime}
\right)^2
+\frac{2k_{\rm F}^5}{E_{\rm F}^{*,3}}
\Biggl[
-\frac{2}{9\rho^2}
-\frac{2M_0^{*\prime}}{3\rho M_0^*}
+2\left(\frac{M_0^{*\prime}}{M_0^*}\right)^2
-\frac{M_0^{*\prime\prime}}{M_0^*}
\Biggr]
\Biggr\}.
\end{align}
Here, primes denote total derivatives with respect to $\rho$,
$E_{\rm F}^*=[{k_{\rm F}^2+M_0^{*,2}}]^{1/2}$, and
$M_0^{*\prime}=-g_\sigma f_0'$,
$M_0^{*\prime\prime}=-g_\sigma f_0''$, since $M_0^\ast=M_{\rm N}-g_\sigma f_0$.

\end{widetext}

\section{$\rho$-Meson Contribution to the Symmetry Energy Curvature}
\label{app:D}

In this appendix, we examine the negative contribution of the $\rho$ meson to the symmetry energy curvature in Eq.\,(\ref{eq:Ksym_rho_analytic}), focusing on the HMT and FFG parameter sets near saturation density.
In SNM, the $\omega$ mean field satisfies
$g_\omega\rho=m_\omega^2w_0+c_\omega g_\omega^4w_0^3$,
with
$Q_\omega=m_\omega^2+3c_\omega g_\omega^4w_0^2$
and
$Q_\rho=m_\rho^2+\Lambda_{\rm V}g_\rho^2g_\omega^2w_0^2$.
To simplify Eq.\,(\ref{eq:Ksym_rho_analytic}), we introduce two dimensionless quantities,
\begin{equation}
\Phi\equiv{c_\omega g_\omega^4w_0^2}/{m_\omega^2},
~~
{\rm x}\equiv{\Lambda_{\rm V}g_\rho^2g_\omega^2w_0^2}/{Q_\rho}.
\end{equation}
Here, the factor $\Phi$ measures the strength of the nonlinear $\omega$ self-interaction, while ${\rm x}$ represents the relative contribution of the $\omega$-$\rho$ cross coupling to $Q_\rho$. For positive $c_\omega$ and $\Lambda_{\rm V}$, both quantities are positive, with ${\rm x}<1$.

Using the $\omega$ field equation, we obtain
\begin{equation}
\frac{g_\omega\rho}{w_0Q_\omega}
=\frac{1+\Phi}{1+3\Phi},
~~
\frac{6c_\omega g_\omega^4w_0^2}{Q_\omega}
=\frac{6\Phi}{1+3\Phi}.
\end{equation}
Similarly, the definition of ${\rm x}$ gives
\begin{equation}
\frac{4\Lambda_{\rm V}g_\rho^2g_\omega^3w_0\rho}
{Q_\omega Q_\rho}
=\frac{4{\rm x}(1+\Phi)}{1+3\Phi}.
\end{equation}
Substituting these relations into Eq.\,(\ref{eq:Ksym_rho_analytic}), we obtain
\begin{equation}
K_{\rm{sym}}^\rho(\rho)
=-\frac{9\Lambda_{\rm V}g_\omega^3g_\rho^4w_0\rho^2}
{Q_\omega Q_\rho^2}\Pi(\Phi,{\rm x}),
\end{equation}
where
\begin{equation}
\Pi(\Phi,{\rm x})
=2+\frac{(1+\Phi)(1-3\Phi)}{(1+3\Phi)^2}
-\frac{4{\rm x}(1+\Phi)}{1+3\Phi}.
\end{equation}
For positive $\Lambda_{\rm V}$, the prefactor is negative, so the sign of $K_{\rm{sym}}^\rho(\rho)$ is determined by $\Pi(\Phi,{\rm x})$. The first two terms together are positive for $\Phi>0$, whereas the last term reduces $\Pi$ as ${\rm x}$ increases. Therefore, the condition $\Pi(\Phi,{\rm x})>0$, which ensures $K_{\rm{sym}}^\rho(\rho)<0$, can be written as
\begin{equation}
{\rm x}<
\frac{3+10\Phi+15\Phi^2}
{4(1+\Phi)(1+3\Phi)}.
\end{equation}
Near saturation density, where $\Phi$ is relatively small, the right-hand side is close to $3/4$. Thus, the $\rho$-meson contribution can remain negative even when ${\rm x}$ is not particularly small.

For the HMT and FFG parameter sets in TAB.\,\ref{tab:isoscalar_params}, we find $\Phi\approx0.034$ and $0.067$, and ${\rm x}\approx0.243$ and $0.318$, respectively, at $\rho_0$. Both values of ${\rm x}$ are well below the corresponding critical values, confirming that the $\rho$-meson contribution to $K_{\rm{sym}}\equiv K_{\rm{sym}}(\rho_0)$ is negative in both cases. The smaller ${\rm x}$ in the HMT case also gives a larger $\Pi$, favoring a more negative contribution, although its magnitude depends on the prefactor as well.

In the limiting case $c_\omega=0$, the nonlinear $\omega$ self-interaction vanishes and $\Phi=0$. The expression for $\Pi$ then simplifies to $
\Pi(0,{\rm x})=3-4{\rm x}$,
so that $K_{\rm{sym}}^\rho(\rho)<0$ requires ${\rm x}<3/4$. This simple limit illustrates that the negative $\rho$-meson contribution does not require a nonlinear $\omega$ self-interaction, but can arise from the $\omega$-$\rho$ cross coupling alone.
Physically, the $\omega$-$\rho$ cross coupling causes $Q_\rho$ to increase with density, modifying the density dependence of the $\rho$-meson contribution to the symmetry energy. This produces a negative contribution to its curvature near saturation density for both parameter sets. The more negative $\rho$-meson contribution in the HMT case further reduces $K_{\rm{sym}}$, while the difference in the total curvature also reflects contributions from the kinetic and other interaction terms.

On the other hand, the $\delta$-meson contribution can be understood more directly by examining its contribution to the symmetry energy rather than starting from the more complicated expression for its curvature. In the FFG case, the $\delta$-meson contribution takes the form of Eq.\,(\ref{eq:Esym_delta_FFG}). The negative sign reflects the attractive contribution of the $\delta$ meson to the symmetry energy.
To understand its density dependence, consider the low-density limit $\theta=k_{\rm F}/M_0^*\ll1$. In this limit, $\widehat D\approx1+\mathcal{O}(\rho^{5/3})$, and the symmetry energy contribution reduces to
\begin{equation}
E_{\rm{sym}}^\delta(\rho)
\approx-\frac{g_\delta^2\rho}{2m_\delta^2}
\left(1-\frac{k_{\rm F}^2}{M_0^{*,2}}\right).
\end{equation}
This simple expansion reveals two competing effects. The leading term is negative and linear in $\rho$, corresponding to the non-relativistic contribution of the attractive isovector-scalar interaction. Since its second density derivative vanishes, it does not contribute to $K_{\rm{sym}}(\rho)$. The next term is positive and proportional to $\rho k_{\rm F}^2$, which scales as $\rho^{5/3}$. Consequently, this relativistic correction produces a positive contribution to the curvature. In particular, the leading low-density behavior is
\begin{equation}
K_{\rm{sym}}^\delta(\rho)
\approx\frac{5g_\delta^2}{m_\delta^2}
\frac{\rho k_{\rm F}^2}{M_0^{*,2}}>0.
\end{equation}

The positive curvature has a simple physical origin. As density increases, relativistic nucleon motion suppresses the scalar density relative to the baryon density, weakening the attractive $\delta$-meson contribution compared with its linear density dependence. This produces a positive curvature even though the contribution to the symmetry energy remains negative.

The same mechanism applies to the SRC-HMT case. At low densities, one still has $B\approx\rho$ and $D\approx1$, so the leading contribution remains $-g_\delta^2\rho/2m_\delta^2$. However, high momentum nucleons and the isospin dependence of SRC modify the relativistic correction proportional to $\rho^{5/3}$, potentially enhancing or reducing the curvature. Near saturation density, the density dependence of the effective mass and scalar field response becomes important, requiring the full relativistic calculation to determine the sign and magnitude of $K_{\rm{sym}}^\delta(\rho)$.

These analyses show that including the $\delta$ meson in the Lagrangian does not necessarily help reduce $K_{\rm{sym}}(\rho)$, despite its negative contribution to the symmetry energy. In contrast, SRC-HMT can enhance the negative $\rho$-meson contribution to the curvature, highlighting the essential role of SRC in producing a softer density dependence of the symmetry energy.

\end{document}